\documentclass[12pt]{article}
\usepackage{graphicx,epsfig,psfig,color,latexsym}
\renewcommand{\theequation}{\arabic{section}.\arabic{equation}}
\newcommand{\bc}{\begin{center}}
\newcommand{\ec}{\end{center}}
\newcommand{\be}{\begin{equation}}
\newcommand{\ee}{\end{equation}}
\newcommand{\bea}{\begin{eqnarray}}
\newcommand{\eea}{\end{eqnarray}}
\newcommand{\ba}{\begin{array}}
\newcommand{\ea}{\end{array}}
\newcommand{\lb}{\label}
\newcommand{\rf}{\ref}
\newcommand{\bfg}{\begin{figure}[htbp]}
\newcommand{\efg}{\end{figure}}
\newcommand{\pr}{Phys. Rev. }
\newcommand{\np}{Nucl. Phys. }
\newcommand{\prl}{Phys. Rev. Lett. }
\newcommand{\prp}{Phys. Rep. }
\newcommand{\ap}{Ann. Phys. (N.Y.) }
\newcommand{\pl}{Phys. Lett. }

\newcommand{\rmp}{Rev. Mod. Phys. }
\newcommand{\nc}{Nuovo Cimento }
\newcommand{\ptp}{Prog. Theor. Phys. }
\newcommand{\zp}{Z. Phys. }

\begin{document}

\begin{flushright}
IPNO-DR-03-02
\end{flushright}
\vspace{0.5 cm}
\bc
{\large \textbf{Bound state equation in the Wilson loop approach\\
with minimal surfaces}}
\vspace{1. cm}

F. Jugeau\footnote{\textit{And Universit\'e de Cergy-Pontoise. 
E-mail: jugeau@ipno.in2p3.fr}}
and H. Sazdjian\footnote{\textit{E-mail: sazdjian@ipno.in2p3.fr}}\\
\textit{Groupe de Physique Th\'eorique,
Institut de Physique Nucl\'eaire\footnote{Unit\'e Mixte de Recherche
8608.},\\
Universit\'e Paris XI,\\
F-91406 Orsay Cedex, France}
\ec
\par
\renewcommand{\thefootnote}{\fnsymbol{footnote}}
\vspace{0.75 cm}

\begin{center}
{\large Abstract}
\end{center}
The large-distance dynamics in quarkonium systems is investigated, in the
large $N_c$ limit, through the saturation of Wilson loop averages by minimal 
surfaces. Using a representation for the quark propagator in the presence 
of the external gluon field based on the use of path-ordered phase factors,
a covariant three-dimensional bound state equation of the Breit--Salpeter
type is derived, in which the interaction potentials are provided by the
energy-momentum vector of the straight segment joining the quark to the
antiquark and carrying a constant linear energy density, equal to the string
tension. The interaction potentials are confining and reduce to the linear
vector potential in the static case and receive, for moving quarks,
contributions from the moments of inertia of the straight
segment. The self-energy parts of the quark propagators induce 
spontaneous breakdown of chiral symmetry with a mechanism identical to
that of the exchange of one Coulomb-gluon. In the nonrelativistic
limit, long range spin-spin potentials are absent; the moments of inertia
of the straight segment provide negative contributions to the spin-orbit
potentials going in the opposite direction to those of the pure 
timelike vector potential. In the ultrarelativistic limit, the mass
spectrum displays linear Regge trajectories with slopes in agreement with
their classical relationship with the string tension.
\par
\vspace{0.5 cm}
\newpage

\section{Introduction} \lb{s1}
\setcounter{equation}{0}

The Wilson loop \cite{w} appears as one of the most efficient tools for
probing the large-distance properties of QCD. It provides a natural 
criterion for confinement through the area law and also participates as
a basic ingredient in the formulation of lattice gauge theories \cite{w,k}.
Equations concerning path-ordered phase factors were first obtained by
Mandelstam \cite{m1} and analyzed in the QCD case by Nambu \cite{nm}.
Loop equations were obtained by Polyakov \cite{p} and Makeenko and Migdal 
\cite{mm1,mm2}. (For reviews, see Refs. \cite{mgd} and \cite{mk}.) The
loop equations actually represent an infinite chain of coupled equations 
relating vacuum expectation values of loop operators having as supports 
different numbers of closed contours emerging from contours with
self-intersections.
These equations must in addition be supplemented with constraint equations 
\cite{m2,g}, the so-called Mandelstam constraints, which are sensitive to the
gauge group structure of the theory. Due to their complexity, the loop
equations have not yet allowed a systematic resolution of QCD in terms of
loop variables. In the two-dimensional case however, explicit expressions of
the Wilson loop averages for various types of contour have been obtained for
$U(N_c)$ gauge theories \cite{kkk}.
Renormalization properties of the Wilson loop averages were studied in
the framework of perturbation theory in Refs. \cite{dv,bnsg}, where it was
shown that the latter are multiplicatively renormalizable.
\par
Considerable simplification is obtained in the large-$N_c$ limit \cite{th1},
corresponding to the planar diagram approximation of the theory. In that
case, apart from the disappearance of many nonleading terms, it is the
factorization property of the Wilson loop average for two disjoint contours
that becomes mostly relevant. Makeenko and Migdal studied the resulting
equations in the above limit pointing out their equivalence with a chain
of Schwinger--Dyson type equations \cite{mm1,mm2}. They showed that for
large contours asymptotic solutions exist correponding to the minimal 
surfaces bounded by the loops.
\par
Concerning physical applications, Eichten and Feinberg, using the area law
for the Wilson loop, could obtain the general expression of spin-dependent
forces for quark-antiquark systems to order $1/c^2$ in terms of color 
electric and magnetic field correlators \cite{ef}. This problem was also
investigated by Gromes \cite{gr}. Later on, Prosperi, Brambilla 
\textit{et al.} completed these results by also obtaining the 
velocity-dependent forces \cite{bmpbbp,bcpbmp,bv1,bp}. On the other hand, 
a wide program of investigations was undertaken by Dosch, Simonov 
\textit{et al.} in the framework of the ``stochastic vacuum model'' 
\cite{dssdg}.
\par
In recent years, the Wilson loop gained renewal of interest in connection
with duality properties of different field theories in different
dimensions manifested through the $AdS$/CFT correspondence \cite{mwdgopr}.
\par
The purpose of the present paper is to investigate the properties of the
Wilson loop concerning the bound state problem in QCD in the large-distance
regime when the latter is probed by minimal surfaces. Among possible
nonperturbative solutions of the loop equations at large distances, minimal
surfaces appear as the most natural ones \cite{mm1,mm2}; they produce in a 
simple way the area law needed for confinement and satisfy the 
factorization property, valid at large $N_c$. A complete solution of the loop
equations should necessarily include short-distance effects for which
minimal surfaces do not seem sufficient alone to provide the appropriate
behavior. Therefore, for that regime perturbation theory results should
appropriately be incorporated in the Wilson loop solution; this aspect of 
the problem will not, however, be analyzed in the present paper. Other
nonperturbative properties of the theory, related to its possible string-like
behavior, might also arise from contributions of fluctuations of surfaces 
around the minimal surface bounded by the loop \cite{eg,l,lsw}.
\par
The main ingredient in our approach is a representation of the quark 
propagator in the presence of an external gluon field as a series
of terms involving free propagators, path-ordered phase factors along
straight lines and their derivatives. That representation generalizes
to the relativistic case the one used in the nonrelativistic limit
\cite{ef}. The gauge invariant quark-antiquark Green function is then 
represented by a series of terms involving Wilson loops having as 
contours skew polygons with an increasing number of sides. Each 
Wilson loop average is then replaced by the contribution of the
corresponding minimal surface bounded by the loop.
Contrary to the usual two-particle Green functions,
the gauge invariant Green function does not manifestly satisfy a genuine
integral equation which might, as in the Bethe--Salpeter equation case 
\cite{sb}, result in a bound state integral equation. In the present case,
the Green function satisfies with the Dirac operators two independent and
compatible equations. Selecting in the large time limit the total momentum
of a bound state \cite{gml} and taking in the center-of-mass frame
the equal-time limit for the two particles allow, with certain 
mathematical assumptions, the grouping of terms into 
a form that leads to a three-dimensional Breit--Salpeter type wave equation
\cite{b,s}, where the potentials are represented by 
the components of the energy-momentum vector of the straight segment joining 
the quark to the antiquark and carrying a constant linear energy density,
equal to the string tension; they involve, apart from the usual confining
linear potential, contributions coming from the moments of inertia of the
above segment, which plays the role of the color flux tube of the 
quarkonium system.
\par
The self-energy parts of the quark propagators, extracted from the
interaction terms, allow the analysis of the chiral symmetry properties of 
the system. The situation here is very similar to that resulting from 
the exchange of one Coulomb-gluon and the latter has been studied in
the literature. The Schwinger--Dyson equation satisfied by the 
self-energy part has a non-trivial solution leading to a spontaneous
breakdown of chiral symmetry.
\par
From the nonrelativistic limit of the wave equation one determines
the hamiltonian of the system to order $1/c^2$. Spin-spin potentials
are absent from the hamiltonian. The contributions of
the color flux tube result in new terms for the orbital angular momentum
and for the spin-orbit potentials. In particular, the momentum of the flux
tube contributes with a negative sign to the spin-orbit potential,
in opposite direction to the contributions of the pure timelike vector
potential, and may account for the phenomenological observations made
for fine splitting.
\par
At high energies, the mass spectrum displays linear Regge trajectories,
the slopes of which tend to satisfy the classical relationship with
the string tension.
\par
The paper is organized as follows. In Sec. 2, we review the equations 
satisfied by the Wilson loop averages and outline the particular status
of minimal surfaces within the set of possible solutions. In Sec. 3, we 
display some basic properties of minimal surfaces. In Sec. 4, the 
representation of the quark propagator in the presence of an external gluon 
field is constructed. Section 5 is devoted to the construction of the 
representation of the gauge invariant two-particle Green function in terms 
of a series of Wilson loops having as boundaries skew polygons. The bound
state equation is derived in Sec. 6. Section 7 is devoted to the extraction
from the interaction terms of the self-energy parts needed for the quark
propagators. Chiral symmetry breaking is studied in Sec. 8. In Sec. 9, 
the main qualitative properties of the bound state spectrum are displayed.
Summary and concluding remarks follow in Sec. 10. In appendix A, the main
properties of minimal surfaces are presented. Appendix B is devoted to the 
determination of the normalization condition of the wave function.
Appendix C gives details about the Breit approximation used for the
resolution of the bound state spectrum.
\par

\section{Loop equations} \lb{s2}
\setcounter{equation}{0}

The starting point is the gauge covariant path-ordered phase factor along 
a line $C_{yx}$ joining point $x$ to point $y$\footnote{Formulas of this 
section are written in Minkowski space.}:
\be \lb{2e1}
U(C_{yx},y,x)\equiv U(y,x)=Pe^{{\displaystyle -ig\int_x^y 
dz^{\mu}A_{\mu}(z)}},
\ee
where $A_{\mu}=\sum_a A_{\mu}^at^a$, $A_{\mu}^a$ ($a=1,\ldots,N_c^2-1$) being
the gluon fields and $t^a$ the generators of the gauge group $SU(N_c)$ in the
fundamental representation, with the normalization
tr$t^at^b=\frac{1}{2}\delta^{ab}$. A more detailed  definition of $U$
is given by the series expansion in the coupling constant $g$; all
equations involving $U$ can be obtained from the latter. Parametrizing
the line $C$ with a parameter $\sigma$, $C=\{x(\sigma)\}$,
$0\leq \sigma \leq 1$, such that $x(0)=x$ and $x(1)=y$, a variation of $C$
induces the following variation of
$U$ [$U(x(\sigma),x(\sigma'))\equiv U(\sigma,\sigma')$,
$A(x(\sigma))\equiv A(\sigma)$]:
\bea \lb{2e2}
\delta U(1,0)&=&-ig\delta x^{\alpha}(1)A_{\alpha}(1)U(1,0)
+igU(1,0)A_{\alpha}(0)\delta x^{\alpha}(0)\nonumber \\
& &+ig\int_0^1d\sigma U(1,\sigma)x^{\prime \beta}(\sigma)F_{\beta\alpha}
(\sigma)\delta x^{\alpha}(\sigma)U(\sigma,0),
\eea
where $x'=\frac{\partial x}{\partial \sigma}$ and $F$ is the field
strength, $F_{\mu\nu}=\partial_{\mu}A_{\nu}-\partial_{\nu}A_{\mu}
+ig[A_{\mu},A_{\nu}]$. The functional derivative
of $U$ with respect to $x(\sigma)$ ($0< \sigma <1$) is then \cite{p}:
\be \lb{2e3}
\frac{\delta U(1,0)}{\delta x^{\alpha}(\sigma)}=
igU(1,\sigma)x^{\prime \beta}(\sigma)F_{\beta\alpha}(\sigma)U(\sigma,0).
\ee
Defining the ordinary derivations with the prescriptions \cite{p}
\bea
\lb{2e4}
\frac{\partial}{\partial x^{\mu}(\sigma)}&=&
\lim_{\varepsilon\rightarrow 0}\int_{\sigma-\varepsilon}^
{\sigma+\varepsilon}d\sigma'\frac{\delta}{\delta x^{\mu}(\sigma')},
\\
\lb{2e5}
\frac{\partial}{\partial x^{\prime \mu}(\sigma)}&=&
\lim_{\varepsilon\rightarrow 0}\int_{\sigma-\varepsilon}^
{\sigma+\varepsilon}d\sigma'(\sigma'-\sigma)
\frac{\delta}{\delta x^{\mu}(\sigma')},
\eea
one obtains:
\be \lb{2e6}
\frac{\partial}{\partial x^{\prime\beta}(\sigma)}
\frac{\delta U(1,0)}{\delta x^{\alpha}(\sigma)}=
igU(1,\sigma)F_{\beta\alpha}(\sigma)U(\sigma,0).
\ee
A derivation with respect to $x^{\mu}(\sigma)$ then yields:
\be \lb{2e7}
\frac{\partial}{\partial x^{\mu}(\sigma)}
\frac{\partial}{\partial x^{\prime\beta}(\sigma)}
\frac{\delta U(1,0)}{\delta x^{\alpha}(\sigma)}=
igU(1,\sigma)(\nabla_{\mu}F_{\beta\alpha}(\sigma))U(\sigma,0),
\ee
where $\nabla$ is the covariant derivative,
$(\nabla F)=(\partial F)+ig[A,F]$.
\par
The Wilson loop, denoted $\Phi(C)$, is defined as the trace in color space
of the path-ordered phase factor (\rf{2e1}) along a closed contour $C$:
\be \lb{2e8}
\Phi(C)=\frac{1}{N_c}\mathrm{tr}Pe^{{\displaystyle -ig\oint_C
dx^{\mu}A_{\mu}(x)}},
\ee
where the factor $1/N_c$ has been put for normalization. It is a gauge
invariant quantity. Its vacuum expectation value is denoted $W(C)$:
\be \lb{2e9}
W(C)=\langle\Phi(C)\rangle_A,\ \ \ \ \ \
W(C_1,C_2)=\langle\Phi(C_1)\Phi(C_2)\rangle_A,
\ee
the averaging being defined in the path integral formalism.
More generally, one meets insertions of local operators $O(x)$ into the 
Wilson loop. Their vacuum expectation values are:
\be \lb{2e9a}
\langle O(x)\rangle_W\equiv \langle \frac{1}{N_c}\mathrm{tr}
P\Big(e^{{\displaystyle -ig\oint_Cdz^{\mu}A_{\mu}(z)}}O(x)\Big)
\rangle_A\bigg|_{x\in C}. 
\ee
\par 
The Wilson loop and its average also satisfy equations analogous to Eqs.
(\rf{2e3}), (\rf{2e6}) and (\rf{2e7}). Considering in Eq. (\rf{2e7}) three
different indices and taking their cyclic permutations one obtains in the 
right-hand side the Bianchi identity. For the Wilson loop average the
equation takes the form
\be \lb{2e10}
\varepsilon^{\nu\mu\beta\alpha}
\frac{\partial}{\partial x^{\mu}}
\frac{\partial}{\partial x^{\prime\beta}}
\frac{\delta W(C)}{\delta x^{\alpha}}=0.
\ee
\par
Contraction of the indices $\mu$ and $\beta$ in Eq. (\rf{2e7}) leads in
the right-hand side to the equation of motion of the gluon field. In the
large $N_c$ limit the quark current can be neglected and the corresponding
term becomes equivalent to the functional derivative
$-\delta/\delta A^{\alpha}$ which acts now on the Wilson loop. 
[Gauge-fixing and ghost terms mutually cancel each other in gauge invariant
quantities \cite{dv,bnsg} and hence can be ignored.] One finds:
\be \lb{2e11}
\frac{\partial}{\partial x^{\beta}}
\frac{\partial}{\partial x'_{\beta}}
\frac{\delta W(C)}{\delta x_{\alpha}}=
i\frac{g^2N_c}{2}\oint_Cdy^{\alpha}\delta^4(y-x)\Big[W(C_{yx},C_{xy})-
\frac{1}{N_c^2}W(C)\Big].
\ee
Because of the delta function, the points $y$ that contribute to the
integral are those that lie in the vicinity of the point x. Two cases may
emerge. First, the contour $C$ may have self-intersection points and point
$x$ may be one of them. In that case, there is a point $y$ which coincides
with $x$ in space but is not the same one on the contour. The contours
$C_{yx}$ and $C_{xy}$ are the complementary contours separated by the
intersection point and contribute as independent closed contours to the
Wilson loop. Second, there is always the contribution of the points $y$
which are close to $x$ in space and on the contour. In that case, one of
the contours $C_{yx}$ or $C_{xy}$ shrinks to zero and the other one becomes
identical to $C$. In the large $N_c$ limit, the Wilson loop average of the
the two contours $C_{yx}$ and $C_{xy}$ factorizes into two loop averages
and Eq. (\rf{2e11}) becomes:
\be \lb{2e12}
\frac{\partial}{\partial x^{\beta}}
\frac{\partial}{\partial x'_{\beta}}
\frac{\delta W(C)}{\delta x_{\alpha}}=
i\frac{g^2N_c}{2}\oint_Cdy^{\alpha}\delta^4(y-x)W(C_{yx})W(C_{xy}).
\ee
[The product $g^2N_c$ is maintained fixed and finite in the above limit
\cite{th1}.]
\par
Equations (\rf{2e10}) and (\rf{2e12}) can be considered as basic
equations of QCD in the large $N_c$ limit in loop space.
The factorized structure of the right-hand side of Eq. (\rf{2e12}) 
puts severe restrictions on the class of its possible solutions. Potential
solutions of that equation with contours without self-intersections may not
be solutions for contours with self-intersections. On the other hand, many 
physical applications related to meson spectroscopy do not require 
consideration of contours with self-intersections (in the static limit).
Due to the huge complexity of the treatment of the problem with general 
types of contour, it seems reasonable, in view of physical applications, 
to limit, at a first stage, the investigations to contours not having 
self-intersections. We shall adopt this limitation in the present paper,
indicating at the end of this section the way of incorporating contours 
with self-intersections. Equation (\rf{2e12}) then becomes: 
\be \lb{2e13}
\frac{\partial}{\partial x^{\beta}}
\frac{\partial}{\partial x'_{\beta}}
\frac{\delta W(C)}{\delta x_{\alpha}}=
i\frac{g^2N_c}{2}\oint_Cdy^{\alpha}\delta^4(y-x)W(C).
\ee
The points $y$ that contribute to the integral are those that lie in the
vicinity of the point $x$ in space and on the contour.
\par
A first class of solution to Eqs. (\rf{2e10}) and (\rf{2e12}) is
provided by perturbation theory \cite{dv,bnsg}. In the right-hand side of
Eq. (\rf{2e12}), $g^2N_c$ represents the unrenormalized coupling constant
and the resolution of the equation is accompanied by the renormalization
of the theory and of the Wilson loop itself. However, perturbation
theory becomes unstable at large distances and the search for
nonperturbative solutions becomes necessary. Among these, minimal surfaces
appear, for several reasons, as natural candidates. First, they easily
reproduce the area law related to confinement \cite{w}; second, they
satisfy, with certain restrictions, the factorization law in the large $N_c$
limit; third, they are connected to the classical action of string theory, 
which in turn is expected to have an implicit relationship with QCD 
\cite{nm,eg,l,lsw}. In the following, we concentrate on the contributions 
of surfaces to Eqs. (\rf{2e10}) and (\rf{2e13}).
\par
Let $S$ be a surface of a given type, that is satisfying a given
equation, and having as contour the closed loop $C$. Let $A$ be its
area. Then a representation of the Wilson loop average can be given by 
the following expression:
\be \lb{2e14}
W(C)=e^{{\displaystyle -i\sigma A(S,C)}},
\ee
where $\sigma$ is a constant, which will be identified with the string
tension. Let us consider Eq. (\rf{2e10}). It represents the action of
some local deformation of the contour at point $x$. However, any local
deformation on the contour also induces an internal deformation of the
surface itself, because the latter is constrained by its defining equation
and by the boundary condition. It turns out that the Bianchi identity 
operator, present in Eq. (\rf{2e10}), is represented in the internal part 
of the surface (and on its contour) by the defining operator of minimal 
surfaces (surfaces having minimal area). Therefore, Eq. (\rf{2e10}) implies
that among surfaces of a given type, only minimal surfaces can be solutions
of it. [The details of this assertion will be presented in the Sec. \rf{s3}.]
This result considerably reduces the class of surface type solutions to
the loop equations. In order to incorporate other types of surface in
representation (\rf{2e14}) there remains the possibility of considering
contributions of an infinite sum of all possible surfaces, which might,
through mutual cancellations, satisfy Eq. (\rf{2e10}). This latter
possibility was considered in Ref. \cite{mm2}. Such a sum can also be
considered as representing fluctuations of surfaces around the minimal
surface and corresponding to string-like contributions \cite{eg,l,lsw}. 
Henceforth, we restrict ourselves to the study of contributions to Eq.
(\rf{2e14}) coming from minimal surfaces.
\par
To study Eq. (\rf{2e13}), one must introduce a short-distance regulator
$a$ since the right-hand side displays a singular behavior. Using for $W(C)$
representation (\rf{2e14}) with a minimal surface one finds that Eq.
(\rf{2e13}) is satisfied provided one has the identification
\be \lb{2e15}
\sigma=\frac{C}{a^2}\frac{g^2N_c}{2},
\ee
where $C$ is a numerical constant. Notice that the use of an arbitrary
trial functional for $W(C)$ would produce in general for the right-hand side
of Eq. (\rf{2e13}) a multiplicative functional depending on the whole contour
$C$, rather than a local factor depending only on $x$ and on its vicinity.
If the unrenormalized coupling constant $g^2N_c$
vanished as $O(a^2)$ with vanishing $a$, then minimal surfaces would
define a theory by their own with a finite dimensional coupling constant
$\sigma$. However, we know from the short-distance behavior of QCD, given
by the perturbative solution, that the unrenormalized coupling constant
vanishes only logarithmically. Therefore condition
(\rf{2e15}) cannot be satisfied in general. One must then interpret
the minimal surface contribution to the Wilson loop average as a part
of a general solution in which it represents its large-distance
behavior, while the other part includes the perturbation theory
contribution representing its short-distance behavior. A matching
condition between the two parts at intermediate distances should then
fix the value of the string tension $\sigma$ in terms of the parameter
$\Lambda$ of QCD. We shall not study this aspect of the problem in the 
present paper, but rather shall assume that such a matching condition 
exists and is fulfilled and shall henceforth consider the properties of the
large-distance behavior of the theory as deduced from the minimal surface
contribution to the Wilson loop average.
\par
Finally, let us comment on the factorization property of minimal surfaces.
The minimal surface of two disconnected closed contours lying sufficiently
far from each other is the sum of the minimal surfaces of each contour.
This ensures the factorization property of the Wilson loop average valid
at large $N_c$. However, when the two contours are close to each other
new global solutions may arise that do not lead to factorization. Thus,
for instance, if the two contours are similar to each other and lie closely
one above the other, forming the two bases of a cylinder, then the global
(or absolute) minimal surface corresponding to the two contours is the area
of the lateral surface of the cylinder rather than the sum of the areas of
its two bases, a result that manifestly does not lead to factorization. In
such cases and even for single contours having complicated forms, the
minimal surface solution should be chosen locally (therefore not 
necessarily the absolute one) with global properties being in agreement with
factorization or with other physical requirements.
\par
Extension of that prescription to the case of contours with 
self-intersections allows minimal surfaces to also satisfy the 
factorization property of the right-hand side of Eq. (\rf{2e12}). 
However, when such contours lie on overlapping surfaces additional 
contributions might be necessary to consider, since internal contours then
belong to two adjacent surfaces. This is in particular the case in 
two-dimensional QCD \cite{kkk}.
\par
In the next section, we shall study properties of minimal surfaces,
showing some of the results mentioned above and deriving results 
needed for later investigations.
\par

\section{Loop equations and minimal surfaces} \lb{s3}
\setcounter{equation}{0}

Let $C$ be a closed contour and $S$ a surface bounded by it. We shall
generally consider as prototypes of contour those having four distinct
sides, with junction points designated by $x_1$, $x_1'$, $x_2'$, $x_2$
(see Fig. \rf{3f1}).
\par
\bfg
\vspace*{0.5 cm}
\bc
\input{3f1.pstex_t}
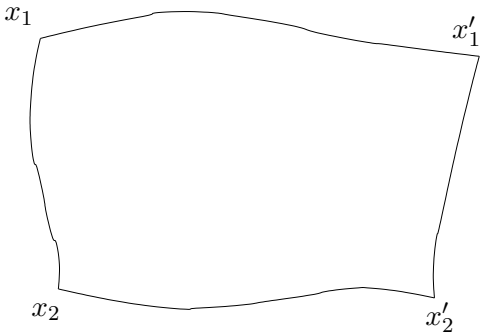
\caption{Prototype of a closed contour with four distinct sides.}
\lb{3f1}
\ec
\efg
The surface bounded by this contour will be parametrized with two
parameters, $\sigma$ and $\tau$, and a point belonging to it will be
represented as $x(\sigma,\tau)$ or $y(\sigma,\tau)$. We adopt for the
partial derivatives the usual notations
\be \lb{3e1}
x'=\frac{\partial x}{\partial \sigma},\ \ \ \ 
\dot x=\frac{\partial x}{\partial \tau}.
\ee
In general, the four sides of the contour will be parametrized as
follows: $\{x_1x_1'\}$ with $\sigma=0$, $\{x_1'x_2'\}$ with $\tau=1$,
$\{x_2'x_2\}$ with $\sigma=1$ and $\{x_2x_1\}$ with $\tau=0$.
The area of the surface is\footnote{Formulas related to surfaces will in
general be written in euclidean space.}:
\be \lb{3e2}
A(C,S)=\int_0^1d\sigma\int_0^1d\tau
\Big(x^{\prime 2}\dot x^2-(x'.\dot x)^2\Big)^{1/2}.
\ee
\par
To avoid the occurrence of possible divergences when dealing with functional
derivatives of the area, it is necessary to introduce a short-distance
regulator in the expression of the area. We adopt for the latter the 
following expression, already suggested in a more general form in Ref.
\cite{mm2}:
\bea \lb{3e3}
& &A=\frac{1}{2}\int d\sigma^{\mu\nu}(x)\int d\sigma^{\mu\nu}(y)F(x-y),
\nonumber\\
& &d\sigma^{\mu\nu}(x)=d\sigma d\tau 
(x^{\prime \mu}\dot x^{\nu}-x^{\prime \nu}\dot x^{\mu}),\nonumber\\
& &F=\frac{a^2}{\pi}\frac{1}{\Big((x-y)^2+a^2\Big)^2},
\eea
where $a$ is the (positive) regulator and goes to zero at the end of
calculations.
In the limit $a\rightarrow 0$, $F$ is actually equal to a two-dimensional
$\delta$-function:
\be \lb{3e3a}
\lim_{a\rightarrow 0}F=
\Big(x^{\prime 2}\dot x^2-(x'.\dot x)^2\Big)^{-1/2}
\delta(\sigma-\sigma')\delta(\tau-\tau').
\ee
This implies that it is only points $x$ and $y$ lying close to each 
other that contribute to the integral and therefore a
limited number of terms of the expansion of $y$ about $x$ have to be 
considered in general. For example, one finds the following limits:
\be \lb{3e4}
\frac{a^2}{\pi}\int d\sigma' d\tau' \frac{1}
{\Big((y(\sigma',\tau')-x(\sigma,\tau))^2+a^2\Big)^2}=
\left\{
\ba{ll}
\Big(x^{\prime 2}\dot x^2-(x'.\dot x)^2\Big)^{-1/2},&
\mathrm{for}\ x\ \mathrm{inside}\ S,\\    
\frac{1}{2}\Big(x^{\prime 2}\dot x^2-(x'.\dot x)^2\Big)^{-1/2},&
\mathrm{for}\ x\ \mathrm{on}\ C.
\ea
\right.\nonumber \\
\ee
\par
We are interested in variations of the area of the surface $S$ when local
variations are introduced on its contour $C$. $S$ being a surface of a
given type, that is satisfying a defining equation, any deformation of
the contour introduces corresponding deformations inside the surface.
Let us for definiteness consider deformations on the line $\tau=0$.
The general deformation of the area is:
\bea \lb{3e5}
\delta A&=&\int d\sigma \delta x^{\alpha}(\sigma,0)
(\delta_{\mu\alpha}x^{\prime \nu}(\sigma,0)-\delta_{\nu\alpha}
x^{\prime \mu}(\sigma,0))\int d\sigma^{\mu\nu}(y)
F(y(\sigma',\tau')-x(\sigma,0)) \nonumber\\
& &+\int d\sigma^{\mu\nu}(y)\delta y^{\lambda}(\sigma',\tau')
\sum_{P(\lambda,\mu,\nu)}\frac{\partial}{\partial y^{\lambda}}
\int d\sigma^{\mu\nu}(z)F(z-y),
\eea
where $P(\lambda,\mu,\nu)$ indicates a cyclic permutation of the
indices $\lambda$, $\mu$, $\nu$ inside the sum. The functional
derivative of $A$ with respect to $x(\sigma,0)$ is then:
\bea \lb{3e6}
\frac{\delta A}{\delta x^{\alpha}(\sigma,0)}&=&
(\delta_{\mu\alpha}x^{\prime \nu}(\sigma,0)-\delta_{\nu\alpha}
x^{\prime \mu}(\sigma,0))\int d\sigma^{\mu\nu}(y)
F(y(\sigma',\tau')-x(\sigma,0)) \nonumber\\
& &+\int d\sigma^{\mu\nu}(y)\frac{\delta y^{\lambda}(\sigma',\tau')}
{\delta x^{\alpha}(\sigma,0)}
\sum_{P(\lambda,\mu,\nu)}\frac{\partial}{\partial y^{\lambda}}
\int d\sigma^{\mu\nu}(z)F(z-y).
\eea
The quantity
$\delta y^{\lambda}(\sigma',\tau')/\delta x^{\alpha}(\sigma,0)$
depends on the type of surface that is considered.
\par
Applying on both sides of Eq. (\rf{3e6}) the operators
$\partial /\partial x^{\prime \beta}(\sigma,0)$ and
$\partial /\partial x^{\gamma}(\sigma,0)$ in successive order, one
obtains:
\bea \lb{3e7}
& &\frac{\partial}{\partial x^{\gamma}(\sigma,0)}
\frac{\partial}{\partial x^{\prime\beta}(\sigma,0)}
\frac{\delta A}{\delta x^{\alpha}(\sigma,0)}=
2\frac{\partial}{\partial x^{\gamma}(\sigma,0)}
\int d\sigma^{\alpha\beta}(y)
F(y(\sigma',\tau')-x(\sigma,0)) \nonumber\\
& &\ \ \ \ \ \ +\int d\sigma^{\mu\nu}(y)
\left(\frac{\partial}{\partial x^{\gamma}(\sigma,0)}
\frac{\partial}{\partial x^{\prime\beta}(\sigma,0)}
\frac{\delta y^{\lambda}(\sigma',\tau')}{\delta x^{\alpha}(\sigma,0)}
\right)\sum_{P(\lambda,\mu,\nu)}\frac{\partial}{\partial y^{\lambda}}
\int d\sigma^{\mu\nu}(z)F(z-y).\nonumber \\
& &
\eea
In order to satisfy the Bianchi identity (\rf{2e10}) with representation
(\rf{2e14}), the sum of the cyclic permutation of the left-hand side of
Eq. (\rf{3e7}) should vanish:
\be \lb{3e8}
\sum_{P(\gamma,\beta,\alpha)}\frac{\partial}{\partial x^{\gamma}(\sigma,0)}
\frac{\partial}{\partial x^{\prime\beta}(\sigma,0)}
\frac{\delta A}{\delta x^{\alpha}(\sigma,0)}=0.
\ee
This involves through the right-hand side of Eq. (\rf{3e7}) the sum
$\sum_{P(\gamma,\beta,\alpha)}\frac{\partial}{\partial x^{\gamma}}
\frac{\partial}{\partial x^{\prime\beta}}\frac{\delta y^{\lambda}}
{\delta x^{\alpha}}$. 
For arbitrary contours, that quantity does not identically vanish.
Therefore, the only solution to Eq. (\rf{3e8}) is the constraint
\be \lb{3e9}
\sum_{P(\lambda,\mu,\nu)}\frac{\partial}{\partial x^{\lambda}}
\int d\sigma^{\mu\nu}(y)F(y-x)=0,\ \ \ x\ \mathrm{inside}\ S\ 
\mathrm{or\ on}\ C,
\ee
which is the defining equation of minimal surfaces.
\par
In order to check Eq. (\rf{2e13}), we limit ourselves to the case of minimal
surfaces. In that case, the second term of the right-hand side of Eq.
(\rf{3e7}) is null and the first term gives after contraction of the indices
$\gamma$ and $\beta$:
\be \lb{3e10}
\frac{\partial}{\partial x^{\beta}}
\frac{\partial}{\partial x^{\prime\beta}}
\frac{\delta A}{\delta x^{\alpha}}=
\frac{1}{a}\frac{x^{\prime\alpha}}{(x^{\prime 2})^{1/2}}+O(a).
\ee
On the other hand, regularizing the delta function of Eq. (\rf{2e13})
according to the prescription 
\be \lb{3e11}
\delta^4(x)\rightarrow -\frac{1}{4\pi^2}\partial^2\frac{1}{x^2+a^2},
\ee
one obtains, with the proper-time parametrization ($x^{\prime 2}=$constant):
\be \lb{3e12}
\int dy^{\alpha}\delta^4(y-x)=\frac{1}{4\pi a}\left(\frac{3}{a^2}
\frac{x^{\prime\alpha}}{(x^{\prime 2})^{1/2}}+
\frac{x^{\prime\prime\prime\alpha}}{(x^{\prime 2})^{3/2}}+
\frac{x^{\prime\alpha}x^{\prime\prime 2}}{(x^{\prime 2})^{5/2}}\right)
+O(a).
\ee
Use of representation (\rf{2e14}) and comparison of both sides of Eq.
(\rf{2e13}) (in its euclidean version) gives for finite $\sigma$ the
identification (\rf{2e15}) with $C=3/(4\pi)$, which necessitates a
quadratically vanishing $g^2N_c$ with $a$ when short-distance interactions
are ignored.
\par
Coming back to the first-order variation of the area [Eq. (\rf{3e6})]
it becomes in the case of a minimal surface:
\bea \lb{3e13}
\frac{\delta A}{\delta x^{\alpha}(\sigma,0)}&=&
(\delta_{\mu\alpha}x^{\prime \nu}(\sigma,0)-\delta_{\nu\alpha}
x^{\prime \mu}(\sigma,0))\int d\sigma^{\mu\nu}(y)
F(y(\sigma',\tau')-x(\sigma,0)), \nonumber \\
&=&-\frac{(x^{\prime 2}\dot x^{\alpha}-x'.\dot x x^{\prime \alpha})}
{\sqrt{x^{\prime 2}\dot x^2-(x'.\dot x)^2}},
\eea
the second equation resulting in the limit $a=0$; it manifestly satisfies 
the orthogonality condition
\be \lb{3e14}
x^{\prime\alpha}(\sigma,0)\frac{\delta A}{\delta x^{\alpha}(\sigma,0)}=0,
\ee
in agreement with Eq. (\rf{2e3}).
\par
Some properties of minimal surfaces are presented in appendix \rf{ap1}.
One main property that we shall use in the present paper concerns the
commutativity property of two successive functional derivatives of the
minimal area on its contour $C$,
\be \lb{3e46a}
\left(\frac{\delta}{\delta x^{\beta}(\sigma')}
\frac{\delta}{\delta x^{\alpha}(\sigma)}-
\frac{\delta}{\delta x^{\alpha}(\sigma)}
\frac{\delta}{\delta x^{\beta}(\sigma')}\right)A(C)=0,
\ee
a feature that is reminiscent of a similar property of the Wilson loop
average:
\be \lb{3e44a}
\left(\frac{\delta}{\delta x^{\beta}(\sigma')}
\frac{\delta}{\delta x^{\alpha}(\sigma)}-
\frac{\delta}{\delta x^{\alpha}(\sigma)}
\frac{\delta}{\delta x^{\beta}(\sigma')}\right)W(C)=0.
\ee
\par

\section{The quark propagator in the external gluon field} \lb{s4}
\setcounter{equation}{0}

In dealing with the quarkonium bound state problem within the Wilson
loop approach, one needs, in the path integral formalism, a
representation of the quark propagator in the presence of an arbitrary
external gluon field, satisfying the equation
\be \lb{4e1a}
\Big(i\gamma.\partial_{(x)}-m-g\gamma.A(x)\Big)S(x,x')=i\delta^4(x-x').
\ee
The usual perturbative representation
$S(x,x')=i(i\gamma.\partial-m-g\gamma.A)^{-1}\delta^4(x-x')$ is not very
convenient here since at each order of the perturbative expansion in
the coupling constant the gauge covariance of the propagator is lost
and the construction of the Wilson loop including contributions of fermion
propagators becomes tricky.
A representation that is well suited to exhibit the Wilson loop
structure of the gauge invariant two-particle Green functions is the
Feynman--Schwinger representation \cite{fschwnwstj}, which represents
the quark propagator as a quantum mechanical path integral. This
representation has however the main drawback that it dissolves the Dirac
operator into the path integral and makes it difficult to obtain from
it an equation that easily displays the properties of fermions. In the 
present work we shall consider a representation that is based on an
explicit use of the free Dirac propagator accompanied by a path-ordered 
phase factor. It will have the advantage of manifestly preserving the 
main properties of fermions.
\par
The building block of the representation is the gauge covariant composite
object, denoted $\widetilde S(x,x')$, made of a free fermion propagator
$S_0(x-x')$ (without color group content) multiplied by the path-ordered phase 
factor $U(x,x')$ [Eq. (\rf{2e1})] taken along the straight segment $xx'$:
\be \lb{4e1}
\Big[\widetilde S(x,x')\Big]_{\ b}^a\equiv 
S_0(x-x')\Big[U(x,x')\Big]_{\ b}^a.
\ee
The advantage of the straight segment over other types of line is that in
the limit $x'\rightarrow x$ $U$ tends to unity in an anambiguous way.
$\widetilde S$ satisfies the following equation with respect to $x$:
\be \lb{4e2}
\Big(i\gamma.\partial_{(x)}-m-g\gamma.A(x)\Big)\widetilde S(x,x')=
i\delta^4(x-x')+i\gamma^{\alpha}\int_0^1d\lambda\,\lambda
\left(\frac{\delta U(x,x')}{\delta x^{\alpha}(\lambda)}\right)S_0(x-x'),
\ee
where the segment $xx'$ has been parametrized with the parameter
$\lambda$ as $x(\lambda)=\lambda x+(1-\lambda)x'$. In the above equation,
the point $x'$ is held fixed; furthermore, the operator
$\delta/\delta x^{\alpha}(\lambda)$ does not act on the explicit boundary
point $x$ of the segment (corresponding to $\lambda=1$, cf. Eqs.
(\rf{2e2})-(\rf{2e3})), this contribution having already been cancelled by
the gluon field term $A$.
A similar equation also holds with respect to $x'$, with $x$ held fixed,
with the Dirac and color group matrices acting from the right.
\par
The quantity $-i(i\gamma.\partial_{(x)}-m-g\gamma.A(x))\delta^4(x-x')$ is
the inverse of the quark propagator $S$ in the presence of the external
gluon field $A$. Reversing Eq. (\rf{4e2}) with respect to $S^{-1}$, one 
obtains an equation for $S$ in terms of $\widetilde S$:
\be \lb{4e3}
S(x,x')=\widetilde S(x,x')-\int d^4x''S(x,x'')\gamma^{\alpha}
\int_0^1d\lambda\,\lambda\frac{\delta}{\delta x^{\alpha}(\lambda)}
\widetilde S(x'',x'),
\ee
where the operator $\delta/\delta x^{\alpha}(\lambda)$ acts on the
factor $U$ of $\widetilde S$, along the internal part of the segment 
$x''x'$, with $x'$ held fixed. Using the equation with $x'$, or making in Eq.
(\rf{4e3}) an integration by parts, one obtains another equivalent equation:
\be \lb{4e4}
S(x,x')=\widetilde S(x,x')+\int d^4x''\int_0^1d\lambda(1-\lambda)
\widetilde S(x,x'')\frac{\stackrel{\leftarrow}{\delta}}
{\delta x^{\alpha}(\lambda)}\gamma^{\alpha}S(x'',x').
\ee
\par
Equations (\rf{4e3}) or (\rf{4e4}) allow us to obtain the propagator
$S$ as an iteration series with respect to $\widetilde S$, which
contains the free fermion propagator, by maintaining at each order
of the iteration its gauge covariance property. For instance, the 
expansion of Eq. (\rf{4e3}) takes the form:
\bea \lb{4e5}
& &S(x,x')=\widetilde S(x,x')-\int d^4y_1\widetilde S(x,y_1)
\gamma^{\alpha_1}\int_0^1d\lambda_1\,\lambda_1\frac{\delta}
{\delta x^{\alpha_1}(\lambda_1)}\widetilde S(y_1,x')\nonumber \\
& &+
\int d^4y_1d^4y_2\int_0^1d\lambda_1\,\lambda_1d\lambda_2\,\lambda_2
\widetilde S(x,y_1)\gamma^{\alpha_1}\Big(\frac{\delta}
{\delta x^{\alpha_1}(\lambda_1)}\widetilde S(y_1,y_2)\Big)
\gamma^{\alpha_2}\Big(\frac{\delta}{\delta x^{\alpha_2}(\lambda_2)}
\widetilde S(y_2,x')\Big)\nonumber \\
& &+\cdots\ ,
\eea
the operator $\delta/\delta x(\lambda_i)$ acting on the inner part of the
segment $y_iy_{i+1}$ of the phase factor $U(y_i,y_{i+1})$.
A verification of Eq. (\rf{4e1a}) 
can be done through the above iteration series. The operator
$\partial/\partial x$ acts on three terms. First, acting on the
free fermion propagator contained in $\widetilde S(x,x'')$, it gives,
with the Dirac operator, a $\delta^4(x-x'')$ type term which then removes
the integration over $x''$. Second, it acts on the boundary point $x$
of $U(x,x'')$ and cancels the gluon field term $A$. Third, it acts
on the inner part of the segment $xx''$ of $U$. This term is then cancelled
by the $\delta^4$-term coming from the next-order term of the 
iteration series under the action of the Dirac operator on the
corresponding free propagator, and so forth.
Equations (\rf{4e3}) and (\rf{4e4}) are generalizations of the
representation used for heavy quarks starting from the static case 
\cite{ef}.
\par
The action of the operator $\delta/\delta x^{\alpha}(\lambda)$ on $U$
can be expressed in terms of an insertion of the field strength 
$F$ [Eq. (\rf{2e3})]. One can check with the first few terms of the
series, using integrations by parts, that one can recover, in perturbation
theory with respect to the coupling constant $g$, the conventional 
perturbative expansion of $S$ in terms of $g$ and $A$.
\par

\section{The two-particle gauge invariant Green function} \lb{s5}
\setcounter{equation}{0}

The next step is to consider the quark-antiquark gauge invariant
Green function, for quarks $q_1$ and $q_2$ with different flavors
and with masses $m_1$ and $m_2$:
\be \lb{5e1}
G(x_1,x_2;x_1',x_2')\equiv \langle \overline \psi_2(x_2)U(x_2,x_1)
\psi_1(x_1)\overline \psi_1(x_1')U(x_1',x_2')\psi_2(x_2')
\rangle_{A,q_1,q_2},
\ee
the averaging being defined in the path-integral formalism.
Here, $U(x_2,x_1)$ is the phase factor (\rf{2e1}) along the
straight segment $x_2x_1$ (and similar\-ly for $U(x_1'x_2')$).
According to the conclusion reached in the final part of
appendix \rf{ap1}, the dynamics of the system concerning its energy
spectrum not containing string-like excitations can be probed by
considering between the quark and the antiquark equal-time
straight segments (in a given reference frame, e.g., the rest
frame of the bound state); deviations of lines $C_{x_2x_1}$ from the
equal-time straight segment contribute only to the wave functional of
the bound state and not to its energy. However, for covariance reasons,
we first consider general straight segments $x_2x_1$ (not necessarily
equal-time) and it is at a later stage that the equal-time limit will
be taken.
\par
Integrating in the large $N_c$ limit with respect to the quark fields,
one obtains:
\be \lb{5e2}
G(x_1,x_2;x_1',x_2')=-\langle \mathrm{tr}_c\,U(x_2,x_1)S_1(x_1,x_1')
U(x_1',x_2')S_2(x_2',x_2)\rangle_A,
\ee
where $S_1$ and $S_2$ are the two quark propagators in the presence
of the external gluon field and $\mathrm{tr}_c$ designates the trace 
with respect to the color group.
\par
The Green function $G$ satisfies the following equation with respect to
the Dirac operator of particle 1 acting on $x_1$:
\bea \lb{5e5}
& &(i\gamma.\partial_{(x_1)}-m_1)G(x_1,x_2;x_1',x_2')=
-i\langle\mathrm{tr}_c\,U(x_2,x_1)\delta^4(x_1-x_1')U(x_1',x_2')
S_2(x_2',x_2)\rangle_A\nonumber \\
& &\ \ \ \ \ \ -i\gamma^{\alpha}\langle\mathrm{tr}_c\int_0^1 
d\sigma(1-\sigma)\frac{\delta U(x_2,x_1)}{\delta x^{\alpha}(\sigma)}
S_1(x_1,x_1')U(x_1',x_2')S_2(x_2',x_2)\rangle_A,
\eea
where the segment $x_2x_1$ has been parametrized with the parameter
$\sigma$ as $x(\sigma)=(1-\sigma)x_1+\sigma x_2$; furthermore, the
operator $\delta/\delta x^{\alpha}$ does not act on the explicit
boundary point $x_1$  of the segment (cf. Eqs. (\rf{2e2})-(\rf{2e3})),
this contribution having already been cancelled by the contribution
of the gluon field $A$ coming from the quark propagator $S_1$.
A similar equation also holds with the Dirac operator of particle 2:
\bea \lb{5e6}
& &G(x_1,x_2;x_1',x_2')(-i\gamma.{\stackrel{\leftarrow}{\partial}}_{(x_2)}
-m_2)=
-i\langle\mathrm{tr}_c\,U(x_2,x_1)S_1(x_1,x_1')U(x_1',x_2')
\delta^4(x_2'-x_2)\rangle_A\nonumber \\
& &\ \ \ \ \ \ +i\langle\mathrm{tr}_c\int_0^1 d\sigma \sigma
\frac{\delta U(x_2,x_1)}{\delta x^{\beta}(\sigma)}S_1(x_1,x_1')
U(x_1',x_2')S_2(x_2',x_2)\rangle_A\gamma^{\beta}.
\eea
\par
Using for $S_1$ and $S_2$ representations (\rf{4e3}) and (\rf{4e4}),
respectively, one obtains for $G$ an expansion in a series of terms
involving an increasing number of straight segments between $x_1$ and
$x_1'$ on the one hand and between $x_2$ and $x_2'$ on the other. With
each segment is associated a path-ordered phase factor $U$; the union
of all such factors, together with $U(x_2,x_1)$ and $U(x_1',x_2')$, forms
a Wilson loop along a skew polygon. We can then represent $G$ in the
following form:
\be \lb{5e3}
G=\sum_{i,j=1}^{\infty}G_{i,j},
\ee
where $G_{i,j}$ represents the contribution of the term of the series
having $(i-1)$ points of integration between $x_1$ and $x_1'$ ($i$
segments) and $(j-1)$ points of integration between $x_2$ and $x_2'$
($j$ segments). We designate by $C_{i,j}$ the contour associated
with the term $G_{i,j}$. A typical configuration for the contour of
$G_{4,3}$ is represented in Fig. \rf{5f1}.
\par
\bfg
\vspace*{0.5 cm}
\bc
\input{5f1.pstex_t}
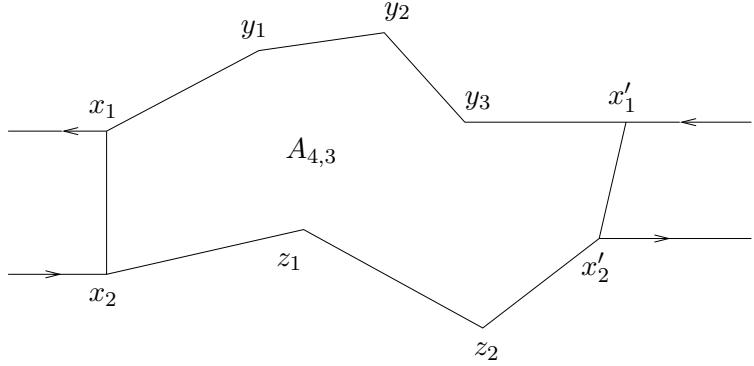
\caption{The contour associated with the term $G_{4,3}$.}
\lb{5f1}
\ec
\efg
Using for the averages of the Wilson loops appearing in the above series
the representation with minimal surfaces, and designating by $A_{i,j}$
the minimal area associated with the contour $C_{i,j}$, one obtains for
the latter the following representation:
\bea \lb{5e4}
G_{i,j}&=&(-1)^iN_c\int d^4y_1\cdots d^4y_{i-1}d^4z_1\cdots d^4z_{j-1}
S_{10}(x_1-y_1)\gamma^{\alpha_1}S_{10}(y_1-y_2)\cdots\nonumber \\
& &\times \gamma^{\alpha_{i-1}}S_{10}(y_{i-1}-x_1')
S_{20}(x_2'-z_{j-1})\gamma^{\beta_{j-1}}S_{20}(z_{j-1}-z_{j-2})
\cdots \gamma^{\beta_1}S_{20}(z_1-x_2)\nonumber \\
& &\times\int_0^1d\tau_1\cdots d\tau_{i-1}\int_0^1d\tau_1'\cdots d\tau_{j-1}'
(1-\tau_1)\cdots (1-\tau_{i-1})(1-\tau_1')\cdots (1-\tau_{j-1}')\nonumber \\
& &\frac{\delta}{\delta x^{\alpha_1}(\tau_1)}\cdots
\frac{\delta}{\delta x^{\alpha_{i-1}}(\tau_{i-1})}
\frac{\delta}{\delta x^{\beta_1}(\tau_1')}\cdots
\frac{\delta}{\delta x^{\beta_{j-1}}(\tau_{j-1}')}\
e^{{\displaystyle -i\sigma A_{i,j}}}.
\eea
Here, the operators $\delta/\delta x(\tau_k)$ and
$\delta/\delta x(\tau_{\ell}')$ act on the surface $A_{i,j}$ through
their action on the segments $y_ky_{k+1}$ and $z_{\ell}z_{\ell+1}$ of the
contour, respectively. (The parametrization of the segments of the line
$x_1x_1'$ is the opposite of that of Sec. \rf{s4}: $\tau$ now increases
along the direction $x_1x_1'$ to be in accordance with that adopted for
the surfaces; cf. comment after Eq. (\rf{3e1}).) $S_{10}$ and
$S_{20}$ are the free quark propagators with masses $m_1$ and
$m_2$, respectively.
\par
Using in Eqs. (\rf{5e5})-(\rf{5e6}) representations (\rf{4e3})-(\rf{4e4})
for the quark propagators, one obtains:
\bea
\lb{5e7}
& &(i\gamma.\partial_{(x_1)}-m_1)G(x_1,x_2;x_1',x_2')=
-i\delta^4(x_1-x_1')\sum_{j=1}^{\infty}G_{0,j}\nonumber \\
& &\ \ \ \ \ \ +i\gamma^{\alpha}\sum_{i,j=1}^{\infty}\int_0^1
d\sigma(1-\sigma)\frac{\delta}{\delta x^{\alpha}(\sigma)}
G_{i,j}\bigg|_{x(\sigma)\in x_1x_2},\\
\lb{5e8}
& &G(x_1,x_2;x_1',x_2')(-i\gamma.{\stackrel{\leftarrow}{\partial}}_{(x_2)}
-m_2)=+i\delta^4(x_2-x_2')\sum_{i=1}^{\infty}G_{i,0}\nonumber \\
& &\ \ \ \ \ \ -i\sum_{i,j=1}^{\infty}\int_0^1d\sigma \sigma
\frac{\delta}{\delta x^{\beta}(\sigma)}G_{i,j}\gamma^{\beta}
\bigg|_{x(\sigma)\in x_1x_2},
\eea
where $G_{0,j}$ and $G_{i,0}$ are particular cases of representation
(\rf{5e4}) in which the particle 1 or 2 propagators have been shrunk to a
delta-function. These equations can also be obtained by making the Dirac
operators act on Eq. (\rf{5e3}) and using in the right-hand side
representation (\rf{5e4}).
\par
The two equations (\rf{5e5}) and (\rf{5e6}) are independent, since the
two Dirac operators concern independent particle variables. They are also
compatible; this is evident from the very fact that $G$ exists and is
given by formula (\rf{5e2}). However, an independent check can also be
done by making the Dirac operator of particle 2 act on Eq. (\rf{5e5}) and
the Dirac operator of particle 1 act on Eq. (\rf{5e6}), using in the
right-hand sides the properties of the propagators and of the phase
factors $U$ and subtracting from each other the two resulting
equations. The result is zero, due mainly to the facts that the final
expressions involve functional derivatives $\delta/\delta x(\sigma)$ and
$\delta/\delta x(\sigma')$ of the phase factor $U(x_2,x_1)$ and these
commute. A similar verification can also be done with Eqs.
(\rf{5e7})-(\rf{5e8}), where now the representation of the Wilson loop
averages by minimal surfaces has been used. One has to use in the
right-hand sides properties of the terms $G_{i,j}$ and
cancellations between contributions of successive $G_{i,j}$s along the
quark and antiquark lines. One ends up with expressions involving
functional derivatives $\delta/\delta x(\sigma)$ and
$\delta/\delta x(\sigma')$ of minimal surfaces along the segment $x_1x_2$. 
The compatibility of the two equations is then due to the fact that two
successive functional derivatives  of a minimal area localized
on its contour commute [Eq. (\rf{3e46a}) and appendix \rf{ap1}].
\par
Eqs. (\rf{5e7})-(\rf{5e8}) are not closed integro-differential equations
for $G$, for once the action of the functional derivatives
$\delta/\delta x$ on the various minimal surfaces has been evaluated one
does not obtain back $G$ on the right-hand sides. This feature makes
difficult the search for a bound state equation in compact form. 
In this respect, if $G$ has a bound state pole in momentum space,
then the right-hand sides of Eqs. (\rf{5e7})-(\rf{5e8})
should also have the same pole; this is possible only if the actions
of the functional derivatives $\delta/\delta x$ on the partial ingredients
$G_{i,j}$ of $G$ yield among other terms common factors that allow coherent
summations of the $G_{i,j}$s to produce again a pole term; otherwise,
each $G_{i,j}$, containing a finite number of free quark propagators, cannot
produce alone such a pole. In $x$-space, the selection of a bound state
is made by taking a large time separation between the pairs of points
$(x_1,x_2)$ on the one hand and $(x_1',x_2')$ on the other \cite{gml}.
\par
The independence of the two equations (\rf{5e7})-(\rf{5e8}) means also 
that they might allow the elimination of the relative time variable of the 
two particles prior to any resolution of an eigenvalue equation and the
reduction of the internal dynamics to a three-dimensional space, a feature 
which was outlined at the beginning of this section according to the results 
obtained at the end of appendix \rf{ap1}. However, such a reduction does
not seem easily manageable on the general forms of Eqs. 
(\rf{5e7})-(\rf{5e8}). Furthermore, arbitrary approximations made in the 
right-hand sides of those equations may destroy their compatibility
property. This is why in the following we shall directly study
the equal-time limit of the system by considering the ``sum'' of the
two equations and then, at a later stage, shall indicate how to determine
its relative time evolution law by considering the ``difference'' of the
two equations.
\par

\section{Bound state equation} \lb{s6}
\setcounter{equation}{0}

In the large $N_c$ limit, the mesonic sector of QCD is composed of
one-particle states, which are bound states of a quark-antiquark pair
and of gluons \cite{th1,vwt}. Taking in the Green function (\rf{5e1}) a large
time separation between the pairs of points $(x_1,x_2)$ and $(x_1',x_2')$
and using the completeness relation one finds:
\be \lb{6e1}
G(x_1,x_2;x_1',x_2')=\sum_n\Phi_n(x_1,x_2)\overline\Phi_n(x_2',x_1'),
\ee
where $\Phi_n$ is the wave functional of the bound state labelled with
the collective quantum numbers $n$:
\bea \lb{6e2}
i\Phi_{n,\alpha_1,\alpha_2}(x_1,x_2)&=&<0|\overline\psi_{2,\alpha_2}(x_2)
U(x_2,x_1)\psi_{1,\alpha_1}(x_1)|n>,\nonumber \\
-i\overline\Phi_{n,\alpha_2',\alpha_1'}(x_2',x_1')&=&
<n|\overline\psi_{1,\alpha_1'}(x_1')U(x_1',x_2')\psi_{2,\alpha_2'}(x_2')|0>.
\eea
Since the lines $C_{x_2x_1}$ and $C_{x_1'x_2'}$ are rigid straight segments
completely determined by their end points, one can consider the wave
functionals as functions of the end point coordinates. Introducing total and
relative coordinates and momenta, 
\bea \lb{6e3}
& &X=\frac{1}{2}(x_1+x_2),\ \ \ x=x_2-x_1,\ \ \ 
P=p_1+p_2,\ \ \ p=\frac{1}{2}(p_2-p_1),\nonumber \\
& &p_{a,\mu}=i\frac{\partial}{\partial x_a^{\mu}},\ \ \ a=1,2,
\eea
and considering a bound state with total momentum $P$, one has:
\be \lb{6e4}
\Phi(x_1,x_2)=e^{-iP.X}\phi(P,x).
\ee
In the large separation time limit, the right-hand side of Eq. (\rf{6e1})
displays a series of oscillating functions in the separation time variable.
By appropriate projections and integrations one can select in this series
the bound state that will survive in the large time limit \cite{gml}.
It should be emphasized that in the above limit only terms that factorize
in G into expressions depending on the line $x_1x_2$ and expressions
depending on the line $x_1'x_2'$ could survive to the selection operation of
the bound state. Terms that still contain expressions joining line $x_1'x_2'$
to line $x_1x_2$ would not contribute to the previous operation and could
be discarded.
\par
It is convenient to consider the total momentum $P$ of the selected bound
state as a reference timelike vector and define transverse and
longitudinal parts of vectors with respect to it:
\bea \lb{6e5}
& &q_{\mu}^T=q_{\mu}-\frac{q.P}{P^2}P_{\mu},\ \ \ 
q_{\mu}^L=(q.\hat P)\hat P_{\mu},\ \ \ 
\hat P_{\mu}=\frac{P_{\mu}}{\sqrt{P^2}},\ \ \ q_L=q.\hat P,\nonumber \\
& & P_L=\sqrt{P^2},\ \ \ q_L\Big|_{\mathrm{c.m.}}=q_0,\ \ \ 
q^{T2}\Big|_{\mathrm{c.m.}}=-\mathbf{q}^2,\ \ \ 
\sqrt{-x^{T2}}\Big|_{c.m.}=r.
\eea
These decompositions are manifestly covariant. To further simplify the
notation we adopt for the Dirac matrices the following convention: they
will be written on the left of the spinor functions with labels 1 or 2
indicating on which particle indices they act, the particle 2 matrices
(the antiquark at $x_2$) acting actually from the right; thus:
\bea \lb{6e6}
\gamma_{1\mu}G&\equiv&(\gamma_{\mu})_{\alpha_1\beta_1}G_{\beta_1\alpha_2,
\alpha_2'\alpha_1'},\ \ \ \ \ \ 
\gamma_{2\mu}G\equiv G_{\alpha_1\beta_2,\alpha_2'\alpha_1'}
(\gamma_{\mu})_{\beta_2\alpha_2},\nonumber \\
\gamma_{2\mu}\gamma_{2\nu}G&\equiv&G_{\alpha_1\beta_2,\alpha_2'\alpha_1'}
(\gamma_{\nu}\gamma_{\mu})_{\beta_2\alpha_2},\ \ \ \ \ \ 
\gamma_{2\mu}\gamma_{25}G\equiv G_{\alpha_1\beta_2,\alpha_2'\alpha_1'}
(\gamma_5\gamma_{\mu})_{\beta_2\alpha_2}.
\eea
Similar definitions also hold when $G$ is replaced by $\Phi$.
Notice that products of $\gamma_2$ matrices act from the right in the 
reverse order.
\par
We next introduce the free Dirac hamiltonians of the two particles:
\be \lb{6e7}
h_{10}=m_1\gamma_{1L}-\gamma_{1L}\gamma_1^T.p_1^T,\ \ \ \ \ \
h_{20}=-m_2\gamma_{2L}-\gamma_{2L}\gamma_2^T.p_2^T.
\ee
Going back to Eqs. (\rf{5e7})-(\rf{5e8}), multiplying the first by
$\gamma_{1L}$, the second by $-\gamma_{2L}$, ad\-ding and subtracting the
two equations one finds:
\bea
\lb{6e8}
& &\Big[(p_{1L}+p_{2L})-(h_{10}+h_{20})\Big]G=
-i\delta^4(x_1-x_1')\gamma_{1L}\sum_{j=1}^{\infty}G_{0,j}
-i\delta^4(x_2-x_2')\gamma_{2L}\sum_{i=1}^{\infty}G_{i,0}
\nonumber \\
& &\ \ \ \ \ +i\Big[\gamma_{1L}\gamma_1^{\alpha}\int_0^1
d\sigma(1-\sigma)\frac{\delta}{\delta x^{\alpha}(\sigma)}
+\gamma_{2L}\gamma_2^{\beta}\int_0^1d\sigma \sigma
\frac{\delta}{\delta x^{\beta}(\sigma)}\Big]\sum_{i,j=1}^{\infty}G_{i,j}
\bigg|_{x(\sigma)\in x_1x_2}, \\
\lb{6e9}
& &\Big[(p_{1L}-p_{2L})-(h_{10}-h_{20})\Big]G=
-i\delta^4(x_1-x_1')\gamma_{1L}\sum_{j=1}^{\infty}G_{0,j}
+i\delta^4(x_2-x_2')\gamma_{2L}\sum_{i=1}^{\infty}G_{i,0}
\nonumber \\
& &\ \ \ \ \ +i\Big[\gamma_{1L}\gamma_1^{\alpha}\int_0^1
d\sigma(1-\sigma)\frac{\delta}{\delta x^{\alpha}(\sigma)}
-\gamma_{2L}\gamma_2^{\beta}\int_0^1d\sigma \sigma
\frac{\delta}{\delta x^{\beta}(\sigma)}\Big]\sum_{i,j=1}^{\infty}G_{i,j}
\bigg|_{x(\sigma)\in x_1x_2}.
\eea
Equation (\rf{6e9}) mainly determines the relative time evolution of
the Green function, while Eq. (\rf{6e8}) mainly determines the
dynamical properties of the two-particle system. Since that equation
does not involve the relative energy operator $(p_{2L}-p_{1L})$ in its
left-hand side, one is entitled to take in it the equal-time limit
$x_L=0$; the equation becomes in that case three-dimensional with respect
to the transverse relative coordinates $x^T$. After determining, within a
given approximation, the dynamical properties of the system from Eq.
(\rf{6e8}) in the limit $x_L=0$, one can go back to Eq. (\rf{6e9}) and
determine, within the same approximation, its relative time evolution law.
We shall henceforth follow this method of approach.
\par
In the large separation time limit between the pairs of points $(x_1,x_2)$
and $(x_1',x_2')$ the delta-functions that are present in the right hand-sides
of Eqs. (\rf{6e8})-(\rf{6e9}) do not contribute and can be ignored. In
order to extract from the right-hand side of Eq. (\rf{6e8}) a bound state
wave function $\Phi$, it is necessary that the actions of the functional
derivatives $\delta/\delta x(\sigma)$ on the various parts $G_{i,j}$
yield among other terms common factors that factorize the Green function
$G$ again. The functional derivative $\delta/\delta x(\sigma)$ acts on a
given $G_{i,j}$ [Eq. (\rf{5e4})] through the exponential factor containing
the minimal area term and yields a term proportional to
$\delta A_{i,j}/\delta x(\sigma)$. This term may itself be acted on by the
other functional derivatives existing in the definition of $G_{i,j}$.
One thus ends up with two types of term. The first contains the first-order
derivative $\delta A_{i,j}/\delta x(\sigma)$, which factorizes the other
derivatives on the left, and the second contains terms involving
higher-order derivatives of $A_{i,j}$, one of the derivatives acting along
the straight segment $x_1x_2$. The dominant part in the large-distance limit
comes from the first type of term; higher-order derivatives of the minimal
area tend to weaken the large-distance behavior (cf. appendix \rf{ap1} for
the second-order derivative). In the following we shall mainly concentrate
on the contribution of the first type of term; terms containing second-order 
derivatives will be considered in Sec. \rf{s7}, in connection with the
self-energy parts of the quark propagators.
\par
The term that is retained with $G_{i,j}$ is an integral over $\sigma$
of a function proportional to $\delta A_{i,j}/\delta x(\sigma)$.
According to Eq. (\rf{3e13}), the latter depends only on the local
properties of the surface at the point $x(\sigma,0)$ lying on the
straight segment $x_1x_2$, namely upon the derivatives $x'(\sigma,0)$ and
$\dot x(\sigma,0)$. Because the line $C_{x_2x_1}$ is a straight segment,
one has $x'(\sigma,0)=x$, independent of the form of the surface. In the
equal-time limit ($x_L=0$) the previous relation becomes
$x'(\sigma,0)=x^T$. The derivative $\dot x(\sigma,0)$
depends, however, on the form of the minimal surface in the vicinity of the
straight segment. Using methods of analysis similar to those used at the end
of appendix \rf{ap1} one can show that in the large separation time limit
$\dot x(\sigma,0)$ tends to a linear function of sigma. In that case,
one can parametrize it as
$\dot x(\sigma,0)=(1-\sigma)\dot x_1+\sigma\dot x_2$, where $\dot x_1$
and $\dot x_2$ are the slopes at the points $x_1$ and $x_2$, respectively,
$\dot x_1=\dot x(0,0)$, $\dot x_2=\dot x(1,0)$.
The latter make still reference to the other end points of the corresponding
segments; for the case of the simplest contour $C_{1,1}$, these are $x_1'$
and $x_2'$. In order to remove any explicit reference to the points of the
remote past an operator representation of the slopes, depending only on 
points $x_1$ and $x_2$ becomes necessary.
\par
To find such a representation, we consider the simplest
contour $C_{1,1}$ corresponding to $G_{1,1}$. Here, one has
$\dot x_1=(x_1'-x_1)$, $\dot x_2=(x_2'-x_2)$, $\dot x_{aL}<0$
with our parametrization and because of the facts that
$x_{aL}\rightarrow +\infty$ and $x_{aL}'\rightarrow -\infty$ ($a=1,2$).
Considering in general the cases $x_L$ and $x_L'$ finite in the above limits,
one can set, modulo negligible terms, $x_{1L}=x_{2L}$. (This is exact in the
equal-time cases $x_L=0$ and $x_L'=0$.) Furthermore, a close examination
of the integrals of the term $\delta A_{1,1}/\delta x(\sigma)$ shows that
the $\dot x_a$ terms ($a=1,2$) appear after integration in the forms
$\dot x_a^{\alpha}/|\dot x_{aL}|$ and through their orthogonal components
to $x_1x_2$. An operator representation of the latter
can be found by making the quark momentum operators $p_{a\mu}$ act on the
term $G_{1,1}$. They generally yield three different contributions. The 
first comes from their action on the corresponding free quark propagator, 
the second from the segment $x_1x_2$ and the third from the segment 
$(x_a'-x_a)$. Since the terms $\dot x_a^{\alpha}/|\dot x_{aL}|$ already 
appear in expressions that are proportional to the string tension $\sigma$,
one can use for the latter terms, as a first approximation, free theory
expressions; in this case, it is sufficient to retain the contributions 
coming from the free quark propagators. The latter, in the large time limit
yield with massive quarks the following dominant behavior:
\be \lb{6e10}
p_{a\mu}S_{a0}(x_a-x_a')\simeq m_a\frac{(x_a-x_a')_{\mu}}
{\sqrt{(x_a-x_a')^2}}S_{a0}(x_a-x_a'),\ \ \ 
(x_a-x_a')_L\rightarrow +\infty,\ \ \ a=1,2,
\ee
from which one deduces:
\be \lb{6e11}
-\frac{\dot x_{a\mu}}{|\dot x_{aL}|}\Longleftrightarrow
\frac{p_{a\mu}}{|p_{aL}|},\ \ \ \ a=1,2,
\ee
where the operators $p_{aL}$ are the free theory expressions of the
individual energies:
\be \lb{6e12}
p_{aL}=h_{a0},\ \ \ \ 
|p_{aL}|=\sqrt{m_a^2-p_a^{T2}}\equiv E_a(p_a^T),\ \ \ \ a=1,2,
\ee
$h_{a0}$ being defined in Eqs. (\rf{6e7}).
\par
This approximation will retain all terms of order up to $O(1/c^2)$ in the
nonrelativistic limit; terms that have been neglected, if they are nonzero,
have contributions in the nonrelativistic limit starting at order $O(1/c^4)$.
\par
We next generalize the above evaluation to the
higher-order contours $C_{i,j}$ appearing in $G_{i,j}$. Here, we make the
assumption that in the large separation time limit between the pairs of
points $(x_1,x_2)$ and $(x_1',x_2')$ the derivatives
$\delta A_{i,j}/\delta x(\sigma)$ can be expanded around the driving
term $\delta A_{1,1}/\delta x(\sigma)$ of the lowest order surface.
Neglecting the higher-order terms of these expansions, one ends up with
the common operators $\delta A_{1,1}/\delta x(\sigma)$ to all factors
$G_{i,j}$, involving the segment $x_1x_2$ and the momentum operators 
$p_{a\mu}$, $a=1,2$. Those terms can then be factorized in front of $G$
and interpreted as potentials.
\par
The bound state equation obtained from Eq. (\rf{6e8}) in the equal-time
limit $x_L=0$ is then:
\be \lb{6e13}
\Big[P_L-(h_{10}+h_{20})-\gamma_{1L}\gamma_1^{\mu}A_{1\mu}
-\gamma_{2L}\gamma_2^{\mu}A_{2\mu}\Big]\psi(P_L,x^T)=0,
\ee
where $\psi$ is the wave function $\phi$ in the equal-time limit,
\be \lb{6e13a}
\psi(P_L,x^T)\equiv\phi(P_L,x_L=0,x^T),
\ee
and the potentials are defined through the equations (in Minkowski space)
\be \lb{6e14}
A_{1\mu}=\sigma\int_0^1d\sigma'(1-\sigma')\frac{\delta A_{1,1}}
{\delta x^{\mu}(\sigma')},\ \ \ \ \ 
A_{2\mu}=\sigma\int_0^1d\sigma'\,\sigma'\frac{\delta A_{1,1}}
{\delta x^{\mu}(\sigma')}. 
\ee
They can be calculated either by using the minkowskian version of Eq.
(\rf{3e13}) and then conditions (\rf{6e11})-(\rf{6e12}), or by first
writing Eqs. (\rf{6e8}) and (\rf{6e13}) in euclidean space, using Eq.
(\rf{3e13}) and then passing to Minkowski space. (In euclidean space
the right-hand sides of Eqs. (\rf{6e14}) contain an additional $(-i)$
factor.) Since $\delta A_{1,1}/\delta x(\sigma')$ is orthogonal to
$x$ [Eq. (\rf{3e14})], the resulting vectors will satisfy this property.
We define transverse vectors with respect to $x$ with a superscript ``$t$'':
\be \lb{6e15}
q_{\mu}^t=q_{\mu}-x_{\mu}\frac{1}{x^2}x.q.
\ee
However, $x$ itself is orthogonal to the total momentum $P$ in the equal-time
limit ($x_L=0$) and reduces to $x^T$. The part of the three-dimensional
relative momentum $p^T$ that is also orthogonal to $x^T$ will be denoted
$p^{Tt}$:
\be \lb{6e16}
p_{\mu}^{Tt}=p_{\mu}^T-x_{\mu}^T\frac{1}{x^{T2}}x^T.p^T.
\ee
This vector enters in the definition of the relative orbital angular
momentum. Defining the corresponding Pauli-Lubanski vector $W_L$ ($L$
referring here to the orbital angular momentum) as
\be \lb{6e17}
W_{L\mu}=\epsilon_{\mu\nu\alpha\beta}P^{\nu}x^{\alpha}p^{\beta},
\ \ \ \ \ \ \epsilon_{0123}=+1,
\ee
one has the relations
\be \lb{6e18}
W_L^2=-P^2\Big(x^{T2}p^{T2}-x^{T\alpha}x^{T\beta}p^T_{\alpha}p^T_{\beta}
-2ix^T.p^T\Big)=-P^2x^{T2}p^{Tt2},\ \ \ \ \ 
-\frac{W_L^2}{P^2}\bigg|_{\mathrm{c.m.}}=\mathbf{L}^2.
\ee
\par
The expression of $A_{1\mu}$ is:
\bea \lb{6e19}
A_{1\mu}&=&-\sigma\sqrt{-x^{T2}}\frac{E_1E_2}{E_1+E_2}
\bigg\{\,\Big[\,g_{\mu L}(\epsilon(p_{2L})-\epsilon(p_{1L}))
\frac{E_1E_2}{(E_1+E_2)^2}(\frac{x^{T2}P^2}{2W_L^2}E_1E_2-1)\nonumber \\
& &\ \ \ -g_{\mu L}\epsilon(p_{1L})\frac{E_1}{E_1+E_2}+
\frac{x^{T2}P^2}{2W_L^2}\frac{E_1E_2}{E_1+E_2}p_{\mu}^{Tt}\,\Big]
\nonumber \\
& &\ \ \ \ \ \times\sqrt{\frac{-x^{T2}P^2}{-W_L^2}}
\left(\,\arcsin\Big(\frac{1}{E_2}\sqrt{\frac{-W_L^2}{-x^{T2}P^2}}\Big)+
\arcsin\Big(\frac{1}{E_1}\sqrt{\frac{-W_L^2}{-x^{T2}P^2}}\Big)\,\right)
\nonumber \\
& &\ \ \ -\Big[\,g_{\mu L}\epsilon(p_{1L})+g_{\mu L}\frac{(E_2-E_1)}
{(E_2+E_1)}(\epsilon(p_{2L})-\epsilon(p_{1L}))
-\frac{1}{E_2}p_{\mu}^{Tt}\,\Big]\nonumber \\
& &\ \ \ \ \ \times\Big(\frac{E_1E_2}{E_1+E_2}\Big)
\Big(\frac{-x^{T2}P^2}{-W_L^2}\Big)
\left(\,\sqrt{1-\frac{-W_L^2}{-x^{T2}P^2E_2^2}}-
\sqrt{1-\frac{-W_L^2}{-x^{T2}P^2E_1^2}}\,\right)\nonumber \\
& &\ \ \ -\frac{1}{2}\Big[\,g_{\mu L}
(\epsilon(p_{2L})-\epsilon(p_{1L}))\frac{E_1E_2}{E_1+E_2}
+p_{\mu}^{Tt}\,\Big]\,\Big(\frac{-x^{T2}P^2}{-W_L^2}\Big)\nonumber \\
& &\ \ \ \ \ \times \left(\,\frac{E_1}{E_1+E_2}\sqrt{1-
\frac{-W_L^2}{-x^{T2}P^2E_2^2}}+\frac{E_2}{E_1+E_2}
\sqrt{1-\frac{-W_L^2}{-x^{T2}P^2E_1^2}}\,\right)\,\bigg\}.
\eea
Here, $\epsilon(p_{1L})$ and $\epsilon(p_{2L})$ are the energy
sign operators of the free quark and the antiquark, respectively:
\be \lb{6e20}
\epsilon(p_{aL})=\frac{h_{a0}}{E_a},\ \ \ \ \ a=1,2,
\ee
$h_{a0}$ and $E_a$ being defined in Eqs. (\rf{6e7}) and (\rf{6e12}).
The expression of $A_{2\mu}$ is obtained from that of $A_{1\mu}$ by an
interchange in the latter of the indices 1 and 2 and a change of sign of
$p^{Tt}$. In particular, the longitunal parts of the potentials $A_1$
and $A_2$ add up in Eq. (\rf{6e13}). One has for their sum the expression:
\bea \lb{6e21}
& &A_{1L}+A_{2L}=\sigma\sqrt{-x^{T2}}\frac{E_1E_2}{E_1+E_2}
\bigg\{\Big(\frac{E_1}{E_1+E_2}\epsilon(p_{1L})+
\frac{E_2}{E_1+E_2}\epsilon(p_{2L})\Big)\nonumber \\
& &\ \ \ \ \ \ \ \ \times\sqrt{\frac{-x^{T2}P^2}{-W_L^2}}
\left(\,\arcsin\Big(\frac{1}{E_2}\sqrt{\frac{-W_L^2}{-x^{T2}P^2}}\Big)+
\arcsin\Big(\frac{1}{E_1}\sqrt{\frac{-W_L^2}{-x^{T2}P^2}}\Big)\,\right)
\nonumber \\
& &+(\epsilon(p_{1L})-\epsilon(p_{2L}))
\Big(\frac{E_1E_2}{E_1+E_2}\Big)\Big(\frac{-x^{T2}P^2}{-W_L^2}\Big)
\left(\,\sqrt{1-\frac{-W_L^2}{-x^{T2}P^2E_2^2}}-
\sqrt{1-\frac{-W_L^2}{-x^{T2}P^2E_1^2}}\,\right)\bigg\}.\nonumber \\
& &
\eea
\par
For sectors of quantum numbers where $W_L^2=0$, the expressions of the 
potentials become:
\bea 
\lb{6e22}
& &A_{1L}+A_{2L}=\frac{1}{2}(\epsilon(p_{1L})+\epsilon(p_{2L}))
\sigma\sqrt{-x^{T2}},\\
\lb{6e23}
& &A_{1\mu}^{Tt}=-\frac{1}{E_1E_2}\Big(\frac{1}{3}(E_1+E_2)-
\frac{1}{2}E_1\Big)p_{\mu}^{Tt}\sigma\sqrt{-x^{T2}},\nonumber \\
& &A_{2\mu}^{Tt}=+\frac{1}{E_1E_2}\Big(\frac{1}{3}(E_1+E_2)-
\frac{1}{2}E_2\Big)p_{\mu}^{Tt}\sigma\sqrt{-x^{T2}}.
\eea
All expressions of the potentials have been written as
classical functions of their arguments, without taking into
account ordering problems. These necessitate a detailed study which will not
be done here. We simply outline some general features that may be useful for
the resolution of the wave equation. (i) Many ordering problems that concern
linear momentum operators do not affect the energy eigenvalues and rather
concern the definition of the kernel of the scalar product of the wave
functions; one can pass from one definition to the other by appropriate
changes of function.
(ii) The square-root and $\arcsin$ functions which involve the variables
$x^{T2}$ and $1/E_a^2$ ($a=1,2$) could be
treated in first approximation by replacing in $E_a$ the radial momentum
operator squared by its mean value in the bound state, or, if the resolution
is done in momentum space by replacing $x^{T2}$ by its mean value.
(iii) A close study
of the chiral properties of the wave equation suggests us to further adopt 
the following rules: the doubly transverse momentum operator $p_{\mu}^{Tt}$,
maintaining its definition of Eq. (\rf{6e16}), and the energy sign operators
$\epsilon(p_{aL})$ $(a=1,2)$ should be placed on the utmost left.
\par
Equation (\rf{6e13}), together with expressions (\rf{6e19}), is very similar
to an equation proposed by Olsson \textit{et al.} on the basis of a model
where quarks are attached at the ends of a straight string or a color flux
tube \cite{ow,lcooowoo}; the difference mainly concerns the energy sign
operators; in Ref. \cite{ow} the equation which was proposed is the
Salpeter equation \cite{s}, in which generally the potential is proportional
to a global energy projecter; here, the energy sign operators, though they
could be expressed through energy projectors, do not match exactly the
projector of the Salpeter equation; the doubly transverse parts of the vector
potentials do even not have energy sign operators. Apart from this slight
difference, however, the physical significance of Eq. (\rf{6e13}) is the
same as that of Ref. \cite{ow}. The vector potentials $A_{a\mu}$ ($a=1,2$)
can be interpreted as representing contributions of the energy-momentum
vector of the straight color flux tube with variable length $r$; its energy
is represented by the sum $A_{1L}+A_{2L}$ [Eq. (\rf{6e21})], while its 
angular momentum contributes through the doubly transverse components 
$A_a^{Tt}$.
\par
The norm of the wave function $\psi$ can be obtained (after a few 
approximations) from Eq. (\rf{6e8}). The details are presented in appendix
\rf{ap2}. The result is:
\be \lb{6e24}
\int d^3x^T\mathrm{tr}\psi^{\dagger}\frac{1}{2}\Big(\frac{h_{1}}{E_1}+
\frac{h_{2}}{E_2}\Big)\psi=2P_LN_c,
\ee
where $h_a$ and $E_a$, $a=1,2$, are the free Dirac hamiltonian and energy of
each particle, including now the self-energy contributions [cf. Sec. \rf{s7}
and Eqs. (\rf{7e9})-(\rf{7e11})].
\par
Finally, let us mention that Eq. (\rf{6e9}) can be used to determine the
relative time evolution of the wave function $\Phi$ [Eq. (\rf{6e2})].
Using in the right-hand side of Eq. (\rf{6e9}) the same types of
approximation as in Eq. (\rf{6e8}), taking the large separation time
limit between the pairs of points $(x_1,x_2)$ and $(x_1',x_2')$ and
passing to the bound state wave function, one can integrate Eq. (\rf{6e9})
for the latter obtaining the relative time dependence in the form of
an ordered exponential function involving the various operators and
potentials appearing in the equation, the initial condition being
given by the function $\psi(P_L,x^T)$ [Eq. (\rf{6e13a})]. We shall not 
need, however, that expression in the present work.
\par

\section{Quark self-energy} \lb{s7}
\setcounter{equation}{0}

In general, there are functional relations between kernels of two-particle
Green functions and self-energies of the constituent particles \cite{bjl}.
In conventional Green function equations (defined without phase factors),
one easily factorizes the self-energy contributions and incorporates them
into the external propagator contributions. Furthermore, self-energy
parts can themselves be evaluated by means of Schwinger--Dyson
equations. In the case of gauge invariant Green functions, however, the
factorization of self-energy contributions becomes a hard task, since the
path-ordered phase factors and the resulting Wilson loops maintain all
interacting pieces into global entities. The presence of self-energies is,
however, necessary for a consistent study of the properties of the
system under given symmetries, transcribed usually  into Ward--Takahashi
identities. In the present case, the symmetry that is of interest is
chiral symmetry. Also, when the dynamical equations of confining 
interactions are expressed in momentum space, the presence of self-energy 
contributions becomes necessary to remove the infrared divergences from 
observable quantities \cite{th2}.
\par
When the interaction kernel of the bound state equation is
represented by the mediation of an effective propagator, one expects
to obtain the self-energy contribution by contracting the lines of the
outgoing particles (assumed to be of the same type) through a single
particle propagator and forming a loop with the kernel. In the present case,
the interaction part of the bound state equation (\rf{6e13}) can be
visualized by multiplying it back by the factor $\gamma_{1L}\gamma_{2L}$.
It has three different tensor parts: the first corresponds to vertices
with the matrices $\gamma_{1L}\gamma_{2L}$; the second and third to
vertices with matrices $\gamma_1^{Tt}.p_1^{Tt}\gamma_{2L}$ and
$\gamma_{1L}\gamma_2^{Tt}.p_2^{Tt}$, respectively. When these terms
are incorporated in a two-point loop and integrated in momentum space, the
non-invariant pieces under spatial rotations disappear and what remains
is simply the part of the interaction with the $\gamma_{1L}\gamma_{2L}$
matrices, inside which also the angular momentum operator has
disappeared (cf. Eq. (\rf{6e22})). This corresponds to the
situation where the interaction kernel is generated by the mediation
of a Coulomb-gluon propagator, proportional in $x$-space to 
$\delta(x_L)\sqrt{-x^{T2}}$. 
\par
The above result also coincides in form with that obtained in two-dimensional
QCD. Here, in the large $N_c$ limit, one has two different but equivalent ways 
of obtaining the bound state equation: either by working with the 
Schwinger--Dyson approach in the axial gauge, or by working with the Wilson
loops with minimal surfaces \cite{kkk}.
(It is even easier to work in the light cone gauge \cite{th2}, but the
latter is less useful for our purpose here.) The Bethe--Salpeter equation
with the one-gluon exchange kernel and the corresponding self-energies
becomes an exact bound state equation and because of the instantaneity of 
the propagator, it yields the Salpeter equation \cite{s}. The procedure
developed in the present paper is also applicable to two-dimensional QCD.
It is sufficient to remove from the results that were obtained the 
rotational motion part. In this case, the interaction potential reduces
to the expression of Eq. (\rf{6e22}) which is nothing but Salpeter's 
kernel. What misses however in the bound state equation is the corresponding
self-energy contributions. We shall now show that Eq. (\rf{6e8}), which
is at the origin of the bound state equation, also yields the quark
self-energies.
\par
Since the formal expressions of the self-energies are the same in both
two and four dimensions in $x$-space, we shall directly work in two
dimensions, by freezing the transverse variables with respect to $x$
and considering the simplified case of a plane. A self-energy contribution
for particle 1 will be recognized as depending only on the variables of
that particle, not making reference to the variables of particle 2, except
for the directions of the total momentum $P$ of the bound state, which is
chosen as the reference timelike direction, and of the relative coordinate
$(x_2-x_1)$. The simplest contribution to the self-energy
of particle 1 comes from the term $G_{2,1}$ in Eq. (\rf{6e8}), a typical
contour of which is shown in Fig. \rf{7f1}.a.
\par
\bfg
\vspace*{0.5 cm}
\bc
\input{7f1.pstex_t}
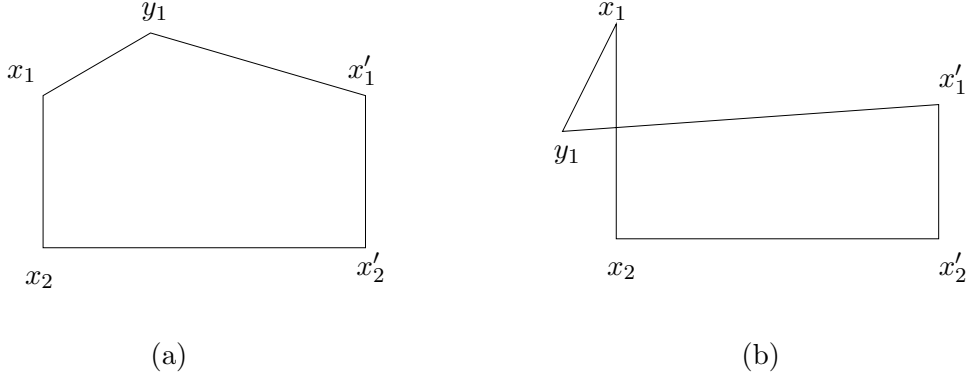
\caption{(a) A typical contour $C_{2,1}$ associated with the term $G_{2,1}$
of the two-particle propagator. (b) A configuration where the segments
$x_1x_2$ and $y_1x_1'$ intersect.}
\lb{7f1}
\ec
\efg
The term that is relevant here is that in which the operator
$\delta/\delta x(\sigma)$ (multiplying the $\gamma_1$ matrices) acts
on the term $\delta A_{2,1}/\delta x(\tau)$ of $G_{2,1}$ [Eq. (\rf{5e4})].
One thus obtains the second-order functional derivative
$\delta^2A_{2,1} /\delta x^{\mu}(\sigma)\delta x^{\alpha}(\tau)$, where
$\tau$ parametrizes the segment $y_1x_1'$ and $\sigma$ the segment
$x_1x_2$. The expression of the second-order functional derivative
of the minimal area has been given in Eq. (\rf{3e15}), where now
$x(\sigma',0)$ and $x(\sigma,0)$ have to be replaced by $x(\sigma)$ and
$x(\tau)$, respectively. Since $x(\sigma)$ and $x(\tau)$ belong to 
different segments, the terms proportional to the explicit delta-functions
can be dropped; furthermore, since the orthogonal variables to the
surface have been frozen, also the last term of that equation can be 
ignored; it is only the second term of the right-hand side of the 
equation that may contribute:
\bea \lb{7e1}
& &\frac{\delta}{\delta x^{\mu}(\sigma)}
\frac{\delta A_{2,1}}{\delta x^{\alpha}(\tau)}=
F\Big(x(\sigma)-x(\tau)\Big)\nonumber \\
& &\ \ \times\Big(\delta_{\lambda\alpha}(x_2-x_1)^{\nu}-
\delta_{\nu\alpha}(x_2-x_1)^{\lambda}\Big)
\Big(\delta_{\lambda\mu}(x_1'-y_1)^{\nu}-
\delta_{\nu\mu}(x_1'-y_1)^{\lambda}\Big),
\eea
where $x(\tau)$ and $x(\sigma)$ are parametrized as
$x(\tau)=y_1+\tau(x_1'-y_1)$ and $x(\sigma)=x_1+\sigma(x_2-x_1)$.
In the limit when the regulator $a$ vanishes, the function $F$, Eq.
(\rf{3e3}), tends to a \textit{two-dimensional} delta-function 
[Eq. (\rf{3e3a})]; therefore, the above second-order derivative is 
nonvanishing only when the two segments
are intersecting (Fig. \rf{7f1}.b); we then have to integrate with respect
to $\sigma$ and $\tau$ with the weight factors $(1-\sigma)$ and $(1-\tau)$
[Eqs. (\rf{5e4}) and (\rf{6e8})]. The calculation can be done by
replacing $F$ by a two-dimensional delta-function involving $\sigma$ and
$\tau$ and the arguments of which can be obtained by first writing the
explicit expression of $(x(\sigma)-x(\tau))^2$ with two-dimensional
components $x^0$ and $x^3$ (in the bound state rest frame) for instance:
\bea \lb{7e2}
F\Big(x(\sigma)-x(\tau)\Big)&=&\delta\Big(x^0(\sigma)-x^0(\tau)\Big)
\delta\Big(x^3(\sigma)-x^3(\tau)\Big)\nonumber \\
&=&\delta(z^0+\sigma x^0-\tau y^0)\,\delta(z^3+\sigma x^3-\tau y^3)
\nonumber \\
&=&-\frac{1}{2x^3y^3}\delta(z^0+\sigma x^0-\tau y^0)
\frac{\partial}{\partial \sigma}\frac{\partial}{\partial \tau}
|z^3+\sigma x^3-\tau y^3|,
\eea
where we have defined $x=x_2-x_1$, $z=x_1-y_1$, $y=x_1'-y_1$.
Actually not the full expression of the above integral is needed, but rather
that part which depends only on the points $x_1$ and $y_!$, which correspond
to the integration endpoints $\sigma=0$ and $\tau=0$. Integrating that
expression by parts and retaining only the latter endpoints, one obtains:
\be \lb{7e3}
\int_0^1 d\sigma d\tau\,(1-\sigma)(1-\tau)F\Big(x(\sigma)-x(\tau)\Big)
=-\frac{1}{2x^3y^3}\delta(z^0)|z^3|+\ldots,
\ee
where the dots stand for the remaining terms that do not contribute to the
self-energy. The contribution of the integral of the second-order 
derivative of the area, Eq. (\rf{7e1}), is then:
\bea \lb{7e4}
& &\int_0^1 d\sigma d\tau\,(1-\sigma)(1-\tau)
\frac{\delta}{\delta x^{\mu}(\sigma)}
\frac{\delta A_{2,1}}{\delta x^{\alpha}(\tau)}=
-\delta_{\mu 0}\delta_{\alpha 0}\delta(z^0)|z^3|\nonumber \\
& &\ \ \ \ \ \ \ \ \ =-\delta_{\mu L}\delta_{\alpha L}
\delta((x_{1L}-y_{1L})\sqrt{-(x_1^T-y_1^T)^2},
\eea
where we have dropped additive terms not depending only on $z$ and
have done the tensor calculation in two dimensions and restored at the
end the covariant expression. 
\par
The presence of the delta-function along
the temporal direction means that the segments $x_1x_2$ and $y_1x_1$
are parallel, or more generally lie in an orthogonal plane to the time
direction (Fig. \rf{7f2}).
\par
\bfg
\vspace*{0.5 cm}
\bc
\input{7f2.pstex_t}
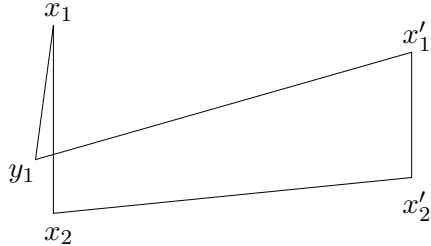
\caption{Configuration of the contour $C_{2,1}$ corresponding to the
self-energy contribution.}
\lb{7f2}
\ec
\efg
In the limit $(x_{1L}-x_{1L}')\rightarrow \infty$, the areas
$A_{2,1}(x_2'x_2x_1y_1x_1'x_2')$ and $A_{1,1}(x_2'x_2y_1x_1'x_2')$ become
almost equal and one may replace in $G_{2,1}$ the former by the latter.
The expression (\rf{7e4}), combined with the multiplicative free quark
propagator $S_{10}(x_1y_1)$, then factorizes $G_{1,1}$ and plays the
role of a self-energy correction. In order to complete the derivation,
one must repeat the same calculations with the higher-order terms
$G_{i,j}$ ($i\geq 3$, $j\geq 1$), where in the second-order derivatives of
the type of (\rf{7e1}), the functional derivative $\delta/\delta x(\tau)$
corresponds to $\delta/\delta x(\tau_1)$ and acts on the second segment
$y_1y_2$ of the corresponding contour (Fig. \rf{5f1}).
One thus finds the first-order self-energy correction (in the string
tension $\sigma$) that factorizes the bound state wave function in Eq. 
(\rf{6e13}); it corresponds to the effective exchange of a 
Coulomb-gluon, with propagator proportional in momentum space to 
$\delta_{\mu L}\delta_{\alpha L}1/((-q^T)^2)^2$.
\par
The higher-order self-energy corrections are obtained by extending the
above procedure to terms that contain higher numbers of second-order
derivatives of the areas $A_{i,j}$. The next correction is provided
by the terms $G_{i,j}$ with $i\geq 4$ and containing two second-order
derivatives of the area. Let us consider for definiteness the term
$G_{4,1}$ corresponding to the contour $C_{4,1}$ (Fig. \rf{7f3}.a).
\par
\bfg
\vspace*{0.5 cm}
\bc
\input{7f3.pstex_t}
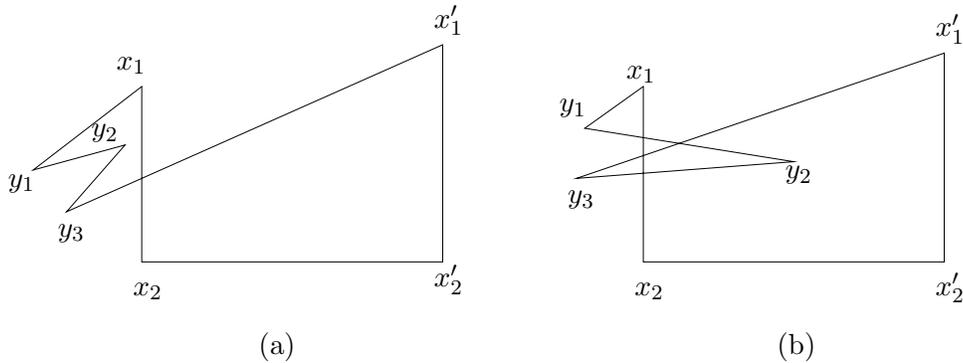
\caption{(a) A configuration of the contour $C_{4,1}$ corresponding to the
second-order self-energy correction. (b) A configuration corresponding
to overlapping integrals.}
\lb{7f3}
\ec
\efg
The self-energy contribution comes from the term containing the
product $\frac{\delta^2 A_{4,1}}{\delta x(\sigma)\delta x(\tau_3)}$
$\frac{\delta^2 A_{4,1}}{\delta x(\tau_1)\delta x(\tau_2)}$, where
$\delta/\delta x(\tau_1)$ acts on the segment $y_1y_2$,
$\delta/\delta x(\tau_2)$ on $y_2y_3$ and
$\delta/\delta x(\tau_3)$ on $y_3x_1'$. The resulting delta-functions
imply intersection of the segments $x_1x_2$ and $y_3x_1'$ and parallelism
of the adjacent segments $y_1y_2$ and $y_2y_3$. The two delta-functions
can be integrated as before in an independent way. The result is
proportional to the product $\delta((x_{1L}-y_{3L})\sqrt{-(x_1^T-y_3^T)^2}
\,\delta((y_{1L}-y_{2L})\sqrt{-(y_1^T-y_2^T)^2}$.
Notice that in the case of overlapping integrals, corresponding to the
situation where two derivatives act on the segments $x_1x_2$ and $y_2y_3$
on the one hand and on the segments $y_1y_2$ and $y_3x_1'$ on the other,
the delta-functions would imply intersections of the segments of
each of the above pairs (Fig. \rf{7f3}.b); integrating the delta-function
of the second pair as before yields the delta-function
$\delta(y_{2L}-y_{3L})$, the implementation of which prevents the
segments $x_1x_2$ and $y_2y_3$ from intersecting and the corresponding
integral vanishes. Also notice that in the two-dimensional limit
that we are considering here (freezing of orthogonal deformations of the
areas) higher-order derivatives of the areas vanish, for they would involve 
derivatives of the function $F$ outside its support.
\par
The generalization to higher orders is now straightforward. Any high-order
contribution will contain in the term $G_{i,j}$ a product of second-order
derivatives of the area $A_{i,j}$, in which one of them is
$\frac{\delta^2 A_{i,j}}{\delta x(\sigma)\delta x(\tau_k)}$ ($k\leq i-1$),
where $\delta/\delta x(\tau_k)$ acts on the segment $y_ky_{k+1}$; the
others are nonoverlapping (disjoint or nested) and act on segments
lying between $x_1$ and $y_{k+1}$, being thus nested within $x_1y_{k+1}$.
The remaining part of the calculation repeats the steps used for the
first-order correction. 
\par
One thus generates the whole series of self-energy corrections that can
be summed into the Schwinger--Dyson integral equation, corresponding to
a Coulomb instantaneous kernel $-\sigma\delta_{\mu L}\delta_{\nu L}
\delta(x_L-y_L)\sqrt{-(x^T-y^T)^2}$. 
\par
Defining the momentum space Fourier transform  of the function 
$\sqrt{-x^{T2}}$ through analytic continuation of the power parameter 
\cite{gsh}, one finds:
\be \lb{7e5}
\int d^3x^Te^{ip^T.x^T}\sqrt{-x^{T2}}=-\frac{8\pi}{(-p^{T2})^2}.
\ee
Denoting by $S(p_L,p^T,m)$ the quark propagator in momentum space with
free mass $m$ tied to the bound state of momentum $P$ and defining the
self-energy contribution through the equation
\be \lb{7e6}
iS^{-1}(p_L,p^T,m)=\gamma.p-m-\Sigma(p^T,m),
\ee
one has the Schwinger--Dyson equation:
\be \lb{7e7}
\Sigma(p^T,m)=8\pi\sigma\gamma_L\int\frac{d^4k}{(2\pi)^4}S(k_L,k^T,m)
\frac{1}{(-(p^T-k^T)^2)^2}\gamma_L.
\ee
Because of the instantaneity of the kernel, the self-energy $\Sigma$
actually depends only on the three-dimensional transverse momentum $p^T$.
The tensor decomposition of $\Sigma$ can be written in the form
\be \lb{7e8}
\Sigma(p^T,m)=\gamma^T.p^TA(-p^{T2},m)+B(-p^{T2},m).
\ee
\par
The self-energies of the quark and of the antiquark have now to be
incorporated in the Dirac energy operators in Eq. (\rf{6e13}) and in the
definitions of the energy sign operators (\rf{6e20}). Defining
\bea 
\lb{7e9}
h_1(p_1^T,m_1)&=&h_{10}+\gamma_{1L}\Sigma(p_1^T,m_1)=
\gamma_{1L}(m_1-\gamma_1^T.p_1^T+\Sigma(p_1^T,m_1))\nonumber \\
&=&\gamma_{1L}\Big((m_1+B_1)-\gamma_1^T.p_1^T(1-A_1)\Big),\nonumber \\
h_2(p_2^T,m_2)&=&h_{20}-\gamma_{2L}\Sigma(-p_2^T,m_2)=
-\gamma_{2L}(m_2+\gamma_2^T.p_2^T+\Sigma(-p_2^T,m_2))\nonumber \\
&=&-\gamma_{2L}\Big((m_2+B_2)+\gamma_2^T.p_2^T(1-A_2)\Big),\\
\lb{7e10}
E_1(-p_1^{T2},m_1)&=&\sqrt{h_1^2}=\sqrt{(m_1+B_1)^2-p_1^{T2}(1-A_1)^2},
\nonumber \\
E_2(-p_2^{T2},m_2)&=&\sqrt{h_2^2}=\sqrt{(m_2+B_2)^2-p_2^{T2}(1-A_2)^2},\\
\lb{7e11}
\epsilon(p_{aL})&=&\frac{h_a}{E_a},\ \ \ \ \  a=1,2,
\eea
equation (\rf{6e13}) becomes
\be \lb{7e12}
\Big[P_L-(h_1+h_2)-\gamma_{1L}\gamma_1^{\mu}A_{1\mu}
-\gamma_{2L}\gamma_2^{\mu}A_{2\mu}\Big]\psi(P_L,x^T)=0,
\ee
where the potentials $A$ are given by expressions (\rf{6e19}) in which
$E_a$ and $\epsilon_a$ ($a=1,2$) are replaced by expressions
(\rf{7e10}) and (\rf{7e11}), respectively; furthermore, the self-energy
functions should also be incorporated in the appearances of the doubly
transverse momentum $p^{Tt}$ and the orbital angular momentum $W_L$.
Since we have replaced $p_1^{Tt}$ and $p_2^{Tt}$ in favor of $p^{Tt}$,
the corresponding substitutions may not seem straightforward. However, the
difficulty is circumvented easily. Each energy factor $E_a$ ($a=1,2$) that
appears explicitly in the potentials is reminiscent of a term of the type
$p_a^{Tt}/E_a$; $p_a^{Tt}$ undergoes the substitution
$p_a^{Tt}\rightarrow (1-A_a)p_a^{Tt}$, where $A_a$ ($a=1,2$) is the 
self-energy function, Eqs. (\rf{7e9}). It is then sufficient to replace
each $E_a$ in the potentials by $E_a/(1-A_a)$, without modifying the 
momentum and orbital angular momentum operators.
\par
The quark propagator is not a gauge invariant quantity and therefore could 
not lead to observable effects. This feature manifests itself through the
self-energy equation (\rf{7e7}), which, due to the infrared singularity
of the integrand, provides an infrared divergent self-energy. On the
other hand, physical observables should be free of infrared singularities.
The treatment of infrared singularities in momentum space depends on the
method of regularization that is adopted. Dimensional regularization, which
has the advantage of preserving symmetry properties, and which we adopt
throughout this work, gives zero for the abovementioned singularities, but
has the disadvantage of hiding the distinction between unobservable and
observable quantities. This is why it is necessary to check at every 
stage of calculations whether physical quantities are indeed free of
infrared singularities by also using an explicit infrared cutoff method.
\par
In the present approach, we have to check that the wave equation
(\rf{7e12}), which describes the physical properties of the bound states,
is free of infrared singularities. To do this, we have to isolate the
singularities coming from the self-energies. The properties of the latter
will be studied in Sec. \rf{s8}; here, we mention the most relevant ones for
our purpose. Designating by $V$ the linear confining potential,
Eq. (\rf{8e6}), it is found that the singularities of the self-energy
parts are contained in the energy factors $E_a$ [Eqs. (\rf{7e10})],
while the ratios $(1-A_a)/(m_a+B_a)$, $a=1,2$, are
free of singularities; from the integral equations satisfied by $E_a$,
Eq. (\rf{8e10}), it is seen that their singularity is represented by the
three-dimensional integral in momentum space of $-V/2$. [Section \rf{s8}
deals with the equal-mass case and mainly with the massless limit; however,
the infrared singularity properties are not affected by the values of the
quark masses and could be abstracted from the equations of that section.]
Therefore, the combinations $(E_a+V/2)$ are free of infrared singularities.
Coming back to the potentials $A_{a\mu}$ [Eqs. (\rf{6e19})] their
large-distance behavior in $x$-space can be studied by expanding the
various functions contained in their expressions in terms of
$(1-A_a)^2\mathbf{L}^2/(E_a^2\mathbf{x}^2)$, $a=1,2$ (in the c.m. frame).
Factorizing the function $\sigma r$, the expansions are of the type
$\sigma r(1+O((1-A_a)^2\mathbf{L}^2/(E_a^2\mathbf{x}^2)))$; therefore the
higher-order terms are nondominant at large distances and could not
lead to infrared singularities. For the present study it is
sufficient to keep the leading terms, which are represented by the
expressions (\rf{6e22})-(\rf{6e23}).
\par
Considering first the contribution of $(A_{1L}+A_{2L})$ and replacing
the energy sign operators by their expressions (\rf{7e11}), we immediately
find that the energy factors and the potential appear with the combinations
$(E_a+V/2)$, which ensure the infrared finiteness of the result.
Next, we consider the contributions of the spacelike potentials
$A_{a\mu}^{Tt}$ [Eqs. (\rf{6e23}), where each $E_a$ should be replaced by
$E_a/(1-A_a)$]. The doubly transverse momentum operator $p^{Tt}$ is equal
in the c.m. frame to
$\mathbf{p}^t=-\frac{1}{\mathbf{x}^2}\mathbf{x}\times\mathbf{L}$.
The effective potential that matters is proportional to
$\sigma r \mathbf{p}^t=-\frac{\sigma}{r}\mathbf{x}\times\mathbf{L}$.
The orbital angular momentum operator $\mathbf{L}$ acts on spherical
harmonics and does not modify the infrared properties of wave functions.
The term $\mathbf{x}/r$ has as Fourier transform a function proportional
to $\mathbf{p}/p^4$, which does not lead to infrared singularities, after
the angular integrations in convolution integrals are done. Therefore,
the wave equation (\rf{7e12}) is free of infrared singularities.
Explicit cases of the above cancellations can also be found in the 
equations presented in Sec. \rf{s8}.
\par

\section{Chiral symmetry breaking} \lb{s8}
\setcounter{equation}{0}

The presence of the self-energy contributions in the bound state
equation (\rf{7e12}) allows us to study the possibility of the
spontaneous breakdown of chiral symmetry. This is intimately related to
the existence of nonperturbative solutions to the Schwinger--Dyson
equation in the chiral limit when the quark mass $m$ tends to zero 
\cite{njl,bjl}.
\par
Using decomposition (\rf{7e8}) and considering the bound state rest
frame, the integral in Eq. (\rf{7e7}) can first be integrated with
respect to the energy variable $k_0$ giving rise to two coupled equations:
\bea 
\lb{8e1}
B(p)&=&-\frac{1}{2}\int\frac{d^3k}{(2\pi)^3}V(\mathbf{p}-\mathbf{k})
\frac{(m+B(k))}{\Big(k^2(1-A(k))^2+(m+B(k))^2\Big)^{1/2}},\\
\lb{8e2}
\mathbf{p}A(k)&=&+\frac{1}{2}\int\frac{d^3k}{(2\pi)^3}
V(\mathbf{p}-\mathbf{k})\frac{\mathbf{k}(1-A(k))}
{\Big(k^2(1-A(k))^2+(m+B(k))^2\Big)^{1/2}},
\eea
where we have defined the potential $V$ as
\be \lb{8e6}
V(\mathbf{p})=-8\pi\sigma\frac{1}{p^4},\ \ \ \ p=|\mathbf{p}|,
\ \ \ \ \ \ \ V(\mathbf{x})=\sigma r,\ \ \ \ r=|\mathbf{x}|.
\ee
In the chiral limit $m=0$ they become:
\bea
\lb{8e3}
B(p)&=&-\frac{1}{2}\int\frac{d^3k}{(2\pi)^3}V(\mathbf{p}-\mathbf{k})
\frac{B(k)}{\Big(k^2(1-A(k))^2+B(k)^2\Big)^{1/2}},\\
\lb{8e4}
\mathbf{p}A(p)&=&+\frac{1}{2}\int\frac{d^3k}{(2\pi)^3}
V(\mathbf{p}-\mathbf{k})\frac{\mathbf{k}(1-A(k))}
{\Big(k^2(1-A(k))^2+B(k)^2\Big)^{1/2}}.
\eea
\par
These equations were extensively studied in the literature; they result
from the assumption that confinement is due to the exchange of Coulomb-gluons
\cite{fgmw,alyopro,ad,aa,lg}. Using variational methods, it has been
shown that the perturbative vacuum state is unstable under quark-antiquark
pair creation. The existence of a new stable vacuum state is ensured by 
the existence of a nontrivial solution to Eqs. (\rf{8e3})-(\rf{8e4}).
We summarize below the main results that have been obtained and outline
some salient features related to physical aspects of the problem.
\par
Equations (\rf{8e3})-(\rf{8e4}) are solved by decomposing the functions
$B$ and $p(1-A)$ along polar combinations, by introducing an angle and
a modulus:
\be \lb{8e5}
B(p)=E(p)\sin\varphi(p),\ \ \ \ \ p(1-A(p))=E(p)\cos\varphi(p).
\ee
Equations (\rf{8e3})-(\rf{8e4}) can be rewritten in the form:
\bea 
\lb{8e7}
E(p)\sin\varphi(p)&=&-\frac{1}{2}\int\frac{d^3k}{(2\pi)^3}
V(\mathbf{p}-\mathbf{k})\sin\varphi(k),\\
\lb{8e8}
E(p)\cos\varphi(p)&=&p-\frac{1}{2}\int\frac{d^3k}{(2\pi)^3}
V(\mathbf{p}-\mathbf{k})\mathbf{\hat p}.\mathbf{\hat k}\cos\varphi(k),
\ \ \ \ \ \ \mathbf{\hat p}=\frac{\mathbf{p}}{p},\ \ 
\mathbf{\hat k}=\frac{\mathbf{k}}{k}.
\eea
The latter equations in turn can be recombined to decouple the function
$\varphi$ from $E$:
\bea
\lb{8e9}
p\sin\varphi(p)&=&-\frac{1}{2}\int\frac{d^3k}{(2\pi)^3}
V(\mathbf{p}-\mathbf{k})
\Big[\sin\varphi(k)\cos\varphi(p)-\mathbf{\hat p.\hat k}
\cos\varphi(k)\sin\varphi(p)\Big],\\
\lb{8e10}
E(p)&=&p\cos\varphi(p)-\frac{1}{2}\int\frac{d^3k}{(2\pi)^3}
V(\mathbf{p}-\mathbf{k})\Big[\sin\varphi(k)\sin\varphi(p)+
\mathbf{\hat p.\hat k}\cos\varphi(k)\cos\varphi(p)\Big].\nonumber \\
& &
\eea
\par
From Eq. (\rf{8e9}) one deduces that the function $\varphi$ is an
infrared finite quantity. From Eq. (\rf{8e10}), after expanding in
the integrand $\mathbf{k}$-dependent terms around $\mathbf{p}$, one deduces
that $E$ is infrared singular and its singularity is represented by the
integral $-\frac{1}{2}\int d^3kV(k)/(2\pi)^3$, a property mentioned and
utilized at the end of Sec. \rf{s7} for the checking of the infrared
finiteness of the wave equation (\rf{7e12}). (We use, however, 
dimensional regularization for analytic calculations throughout this work.)
\par
The function $\sin\varphi(p)$ wil be identified later with the Goldstone
boson wave function. Therefore, it should be a normalizable and
presumably nodeless function. An analysis of the above equations shows 
that the function $\varphi$ behaves at infinity as $p^{-5}$ and tends at
the origin to $\pi/2$. The solution that is found is indeed a monotonically
decreasing function with the above properties. On the other hand, due
to the infrared singularity of the potential $V$, the energy function $E$
vanishes at some finite value $p_0$ of $p$ and becomes negative for 
$p<p_0$; it tends to a finite negative value at the origin and behaves as
$p$ at infinity. The functions $B$ and $p(1-A)$ vanish simultaneously at
$p_0$; they are negative for $p<p_0$, with the limiting values $B(0)=E(0)$
and $\lim_{p=0}p(1-A)=0$; asymptotically, $B$ behaves as $p^{-4}$ and $A$
as $p^{-2}$.
\par
An order parameter for chiral symmetry breaking is the quark condensate
$<\overline\psi\psi>$. For a given type of quark, say $u$, it can be
defined as minus the trace (in color and Dirac spinor spaces) of the quark 
propagator at the origin in $x$-space:
\be \lb{8e11}
<\overline uu>=-\mathrm{tr_{c,sp}}S(x)\Big|_{x=0}=-N_c\int
\frac{d^4p}{(2\pi)^4}\mathrm{tr_{sp}}S(p).
\ee
Using definition (\rf{7e6}) and decomposition (\rf{7e8}), one finds:
\be \lb{8e12}
<\overline uu>=-2N_c\int\frac{d^3p}{(2\pi)^3}\sin\varphi(p).
\ee
From Eq. (\rf{8e7}) one also sees that the quark condensate enters in
the asymptotic behavior of the mass term of the quark propagator in the
chiral limit:
\be \lb{8e12a}
p\sin\varphi(p)_{\stackrel{{\displaystyle \simeq}}{p\rightarrow \infty}}
-\frac{2\pi\sigma}{N_c}\frac{<\overline uu>}{p^4}.
\ee
This relation is the analog of that obtained in perturbation
theory from operator product expansion and the renormalization group
analysis \cite{pl}.
\par
We next turn to the bound state equation (\rf{7e12}). A consistent study of
the chiral limit should be done starting from the situation where the quark
masses are different from zero and then taking the limits $m_1=m_2=0$. In
the present case, we directly consider the equal mass case $m_1=m_2=m$,
corresponding to the light quark sectors $u$ and $d$ without isospin
breaking.
\par
In general, the resolution of Eq. (\rf{7e12}) proceeds by first
decomposing the wave function $\psi$ along a basis of $2\times 2$ matrices
spanned by the matrices $\gamma_L$ and $\gamma_5$. In the following, we
consider the rest frame of the bound state ($\mathbf{P=0}$). One has the
decomposition
\be \lb{8e13}
\psi=\psi_1+\gamma_0\psi_2+\gamma_5\psi_3+\gamma_0\gamma_5\psi_4,
\ee
where the functions $\psi_a$, $a=1,\ldots,4$ are themeselves $2\times 2$
matrices on which act the spin Pauli matrices $\mbox{\boldmath$\sigma$}$.
The quantum numbers of the state are usually defined with respect to the
components that survive in the nonrelativistic limit; these are $\psi_3$
and $\psi_4$, which have the same quantum numbers. For the present problem,
we are considering the sector characterized by the following quantum
numbers: total spin $s=0$, orbital angular momentum $\ell=0$, total
angular momentum $j=0$. We adopt the rule of writing the energy sign 
operators present in the potentials [Eqs. (\rf{6e19}) and 
(\rf{7e10})-(\rf{7e11})] as well as the factors $((1-A)/E)\mathbf{p}^t$ on
the utmost left. The doubly transverse momentum operator $\mathbf{p}^t$ 
[Eq. (\rf{6e16})] annihilates $S$-states; it commutes with $\mathbf{x}^2$ 
and with all scalar operators that act on $S$-states in $x$-space. In the
equal-mass case, the transverse potentials
$\mathbf{A}_1^t$ and $\mathbf{A}_2^t$ are opposite to each other; defining
\be \lb{8e14}
\mathbf{A}_2^t=\frac{(1-A)}{E}\mathbf{p}^t\widetilde A=-\mathbf{A}_1^t,
\ee
using definition (\rf{8e6}), introducing the spin
\textit{operators} $\mathbf{s}_1$ and $\mathbf{s}_2$ of particles 1 and 2,
and removing indices 1 and 2 from various factors which are equal, we obtain
the following four equations:
\bea 
\lb{8e15}
& &P_0\psi_1+2(1-A)(\mathbf{s}_1-\mathbf{s}_2).\mathbf{p}
\Big(1+\frac{1}{2E}V\Big)\psi_3=0,\\
\lb{8e16}
& &P_0\psi_2=0,\\
\lb{8e17}
& &P_0\psi_3-2(m+B)\Big(1+\frac{1}{2E}V\Big)\psi_4+2(1-A)
(\mathbf{s}_1-\mathbf{s}_2).\mathbf{p}\Big(1+\frac{1}{2E}V\Big)\psi_1
\nonumber \\
& &\ \ \ \ \ \ \ \ \ \ -2\frac{(1-A)}{E}(\mathbf{s}_1-\mathbf{s}_2).
\mathbf{p}^t\widetilde A\psi_1=0,\\
\lb{8e18}
& &P_0\psi_4-2(m+B)\Big(1+\frac{1}{2E}V\Big)\psi_3=0.
\eea
We have neglected in Eq. (\rf{8e17}) the internal $\mathbf{L}^2$ dependent
parts, present in the potentials $(A_{1L}+A_{2L})$ and $\mathbf{A}^t$ 
acting on $\psi_1$, which is a $P$-state, and used the limits 
(\rf{6e22})-(\rf{6e23}); these neglected parts do not seem to play a 
crucial role in the following calculations; with the decomposition 
(\rf{8e14}), $\widetilde A=\sigma r/6$.
\par
From Eqs (\rf{8e15}) and (\rf{8e18}) one deduces the relation:
\be \lb{8e19}
(m+B)\psi_1+(1-A)(\mathbf{s}_1-\mathbf{s}_2).\mathbf{p}\psi_4=0.
\ee
This equation, which is valid in general for $m\neq 0$ and $P_0\neq 0$,
should also be satisfied in the limit $m\rightarrow 0$ and for
$P_0\rightarrow 0$, if the Goldstone boson properties are expected to
vary smoothly under explicit chiral symmetry breaking.
Then, the ground state solution of the above equations in the limit
$m=0$ with zero energy is:
\be \lb{8e20}
\psi_1=\psi_2=\psi_4=0,\ \ \ \ \psi_3\neq 0,
\ee
with $\psi_3$ satisfying the equation
\be \lb{8e21}
E(p)\psi_3(p)=-\frac{1}{2}\int \frac{d^3k}{(2\pi)^3}
V(\mathbf{p}-\mathbf{k})\psi_3(k).
\ee
This equation is similar in form to Eq. (\rf{8e7}), indicating that 
$\psi_3$ is proportional to the self-energy function $\sin\varphi$:
\be \lb{8e22}
\psi_3(p)=C\sin\varphi(p),
\ee
with $C$ a constant.
\par
In order to calculate the pion decay constant $F_{\pi}$, one must again
consider the case where $m\neq 0$. The pion decay constant is defined as:
\be \lb{8e23}
<0|\overline d(0)\gamma_{\mu}\gamma_5u(0)|\pi^+(P)>=i\sqrt{2}F_{\pi}P_{\mu}.
\ee
This is related to the component $\psi_4$:
\be \lb{8e24}
\int \frac{d^3k}{(2\pi)^3}\psi_4(k)=\frac{\sqrt{2}}{4}F_{\pi}P_0.
\ee
The normalization condition (\rf{6e24}) yields, after elimination of 
$\psi_1$ through Eq. (\rf{8e19}) and replacement of $\psi_3$ by its 
expression (\rf{8e22}):
\be \lb{8e25}
8C\int\frac{d^3k}{(2\pi)^3}\psi_4(k)=2P_0N_c.
\ee
Using Eq. (\rf{8e24}), one finds:
\be \lb{8e26}
CF_{\pi}=\frac{N_c}{\sqrt{2}}.
\ee
This result is actually a consequence of the Ward-Takahashi identity 
relative to the Green function 
$<T\overline u(x)\gamma_{\mu}\gamma_5d(x)i\overline d(0)\gamma_5u(0)>$, 
which implies (in the chiral limit) the equation
$\sqrt{2}F_{\pi}<0|i\overline d(0)\gamma_5u(0)|\pi^+>=2<\overline uu>$;
this is equivalent to the relation
\be \lb{8e27}
\int \frac{d^3k}{(2\pi)^3}\psi_3(k)=-\frac{<\overline uu>}
{2\sqrt{2}F_{\pi}}.
\ee
Using Eq. (\rf{8e12}), one obtains Eq. (\rf{8e26}).
\par
The normalization condition (\rf{8e25}) can also be analyzed by using
Eq. (\rf{8e18}). Integrating that equation with respect to $\mathbf{p}$
and expanding in the second term all quantities with respect to $m$ to 
first-order in $m$ and then using the integral equations satisfied by 
the first-order self-energy functions, one ends up with the relation
\be \lb{8e28}
\int \frac{d^3k}{(2\pi)^3}\psi_4=\frac{2m}{P_0}
\int \frac{d^3k}{(2\pi)^3}\psi_3,
\ee
which is equivalent to the well-known relation of Gell-Mann, Oakes and
Renner \cite{gmor}: $m_{\pi}^2F_{\pi}^2=-2m<\overline uu>$.
\par
In order to be able to calculate $F_{\pi}$, one needs to know the function
$\psi_4$. This information comes from Eq. (\rf{8e17}). Eliminating $\psi_1$
through Eq. (\rf{8e19}) and noticing that $\psi_4$ iself is proprtional to
$P_0$, one can take in all factors multiplying $\psi_4$ the chiral limit.
Defining
\be \lb{8e29}
\psi_4\equiv P_0\widetilde \psi_4\equiv P_0\frac{1}{2}g\psi_3,
\ee
the function $g(p)$ then satisfies the integral equation
\bea \lb{8e30}
& &E(p)g(p)=\sin\varphi(p)\nonumber \\
& &\ \ \ \ -\frac{1}{2}\int\frac{d^3k}{(2\pi)^3}
V(\mathbf{p}-\mathbf{k})\Big[\sin\varphi(k)\sin\varphi(p)
+\mathbf{\hat p}.\mathbf{\hat k}\cos\varphi(k)\cos\varphi(p)\Big]g(k)
\nonumber \\
& &\ \ \ \ +\frac{1}{3p}\cos\varphi(p)\int\frac{d^3k}{(2\pi)^3}
V(\mathbf{p}-\mathbf{k})(\mathbf{p}-\mathbf{k}).\hat\mathbf{k}
\cos\varphi(k)g(k).
\eea
Except for the last term, which comes from the contribution of the
rotational motion of the flux tube, this equation is the same as a
corresponding one derived in Refs. \cite{fgmw,ad}.
The infrared singular part of $E$ cancels the singular part of the
first integral and the function $g$ appears as infrared finite. 
This is also seen more explicitly by using Eqs. (\rf{8e9})-(\rf{8e10});
after a few algebraic manipulations one can cast Eq. (\rf{8e30}) into the
form \cite{ad}
\bea \lb{8e31}
& &pg(p)\cos\varphi(p)=\sin\varphi(p)\nonumber \\
& &\ \ \ \ +\frac{1}{2}\int \frac{d^3k}{(2\pi)^3}V(\mathbf{p}-\mathbf{k})
\Big[\sin\varphi(k)\sin\varphi(p)+\mathbf{\hat p.\hat k}
\cos\varphi(k)\cos\varphi(p)\Big]\Big(g(p)-g(k)\Big)\nonumber \\
& &\ \ \ \ +\frac{1}{3p}\cos\varphi(p)\int\frac{d^3k}{(2\pi)^3}
V(\mathbf{p}-\mathbf{k})(\mathbf{p}-\mathbf{k}).\hat\mathbf{k}
\cos\varphi(k)g(k).
\eea
which is free of infrared singularities. The function $g$ has the same
asymptotic behavior as $\sin\varphi$ and turns out to be nodeless.
Integrating this equation with respect to $\mathbf{p}$ yields the relation
\bea \lb{8e32}
& &\int \frac{d^3k}{(2\pi)^3}kg(k)\cos\varphi(k)=
\int \frac{d^3k}{(2\pi)^3}\sin\varphi(k)\nonumber \\
& &\ \ \ \ \ +\int\frac{d^3p}{(2\pi)^3}\frac{1}{3p}\cos\varphi(p)
\int\frac{d^3k}{(2\pi)^3}
V(\mathbf{p}-\mathbf{k})(\mathbf{p}-\mathbf{k}).\hat\mathbf{k}
\cos\varphi(k)g(k),
\eea
which can be considered as a consistency check for various approximations
and numerical calculations. The resolution of Eqs. (\rf{8e30}) or
(\rf{8e31}) provides the function $g$ and hence $\psi_4$.
\par
A rough evaluation of the orders of magnitude of the quark condensate and
of the pion decay constant can be done using analytic approximations.
Taking into account the properties of the function $\varphi$ at infinity
and near the origin and its nodeless character, the following
approximations can be used inside the integrals:
\be \lb{8e33}
\sin\varphi(k)\simeq \frac{1}{(1+k^2/(b\sigma))^{5/2}},\ \ \ \ \ 
\cos\varphi(k)\simeq \frac{k}{\sqrt{b\sigma}}\bigg\{
\frac{1}{(1+k^2/(b\sigma))^{1/2}}+
\frac{\sqrt{5}-1}{(1+k^2/(b\sigma))^{5/2}}\bigg\},
\ee
where $b$ is a constant to be determined. For the numerical
applications we use $\sigma=0.18$ GeV$^2$, $\sqrt{\sigma}=424$ MeV.
With these approximations, the integrals in Eqs. (\rf{8e7})-(\rf{8e8})
can be evaluated analytically (with dimensional regularization) yielding
hypergeometric functions. The parameter $b$ is then determined by the 
requirement that the left-hand sides have the same zero. We find $b=0.31$.
The evaluation of the integral in Eq. (\rf{8e12}) (with approximation 
(\rf{8e33}) and $N_c=3$) gives $<\overline uu>=-(115\ \mathrm{MeV})^3$, in
agreement with the results found in Refs. \cite{ad,lg}, 
$<\overline uu>\simeq -(100\ \mathrm{MeV})^3$. 
\par
Since the function $g$ has similar asymptotic properties as $\sin\varphi$,
we approximate it as:
\be \lb{8e34}
g(p)\simeq \frac{\gamma}{\sqrt{b\sigma}} \sin\varphi(p),
\ee
where $\gamma$ is a constant; it is determined from Eq. (\rf{8e32}). At a
first stage we neglect the last integral; we find $\gamma=0.52$. Equations
(\rf{8e25}), (\rf{8e26}) and (\rf{8e29}) then
give $F_{\pi}=16$ MeV, with the same order of magnitude as the results
found in Refs. \cite{ad,lg}, $F_{\pi}\simeq 11$ MeV. Consideration of the
last integral in Eq. (\rf{8e32}) increases the value of $\gamma$ by an
amount of 16\% and brings $F_{\pi}$ to 18 MeV; it does not therefore change 
its order of magnitude. The experimental value
of $F_{\pi}$ is nearly 93 MeV. The quark condensate $<\overline uu>$ is not
a directly measurable quantity; QCD sum rules \cite{n} give the
prediction $<\overline uu>\simeq -(225\ \mathrm{MeV})^3$ at the scale of
1 GeV. 
\par
In Ref. \cite{alyopro}, numerical calculations have been done with the
harmonic oscillator potential, which makes it difficult to compare
predictions. In Refs. \cite{fgmw,aa}, rather large values of $F_{\pi}$ and
$<\overline uu>$ are presented, but these are obtained by fitting parameters 
in order to fix the value of the constituent quark mass at 300 MeV, 
corresponding to big values of the string tension (or of its equivalents).
\par
The relative smallness (about one order of magnitude) of the
theoretical predictions for $<\overline uu>$ and $F_{\pi}$ found here
and in Refs. \cite{ad,lg} seems to indicate that short-distance
forces should have sizable contributions in these quantities; this is
corroborated by the facts that, first, they act in momentum space with a
potential having the same sign as the confining one, and second, they induce
much slower asymptotic behavior to the function $\varphi$. A definite
conclusion about the quantitative aspects of chiral symmetry breaking could
be reached only when both kinds of forces, long-range and short-range,
are considered.
\par

\section{Properties of the bound state spectrum} \lb{s9}
\setcounter{equation}{0}

We turn in this section to a study of the main qualitative properties of 
the bound state spectrum that emerge from the wave equation, leaving to a
separate work a more quantitative study of it, in particular when 
short-distance effects are incorporated.
\par

\subsection{One-particle limit} \lb{s9s1}

We begin with the case when one of the particles becomes infinitely massive;
this is achieved by taking one of the masses, $m_2$, say, to infinity.
Defining $P_0=m_2+p_{10}$, $\mathbf{x}=\mathbf{x}_1-\mathbf{x}_2$,
$\mathbf{p}=(\mathbf{p}_1-\mathbf{p}_2)/2=\mathbf{p}_1$, the wave equation
(\rf{7e12}) reduces to a Dirac type equation for the quark in the presence 
of a vector static potential:
\be \lb{9e1}
\Big[p_{10}-h_1-A_0+\gamma_{10}\mbox{\boldmath$\gamma$}.\mathbf{A}\Big]
\psi=0,
\ee
where the potentials, in the classical limit, are:
\bea 
\lb{9e2}
A_0&=&\sigma r\bigg\{\sqrt{\frac{E_1^2\mathbf{x}^2}{\mathbf{L}^2}}
\arcsin\sqrt{\frac{\mathbf{L}^2}{E_1^2\mathbf{x}^2}}-(1-\epsilon(p_{10}))
\frac{E_1^2\mathbf{x}^2}{\mathbf{L}^2}
\Big(1-\sqrt{1-\frac{\mathbf{L}^2}{E_1^2\mathbf{x}^2}}\Big)\bigg\},\\
\lb{9e3}
\mathbf{A}&=&\frac{\sigma r}{2}\Big(\frac{\mathbf{p}^t}{E_1}\Big)
\frac{E_1^2\mathbf{x}^2}{\mathbf{L}^2}\bigg\{
\sqrt{\frac{E_1^2\mathbf{x}^2}{\mathbf{L}^2}}
\arcsin\sqrt{\frac{\mathbf{L}^2}{E_1^2\mathbf{x}^2}}
-\sqrt{1-\frac{\mathbf{L}^2}{E_1^2\mathbf{x}^2}}\bigg\}.
\eea
[$r=|\mathbf{x}|$, $\mathbf{p}^t$ is defined in Eq. (\rf{6e16}),
$\mathbf{L}$ is the orbital angular momentum operator, $h_1$,
$E_1$ and $\epsilon(p_{10})$ are defined in Eqs. (\rf{7e9})-(\rf{7e11}).]
They represent the energy-momentum of a straight segment of length $r$
with linear energy density $\sigma$ turning around the point $x_2$ (the
position of the heavy antiquark).
\par
Taking further the limit of a heavy quark, the potentials become to order
$1/c^2$:
\be \lb{9e4}
A_0=\sigma r\Big(1+\frac{\mathbf{L}^2}{6m_1^2\mathbf{x}^2}\Big),\ \ \ \ 
\ \ \mathbf{A}=\frac{\sigma r}{3}\frac{\mathbf{p}^t}{m_1}.
\ee
\par

\subsection{Nonrelativistic limit} \lb{s9s2}

Next, we consider the case of two heavy quarks. The potential $A_{1\mu}$
[Eq. (\rf{6e19})] becomes (in the c.m. frame) to order $1/c^2$:
\bea 
\lb{9e5}
A_{10}&=&\sigma r\Big[\frac{1}{2}+\frac{1}{6}\Big(\frac{m_1}{m_1+m_2}\Big)
\Big(\frac{1}{m_1^2}+\frac{1}{m_2^2}-\frac{1}{m_1m_2}\Big)
\frac{\mathbf{L}^2}{\mathbf{x}^2}\nonumber \\
& &\ \ \ \ \ \ \ \ \ \ +\frac{1}{8}\Big(\frac{m_1m_2}{m_1+m_2}\Big)^2
\Big(\frac{1}{m_1^4}-\frac{1}{m_2^4}\Big)\frac{\mathbf{L}^2}{\mathbf{x}^2}
\Big],\\
\lb{9e6}
\mathbf{A}_1&=&-\frac{\sigma r}{2}\mathbf{p}^t
\Big(\frac{2}{3m_1}-\frac{1}{3m_2}\Big),
\eea
the definitions of $x$ and $p$ being the same as in Eqs. (\rf{6e3}). $A_2$
is obtained from $A_1$ by exchange of indices 1 and 2 and replacement of
$\mathbf{p}$ by $-\mathbf{p}$. The self-energy $\Sigma$ [Eqs. (\rf{7e8}),
(\rf{8e1})-(\rf{8e2})] behaves for large $m$ as:
\be \lb{9e7}
\Sigma(\mathbf{p},m)=-\frac{2\sigma}{\pi m}+O(1/m^2).
\ee
\par
After having subtracted the masses and made a few changes of function to 
reach an explicitly hermitian form, the nonrelativistic hamiltonian, to order
$1/c^2$ takes the following form (in the c.m. frame):
\bea \lb{9e8}
H&=&\frac{\mathbf{p}^2}{2\mu}+\sigma r-\frac{2\hbar\sigma}{\pi}
\Big(\frac{1}{m_1}+\frac{1}{m_2}\Big)
-\frac{1}{8}\Big(\frac{1}{m_1^3}+\frac{1}{m_2^3}\Big)(\mathbf{p}^2)^2
+\frac{\hbar^2}{4}\Big(\frac{1}{m_1^2}+\frac{1}{m_2^2}\Big)
\frac{\sigma}{r}\nonumber \\
& &-\frac{\sigma}{6r}\Big(\frac{1}{m_1^2}+\frac{1}{m_2^2}-\frac{1}{m_1m_2}
\Big)(\mathbf{L}^2+2\hbar^2)
+\frac{\sigma}{2r}\Big(\frac{\mathbf{L}.\mathbf{s}_1}
{m_1^2}+\frac{\mathbf{L}.\mathbf{s}_2}{m_2^2}\Big)\nonumber \\
& &-\frac{2\sigma}{3r}\Big(\frac{1}{m_1^2}-\frac{1}{2m_1m_2}\Big)
\mathbf{L}.\mathbf{s}_1-\frac{2\sigma}{3r}\Big(\frac{1}{m_2^2}-
\frac{1}{2m_1m_2}\Big)\mathbf{L}.\mathbf{s}_2.
\eea
[$\mu=m_1m_2/(m_1+m_2)$, $\mathbf{s}_1$ and $\mathbf{s}_2$ are the spin
operators of the quark and of the antiquark.]
Several remarks can be made at this stage. First, the hamiltonian is
independent of spin-spin interactions; this is already evident from
the wave equation (\rf{7e12}) where no direct interactions involving
the spacelike $\gamma$-matrices of both quarks exist. The absence of
long-range spin-spin interactions is compatible with experimental data.
Second, we notice the presence of purely orbital angular momentum dependent
pieces (proportional to $\mathbf{L}^2$), the origin of which is related to
the contribution to the rotational motion of the system of the moment of
inertia of the flux tube, represented by the straight segment joining the
quark to the antiquark. The corresponding centrifugal energy
produces a global plus sign in front of those terms; the minus sign
results from the additional contributions of the momentum of the flux tube,
which also couples, through the spacelike potentials $\mathbf{A}_1$
and $\mathbf{A}_2$ to the quarks [Eq. (\rf{9e6})]. Those terms were also
obtained in Refs. \cite{bmpbbp} and \cite{ow,lcooowoo}. Third, we have
two kinds of spin-orbit term. The first, appearing after the term in
$\mathbf{L}^2$, comes from the contribution of a conventional timelike
vector interaction represented by the potential $\sigma r$, which is the
dominant part of the combination $A_{10}+A_{20}$ [Eq. (\rf{9e5})].
The second type of contribution, provided by the last two terms 
of Eq. (\rf{9e8}), comes from the contributions of the direct
interactions of the momentum of the flux tube with the quarks, represented
by the spacelike potentials $\mathbf{A}_1$ and $\mathbf{A}_2$ 
[Eq. (\rf{9e6})]. The latter terms induce negative signs to the 
spin-orbit couplings, in opposite direction to the former one, a feature 
which is also observed on phenomenological grounds for the large-distance
effects in fine splitting.
\par
Adopting the notations of Ref. \cite{ef} for the potentials corresponding
to spin-orbit ($V_1$ and $V_2$) and to spin-spin interactions ($V_3$ and
$V_4$), we have:
\be \lb{9e9}
V_1=-\frac{2}{3}\sigma r,\ \ \ \ V_2=+\frac{1}{3}\sigma r,\ \ \ \ 
V_3=0,\ \ \ \ \ V_4=0.
\ee
Designating by $V$ the nonrelativistic confining potential [Eq. (\rf{8e6})],
the potentials $V$, $V_1$ and $V_2$ satisfy the Gromes relation \cite{gr}
\be \lb{9e10}
V+V_1-V_2=0,
\ee
which is a consequence of Lorentz covariance.
\par
The expressions of potentials $V_1$ and $V_2$ do not satisfy, however, 
Buchm\"uller's conjecture about the spin-orbit terms \cite{bm}, according
to which confinement is due to a pure color electric field in the 
\textit{co-moving frame} of the two quarks, thus reducing the spin-orbit
potential to a Thomas precession term and entailing $V_1=-\sigma r$
and $V_2=0$. 
Leaving aside the question of a phenomenological determination
of these potentials, we notice that lattice calculations \cite{hm,bsw},
which favor the negative sign of $V_1$, do not, however, clearly
distinguish between the two possibilities above, due to the existing 
uncertainties.
In order to clarify the structure of the spin-orbit potentials,
it is necessary to investigate the contributions of higher-order terms
not taken into account in the potentials (\rf{6e19}) (of the type of those 
considered for the extraction of the self-energy terms). Another 
independent check consists in calculating the field correlators appearing
in the nonrelativistic expansion of Ref. \cite{ef}. 
\par
Reviews about bound state problems of quarks can be found in Refs. 
\cite{lsg} and \cite{bl}. 
\par

\subsection{Regge trajectories} \lb{s9s3}

Finally, we consider the high energy behavior of the mass spectrum for
ultrarelativistic systems. It has been known for a long time that the
linear confining potential of the static case produces in the
ultrarelativistic limit of massless quarks, through the Salpeter
equation, linear Regge trajectories. The inclusion of a flux tube,
represented by a straight string, modifies the relationship of the
angular momentum with the total mass of the system and increases the
slope $\alpha'$ of the Regge trajectories by an amount of 15-20\%,
enforcing the classical relation $\alpha'=(2\pi\sigma)^{-1}$ with the
string tension $\sigma$ \cite{id,lcooowoo}. 
\par
We have checked these properties on Eq. (\rf{7e12}), by solving it in
an approximate way that preserves its main qualitative features. The
approximation that we use is the Breit approximation \cite{b}, which
consists in transforming the wave equation into a local equation in
$x$-space. The main lines of the approximation are presented in appendix
\rf{ap3}. The resulting Regge trajectories for the $\pi$ and $\rho$
families are presented in Fig. \rf{9f1}, where for comparison we have 
also presented the trajectories of the linear confining potential
corresponding to the situation where $A_{10}+A_{20}=\sigma r$ and
$\mathbf{A}_1=\mathbf{A}_2=\mathbf{0}$ in Eq. (\rf{7e12}). One 
verifies, first, the linearity of the trajectories and, second, the 
increase of the slopes when the flux tube is present. Similar 
trajectories and behaviors are also obtained with the $a_0$ and $a_1$
families.
\par
There are uncertainties of a few percent in our results coming from the way
of calculating the mean values of operators present in the flux tube
functions, depending on whether one evaluates them on the full function
$\psi$ or only on the basic component $\psi_3$ [Eq. (\rf{8e13})] and
whether one uses for the latter a free norm or the norm relative to the
Pauli--Schr\"odinger equation that it satisfies. 
\par
\bfg
\vspace*{0.5 cm}
\bc
\setlength{\unitlength}{0.240900pt}
\ifx\plotpoint\undefined\newsavebox{\plotpoint}\fi
\sbox{\plotpoint}{\rule[-0.200pt]{0.400pt}{0.400pt}}%
\begin{picture}(1650,1170)(0,0)
\font\gnuplot=cmr10 at 10pt
\gnuplot
\sbox{\plotpoint}{\rule[-0.200pt]{0.400pt}{0.400pt}}%
\put(121.0,123.0){\rule[-0.200pt]{4.818pt}{0.400pt}}
\put(101,123){\makebox(0,0)[r]{0}}
\put(1569.0,123.0){\rule[-0.200pt]{4.818pt}{0.400pt}}
\put(121.0,306.0){\rule[-0.200pt]{4.818pt}{0.400pt}}
\put(101,306){\makebox(0,0)[r]{1}}
\put(1569.0,306.0){\rule[-0.200pt]{4.818pt}{0.400pt}}
\put(121.0,489.0){\rule[-0.200pt]{4.818pt}{0.400pt}}
\put(101,489){\makebox(0,0)[r]{2}}
\put(1569.0,489.0){\rule[-0.200pt]{4.818pt}{0.400pt}}
\put(121.0,672.0){\rule[-0.200pt]{4.818pt}{0.400pt}}
\put(101,672){\makebox(0,0)[r]{3}}
\put(1569.0,672.0){\rule[-0.200pt]{4.818pt}{0.400pt}}
\put(121.0,855.0){\rule[-0.200pt]{4.818pt}{0.400pt}}
\put(101,855){\makebox(0,0)[r]{4}}
\put(1569.0,855.0){\rule[-0.200pt]{4.818pt}{0.400pt}}
\put(121.0,1038.0){\rule[-0.200pt]{4.818pt}{0.400pt}}
\put(101,1038){\makebox(0,0)[r]{5}}
\put(1569.0,1038.0){\rule[-0.200pt]{4.818pt}{0.400pt}}
\put(121.0,123.0){\rule[-0.200pt]{0.400pt}{4.818pt}}
\put(121,82){\makebox(0,0){0}}
\put(121.0,1110.0){\rule[-0.200pt]{0.400pt}{4.818pt}}
\put(284.0,123.0){\rule[-0.200pt]{0.400pt}{4.818pt}}
\put(284,82){\makebox(0,0){1}}
\put(284.0,1110.0){\rule[-0.200pt]{0.400pt}{4.818pt}}
\put(447.0,123.0){\rule[-0.200pt]{0.400pt}{4.818pt}}
\put(447,82){\makebox(0,0){2}}
\put(447.0,1110.0){\rule[-0.200pt]{0.400pt}{4.818pt}}
\put(610.0,123.0){\rule[-0.200pt]{0.400pt}{4.818pt}}
\put(610,82){\makebox(0,0){3}}
\put(610.0,1110.0){\rule[-0.200pt]{0.400pt}{4.818pt}}
\put(773.0,123.0){\rule[-0.200pt]{0.400pt}{4.818pt}}
\put(773,82){\makebox(0,0){4}}
\put(773.0,1110.0){\rule[-0.200pt]{0.400pt}{4.818pt}}
\put(937.0,123.0){\rule[-0.200pt]{0.400pt}{4.818pt}}
\put(937,82){\makebox(0,0){5}}
\put(937.0,1110.0){\rule[-0.200pt]{0.400pt}{4.818pt}}
\put(1100.0,123.0){\rule[-0.200pt]{0.400pt}{4.818pt}}
\put(1100,82){\makebox(0,0){6}}
\put(1100.0,1110.0){\rule[-0.200pt]{0.400pt}{4.818pt}}
\put(1263.0,123.0){\rule[-0.200pt]{0.400pt}{4.818pt}}
\put(1263,82){\makebox(0,0){7}}
\put(1263.0,1110.0){\rule[-0.200pt]{0.400pt}{4.818pt}}
\put(1426.0,123.0){\rule[-0.200pt]{0.400pt}{4.818pt}}
\put(1426,82){\makebox(0,0){8}}
\put(1426.0,1110.0){\rule[-0.200pt]{0.400pt}{4.818pt}}
\put(1589.0,123.0){\rule[-0.200pt]{0.400pt}{4.818pt}}
\put(1589,82){\makebox(0,0){9}}
\put(1589.0,1110.0){\rule[-0.200pt]{0.400pt}{4.818pt}}
\put(121.0,123.0){\rule[-0.200pt]{353.641pt}{0.400pt}}
\put(1589.0,123.0){\rule[-0.200pt]{0.400pt}{242.586pt}}
\put(121.0,1130.0){\rule[-0.200pt]{353.641pt}{0.400pt}}
\put(40,626){\makebox(0,0){$J$}}
\put(855,21){\makebox(0,0){$P^2\ (\mathrm{GeV}^2)$}}
\put(121.0,123.0){\rule[-0.200pt]{0.400pt}{242.586pt}}
\put(121,123){\usebox{\plotpoint}}
\multiput(121.00,123.58)(0.500,0.500){363}{\rule{0.500pt}{0.120pt}}
\multiput(121.00,122.17)(181.962,183.000){2}{\rule{0.250pt}{0.400pt}}
\multiput(304.00,306.58)(0.505,0.500){363}{\rule{0.504pt}{0.120pt}}
\multiput(304.00,305.17)(183.953,183.000){2}{\rule{0.252pt}{0.400pt}}
\multiput(489.00,489.58)(0.508,0.500){363}{\rule{0.507pt}{0.120pt}}
\multiput(489.00,488.17)(184.949,183.000){2}{\rule{0.253pt}{0.400pt}}
\multiput(675.00,672.58)(0.516,0.500){363}{\rule{0.513pt}{0.120pt}}
\multiput(675.00,671.17)(187.935,183.000){2}{\rule{0.257pt}{0.400pt}}
\multiput(864.00,855.58)(0.519,0.500){363}{\rule{0.515pt}{0.120pt}}
\multiput(864.00,854.17)(188.930,183.000){2}{\rule{0.258pt}{0.400pt}}
\put(121,123){\circle*{12}}
\put(304,306){\circle*{12}}
\put(489,489){\circle*{12}}
\put(675,672){\circle*{12}}
\put(864,855){\circle*{12}}
\put(1054,1038){\circle*{12}}
\sbox{\plotpoint}{\rule[-0.500pt]{1.000pt}{1.000pt}}%
\put(121,123){\usebox{\plotpoint}}
\multiput(121,123)(15.171,14.165){13}{\usebox{\plotpoint}}
\multiput(317,306)(15.648,13.636){14}{\usebox{\plotpoint}}
\multiput(527,489)(15.927,13.309){14}{\usebox{\plotpoint}}
\multiput(746,672)(16.045,13.167){13}{\usebox{\plotpoint}}
\multiput(969,855)(16.130,13.061){14}{\usebox{\plotpoint}}
\put(1195,1038){\usebox{\plotpoint}}
\put(121,123){\makebox(0,0){$\star$}}
\put(317,306){\makebox(0,0){$\star$}}
\put(527,489){\makebox(0,0){$\star$}}
\put(746,672){\makebox(0,0){$\star$}}
\put(969,855){\makebox(0,0){$\star$}}
\put(1195,1038){\makebox(0,0){$\star$}}
\sbox{\plotpoint}{\rule[-0.200pt]{0.400pt}{0.400pt}}%
\put(475,123){\usebox{\plotpoint}}
\multiput(475.00,123.58)(0.530,0.500){363}{\rule{0.524pt}{0.120pt}}
\multiput(475.00,122.17)(192.912,183.000){2}{\rule{0.262pt}{0.400pt}}
\multiput(669.00,306.58)(0.524,0.500){363}{\rule{0.520pt}{0.120pt}}
\multiput(669.00,305.17)(190.921,183.000){2}{\rule{0.260pt}{0.400pt}}
\multiput(861.00,489.58)(0.524,0.500){363}{\rule{0.520pt}{0.120pt}}
\multiput(861.00,488.17)(190.921,183.000){2}{\rule{0.260pt}{0.400pt}}
\multiput(1053.00,672.58)(0.524,0.500){363}{\rule{0.520pt}{0.120pt}}
\multiput(1053.00,671.17)(190.921,183.000){2}{\rule{0.260pt}{0.400pt}}
\multiput(1245.00,855.58)(0.524,0.500){363}{\rule{0.520pt}{0.120pt}}
\multiput(1245.00,854.17)(190.921,183.000){2}{\rule{0.260pt}{0.400pt}}
\put(475,123){\circle*{12}}
\put(669,306){\circle*{12}}
\put(861,489){\circle*{12}}
\put(1053,672){\circle*{12}}
\put(1245,855){\circle*{12}}
\put(1437,1038){\circle*{12}}
\sbox{\plotpoint}{\rule[-0.500pt]{1.000pt}{1.000pt}}%
\put(475,123){\usebox{\plotpoint}}
\multiput(475,123)(15.242,14.088){13}{\usebox{\plotpoint}}
\multiput(673,306)(15.615,13.673){14}{\usebox{\plotpoint}}
\multiput(882,489)(15.836,13.417){14}{\usebox{\plotpoint}}
\multiput(1098,672)(15.986,13.237){13}{\usebox{\plotpoint}}
\multiput(1319,855)(16.073,13.131){14}{\usebox{\plotpoint}}
\put(1543,1038){\usebox{\plotpoint}}
\put(475,123){\makebox(0,0){$\star$}}
\put(673,306){\makebox(0,0){$\star$}}
\put(882,489){\makebox(0,0){$\star$}}
\put(1098,672){\makebox(0,0){$\star$}}
\put(1319,855){\makebox(0,0){$\star$}}
\put(1543,1038){\makebox(0,0){$\star$}}
\sbox{\plotpoint}{\rule[-0.200pt]{0.400pt}{0.400pt}}%
\put(140,306){\usebox{\plotpoint}}
\multiput(140.00,306.58)(0.609,0.500){363}{\rule{0.587pt}{0.120pt}}
\multiput(140.00,305.17)(221.781,183.000){2}{\rule{0.294pt}{0.400pt}}
\multiput(363.00,489.58)(0.609,0.500){363}{\rule{0.587pt}{0.120pt}}
\multiput(363.00,488.17)(221.781,183.000){2}{\rule{0.294pt}{0.400pt}}
\multiput(586.00,672.58)(0.538,0.500){363}{\rule{0.531pt}{0.120pt}}
\multiput(586.00,671.17)(195.899,183.000){2}{\rule{0.265pt}{0.400pt}}
\multiput(783.00,855.58)(0.546,0.500){363}{\rule{0.537pt}{0.120pt}}
\multiput(783.00,854.17)(198.885,183.000){2}{\rule{0.269pt}{0.400pt}}
\put(140,306){\circle*{12}}
\put(363,489){\circle*{12}}
\put(586,672){\circle*{12}}
\put(783,855){\circle*{12}}
\put(983,1038){\circle*{12}}
\sbox{\plotpoint}{\rule[-0.500pt]{1.000pt}{1.000pt}}%
\put(145,306){\usebox{\plotpoint}}
\multiput(145,306)(16.771,12.228){15}{\usebox{\plotpoint}}
\multiput(396,489)(16.748,12.260){15}{\usebox{\plotpoint}}
\multiput(646,672)(16.653,12.388){15}{\usebox{\plotpoint}}
\multiput(892,855)(16.580,12.486){15}{\usebox{\plotpoint}}
\put(1135,1038){\usebox{\plotpoint}}
\put(145,306){\makebox(0,0){$\star$}}
\put(396,489){\makebox(0,0){$\star$}}
\put(646,672){\makebox(0,0){$\star$}}
\put(892,855){\makebox(0,0){$\star$}}
\put(1135,1038){\makebox(0,0){$\star$}}
\sbox{\plotpoint}{\rule[-0.200pt]{0.400pt}{0.400pt}}%
\put(495,306){\usebox{\plotpoint}}
\multiput(495.00,306.58)(0.604,0.500){363}{\rule{0.583pt}{0.120pt}}
\multiput(495.00,305.17)(219.790,183.000){2}{\rule{0.292pt}{0.400pt}}
\multiput(716.00,489.58)(0.596,0.500){363}{\rule{0.577pt}{0.120pt}}
\multiput(716.00,488.17)(216.803,183.000){2}{\rule{0.288pt}{0.400pt}}
\multiput(934.00,672.58)(0.604,0.500){363}{\rule{0.583pt}{0.120pt}}
\multiput(934.00,671.17)(219.790,183.000){2}{\rule{0.292pt}{0.400pt}}
\multiput(1155.00,855.58)(0.565,0.500){363}{\rule{0.552pt}{0.120pt}}
\multiput(1155.00,854.17)(205.853,183.000){2}{\rule{0.276pt}{0.400pt}}
\put(495,306){\circle*{12}}
\put(716,489){\circle*{12}}
\put(934,672){\circle*{12}}
\put(1155,855){\circle*{12}}
\put(1362,1038){\circle*{12}}
\sbox{\plotpoint}{\rule[-0.500pt]{1.000pt}{1.000pt}}%
\put(495,306){\usebox{\plotpoint}}
\multiput(495,306)(16.505,12.585){15}{\usebox{\plotpoint}}
\multiput(735,489)(16.580,12.486){15}{\usebox{\plotpoint}}
\multiput(978,672)(16.580,12.486){14}{\usebox{\plotpoint}}
\multiput(1221,855)(16.580,12.486){15}{\usebox{\plotpoint}}
\put(1464,1038){\usebox{\plotpoint}}
\put(495,306){\makebox(0,0){$\star$}}
\put(735,489){\makebox(0,0){$\star$}}
\put(978,672){\makebox(0,0){$\star$}}
\put(1221,855){\makebox(0,0){$\star$}}
\put(1464,1038){\makebox(0,0){$\star$}}
\end{picture}
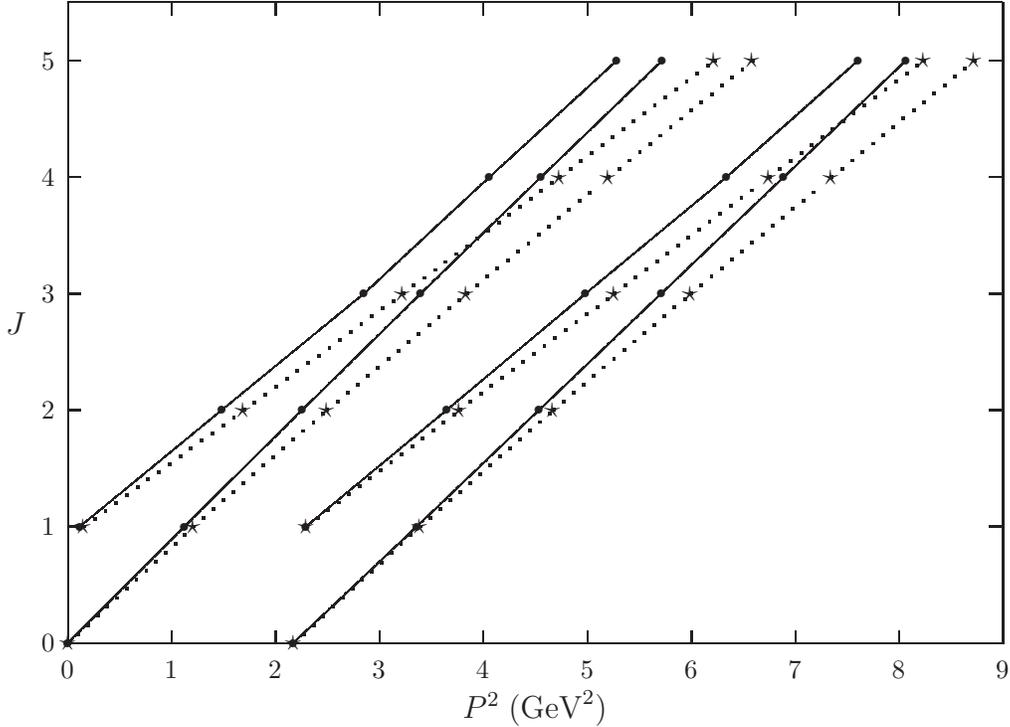
\caption{Regge trajectories of the $\pi$ and $\rho$ families and their
first daughter trajectories. Full lines correspond to the flux tube
potential, dotted lines to the linear potential. Inputs: $m_1=m_2=0$,
$\sigma=0.18$ GeV$^2$.}
\lb{9f1}
\ec
\efg
On phenomenological grounds, one actually determines the value of the
string tension $\sigma$ from the experimental Regge trajectories using
the string theory relation $\alpha'=(2\pi\sigma)^{-1}$. From the slope
of the $\pi$-family trajectory one obtains $\sigma=0.22$ GeV$^2$, while
from the $\rho$-family trajectory one obtains $\sigma=0.18$ GeV$^2$
\cite{bl}. A precise comparison of our results with those data necessitates
a more elaborate resolution of the wave equation and presumably the
inclusion of the short-distance potentials.
\par

\section{Summary and concluding remarks} \lb{s10}
\setcounter{equation}{0}

We have investigated, in the large $N_c$ limit, the large-distance
dynamics of QCD within the quarkonium system through the saturation of 
Wilson loop averages by minimal surfaces. The dynamics is described by a wave 
equation of the Breit--Salpeter type, in which the interaction potentials
are provided by the energy-momentum vector of the straight segment 
joining the quark to the antiquark and carrying a constant linear energy
density, equal to the string tension; the latter represents the effective
contribution of the color flux tube of the quarkonium system. In the
static case, confinement is realized by the usual linear potential, while
in the general case of moving quarks additional contributions come from
the moments of inertia of the straight segment .
\par
Taking into account the self-energy parts of the quark propagators, it was
shown that chiral symmetry is spontaneously broken with a mechanism
identical to that of the exchange of one Coulomb-gluon.
\par
In the nonrelativistic limit, long range spin-spin potentials are absent,
while the moments of inertia of the flux tube bring contributions to the
orbital angular momentum and spin-orbit potentials with negative signs, in
opposite direction, for the latter case, to the spin-orbit term of the pure
timelike vector potential.
\par
In the ultrarelativistic limit, the mass spectrum displays linear Regge
trajectories, the slopes of which tend to satisfy the classical relation
with the string tension obtained in straight string theories.
\par
The potentials that have been isolated in the present work do not
represent the complete interaction kernel of the wave equation, but only
the large-distance dominant contributions to it. The determination of
the higher-order terms necessitates further study. Also, mathematical 
assumptions made for neglecting certain terms or approximating 
others in the large separation time limit need to be analyzed in more
detail.
\par
The formalism that has been developed here, centering the Wilson loop
representation on minimal surfaces, ignores the contributions of 
short-distance effects provided by perturbation theory. A complete 
resolution of QCD requires the presence of both large- and short-distance
effects. A corresponding resolution of the loop equations is far from 
being trivial, although approximate solutions might still provide a 
practical framework for a quantitative investigation of the spectroscopic
properties of the theory.
\par
\vspace{0.25 cm}
\noindent
\textbf{Acknowledgements}
\par
We thank J. Stern for stimulating discussions. The numerical calculation
of the mass spectrum used a program based on the inverse iterative method 
provided to us by J.-L. Ballot to whom we are grateful. This work
was supported in part by the European Community network EURIDICE under 
Contract No. HPRN-CT-2002-00311.
\par

\appendix
\renewcommand{\theequation}{\Alph{section}.\arabic{equation}}

\section{Properties of minimal surfaces} \lb{ap1}
\setcounter{equation}{0}

An important quantity in loop dynamics is the second-order functional
derivative of the area at two different points on the contour.
For definiteness we consider the two points on the line $\tau=0$
(line $x_1x_2$ of the contour $C$ of Fig. \rf{3f1}). For a minimal surface,
the variations of the contour maintain at all stages the minimality
property of the area; therefore the second term of the right-hand side 
of Eq. (\rf{3e6}) and all its variations are null and can be ignored. 
The second-order functional derivative of the area is then:
\bea \lb{3e15}
& &\frac{\delta}{\delta x^{\beta}(\sigma',0)}
\frac{\delta A}{\delta x^{\alpha}(\sigma,0)}=
(\delta_{\mu\alpha}\delta_{\nu\beta}-\delta_{\mu\beta}\delta_{\nu\alpha})
\frac{\partial}{\partial \sigma}\delta(\sigma-\sigma')
\int d\sigma^{\mu\nu}(y)F(y-x(\sigma,0))\nonumber \\
& &\ \ +\Big(\delta_{\mu\alpha}x^{\prime\nu}(\sigma,0)-
\delta_{\nu\alpha}x^{\prime\mu}(\sigma,0)\Big)
\Big(\delta_{\mu\beta}x^{\prime\nu}(\sigma',0)-
\delta_{\nu\beta}x^{\prime\mu}(\sigma',0)\Big)
F\Big(x(\sigma,0)-x(\sigma',0)\Big)\nonumber \\
& &\ \ +\Big(\delta_{\mu\alpha}x^{\prime\nu}(\sigma,0)-
\delta_{\nu\alpha}x^{\prime\mu}(\sigma,0)\Big)
\delta(\sigma-\sigma')\frac{\partial}{\partial x^{\beta}(\sigma,0)}
\int d\sigma^{\mu\nu}(y)F(y-x(\sigma,0))\nonumber \\
& &\ \ -\Big(\delta_{\mu\alpha}x^{\prime\nu}(\sigma,0)-
\delta_{\nu\alpha}x^{\prime\mu}(\sigma,0)\Big)\int d\sigma''d\tau''
\frac{\delta y^{\lambda}(\sigma'',\tau'')}{\delta x^{\beta}(\sigma',0)}
\nonumber \\
& &\ \ \times\Big[(y^{\prime\gamma}\dot y^{\nu}-y^{\prime\nu}\dot y^{\gamma})
\delta_{\mu\lambda}+(y^{\prime\mu}\dot y^{\gamma}-
y^{\prime\gamma}\dot y^{\mu})\delta_{\nu\lambda}+
(y^{\prime\nu}\dot y^{\mu}-y^{\prime\mu}\dot y^{\nu})\delta_{\gamma\lambda}
\Big]\frac{\partial}{\partial y^{\gamma}}F(y-x(\sigma,0)).\nonumber \\
& &
\eea
\par
The evaluation of the the last integral can be done by expanding
$\delta y^{\lambda}/\delta x^{\beta}(\sigma',0)$ about the point 
$x(\sigma,0)$ that appears in $F$:
\be \lb{3e16}
\frac{\delta y^{\lambda}(\sigma'',\tau'')}{\delta x^{\beta}(\sigma',0)}
=\frac{\delta y^{\lambda}(\sigma,0)}{\delta x^{\beta}(\sigma',0)}
+(\sigma''-\sigma)\frac{\partial}{\partial \sigma}
\frac{\delta y^{\lambda}(\sigma,0)}{\delta x^{\beta}(\sigma',0)}
+\tau''\frac{\partial}{\partial \tau}
\frac{\delta y^{\lambda}(\sigma,\tau)}{\delta x^{\beta}(\sigma',0)}
\bigg|_{\tau=0}+\ldots\ ,
\ee
where the dots stand for terms that do not contribute in the limit
$a=0$. One notices that the first term in the right-hand side of Eq.
(\rf{3e16}) is a derivative that lies on the contour and hence is equal to
$\delta_{\mu\nu}\delta(\sigma-\sigma')$:
\be \lb{3e17}
\frac{\delta y^{\lambda}(\sigma,0)}{\delta x^{\beta}(\sigma',0)}
\equiv \frac{\delta x^{\lambda}(\sigma,0)}{\delta x^{\beta}(\sigma',0)}
=\delta_{\beta\lambda}\delta(\sigma-\sigma').
\ee
It is also convenient, to complete the calculations, to stick to 
the following diagonal constant metric:
\be \lb{3e18}
x'.\dot x=0,\ \ \ \ x^{\prime 2}=\mathrm{const.},\ \ \ \
\dot x^2=\mathrm{const.},\ \ \ \ 
\lambda\equiv \sqrt{\frac{x^{\prime 2}}{\dot x^2}}.
\ee
The result is:
\bea \lb{3e19}
& &\frac{\delta}{\delta x^{\beta}(\sigma',0)}
\frac{\delta A}{\delta x^{\alpha}(\sigma,0)}=
2\frac{\partial}{\partial \sigma}\delta(\sigma-\sigma')
\int d\sigma^{\alpha\beta}(y)F(y-x(\sigma,0))\nonumber \\
& &\ \ \ \ \ +2\Big(\delta_{\alpha\beta}x'(\sigma,0).x'(\sigma',0)
-x^{\prime\beta}(\sigma,0)x^{\prime\alpha}(\sigma',0)\Big)
F\Big(x(\sigma,0)-x(\sigma',0)\Big)\nonumber \\
& &\ \ \ \ \ -\delta(\sigma-\sigma')\frac{1}{a}\sqrt{x^{\prime 2}}
\Big(\delta_{\alpha\beta}-
\frac{x^{\prime\alpha}x^{\prime\beta}}{x^{\prime 2}}\Big)
\nonumber \\
& &\ \ \ \ \ -\sqrt{\frac{x^{\prime 2}}{\dot x^2}}
\frac{\partial}{\partial \tau}
\frac{\delta y^{\lambda}(\sigma,\tau)}{\delta x^{\beta}(\sigma',0)}
\bigg|_{\tau=0}\Big(\delta_{\lambda\alpha}
-\frac{x^{\prime \lambda}(\sigma,0)x^{\prime \alpha}(\sigma,0)}
{x^{\prime 2}(\sigma,0)}-\frac{\dot x^{\lambda}(\sigma,0)
\dot x^{\alpha}(\sigma,0)}{\dot x^2(\sigma,0)}\Big).
\eea
The two points $x(\sigma,0)$ and $x(\sigma',0)$ lying on the boundary,
the function $F\Big(x(\sigma,0)-x(\sigma',0)\Big)$ is equivalent,
in the limit $a=0$, to 
$\frac{1}{2a\sqrt{x^{\prime 2}}}\delta(\sigma-\sigma')$ and thus cancels 
the singular term of the third line. (We assume here that the line under
consideration does not have self-intersections neither displays
backtracking.) 
There remains to evaluate the quantity $\frac{\partial}{\partial \tau}
\frac{\delta y^{\lambda}(\sigma,\tau)}{\delta x^{\beta}(\sigma',0)}
\bigg|_{\tau=0}$. This in turn requires a more detailed knowledge
of the derivative $\frac{\delta y^{\lambda}(\sigma,\tau)}
{\delta x^{\beta}(\sigma',0)}$. That quantity does not have a compact
expression for general contours. Studies of its main properties can be 
found in Refs. \cite{c} and \cite{lsw}. In the formulation that we present
below, we follow partly the covariant approach adopted in Ref. \cite{lsw}
and put emphasis on the aspects that will be needed in the present 
work.
\par
Let us first point out the fact that a variation $\delta x$ on the contour
that lies in the tangent plane to the minimal surface does not modify the 
internal part of the minimal surface but only adds to it a new small
piece on its boundary within the tangent plane; the variation thus amounts
to a simple extension of the boundary on the same surface. In that case
$\delta y$ should lie in the tangent plane to the minimal surface at $y$,
being a linear combination of $y'$ and $\dot y$ and corresponding to
a reparametrization of the surface. In this respect, it can be checked that
if $\delta y^{\lambda}$ contains terms proportional to $y^{\prime\lambda}$ or
to $\dot y^{\lambda}$ then the contributions of these in Eq. (\rf{3e15}) are
identically zero. Therefore, the nontrivial contributions to Eq. (\rf{3e15})
come only from the transverse variations of $\delta y$ with respect to the
surface, which are themselves due to transverse variations $\delta x$ on the
contour with respect to that surface. We shall therefore concentrate on 
these kinds of variation.
\par
To proceed further and to exhibit some general features, we shall use
for a while covariant notations. The parameters of the surface will be
denoted $\xi^a$, $a=1,2$, and the metric $g_{ab}$:
\be \lb{3e20}
g_{ab}=\frac{\partial x^{\mu}}{\partial \xi^a}
\frac{\partial x^{\mu}}{\partial \xi^b},\ \ \ \ \ a,b=1,2.
\ee
The entries of the inverse of $g$ are denoted $g^{ab}$ and raising and
lowering of indices are done with these tensors. The determinant of $g$
is denoted $|g|$. We shall not meet below singular quantities and therefore,
as long as not needed, we shall not use regulated expressions. The
area is then $A=\int d^2\xi|g|^{1/2}$ and the equation of the minimal 
surface is: 
\be \lb{3e21}
|g|^{-1/2}\partial_a\left(|g|^{1/2}g^{ab}\partial_by^{\mu}(\xi)\right)=0.
\ee
The determinant $|g|$ remains constant on the minimal surface. 
\par
Next, we introduce variations $\delta y$ on the minimal surface, including
the boundaries. The equation of the new minimal surface is obtained by
replacing in Eq. (\rf{3e21}) $y$ by $y+\delta y$. The linearized
equation in $\delta y$ is:
\be \lb{3e22}
\Delta_{\mu\nu}\delta y^{\nu}\equiv |g|^{-1/2}\partial_a\left(|g|^{1/2}
M_{\mu\nu}^{ab}\partial_b\delta y^{\nu}(\xi)\right)=0,
\ee
where one has defined
\be \lb{3e23}
M_{\mu\nu}^{ab}=\delta_{\mu\nu}g^{ab}+\partial^ay_{\mu}\partial^by_{\nu}-
\partial^ay_{\nu}\partial^by_{\mu}-g^{ab}\partial_cy_{\nu}\partial^cy_{\mu}.
\ee
The operator $\Delta_{\mu\nu}$ is transverse: it annihilates any tangential
variation $\delta y^{\nu}=\omega^c\partial_cy^{\nu}$: 
\be \lb{3e24}
\Delta_{\mu\nu}\omega^c\partial_cy^{\nu}=0.
\ee
One next introduces the Green function of the operator $\Delta_{\mu\nu}$
subject to the boundary condition that it vanishes along the contour $C$:
\bea \lb{3e25}
\Delta_{\mu\nu}(\xi)G_{\nu\rho}(\xi,\eta)&=&|g|^{-1/2}\Big(\delta_{\mu\rho}
-Q_{\mu\rho}(\xi)\Big)\delta^2(\xi-\eta),\nonumber \\
G_{\nu\rho}(\xi,\eta)\Big|_{\xi\in C}&=&
G_{\nu\rho}(\xi,\eta)\Big|_{\eta\in C}=0,
\eea
where $Q_{\mu\rho}(\xi)$ is the projector on the tangent plane of the
initial minimal surface at $y(\xi)$:
\be \lb{3e26}
Q_{\mu\rho}(\xi)=\partial_ay_{\mu}(\xi)g^{ab}\partial_by_{\rho}(\xi).
\ee
Using then the Green identity, the symmetry property
$M_{\mu\nu}^{ab}=M_{\nu\mu}^{ba}$, the Stokes theorem and the
transversality of $\delta y$, one obtains for the transverse variations 
$\delta y$ the following successive expressions:
\bea \lb{3e27}
\delta y_{\mu}(\eta)&=&\int d^2\xi|g|^{1/2}\Big[\delta y_{\rho}(\xi)
\Delta_{\rho\nu}(\xi)G_{\nu\mu}(\xi,\eta)-\Big(\Delta_{\rho\nu}(\xi)
\delta y_{\nu}(\xi)\Big)G_{\rho\mu}(\xi,\eta)\Big]\nonumber \\
&=&\int d^2\xi\partial_a\Big[\delta y_{\rho}(\xi)|g|^{1/2}
M_{\rho\nu}^{ab}(\xi)\partial_bG_{\nu\rho}(\xi,\eta)-
|g|^{1/2}M_{\rho\nu}^{ab}\Big(\partial_b\delta y_{\nu}(\xi)\Big)
G_{\rho\mu}(\xi,\eta)\Big]\nonumber \\
&=&-\oint_Cd\xi^c\epsilon_{ca}|g|^{1/2}g^{ab}\delta y_{\nu}(\xi)
\partial_bG_{\nu\mu}(\xi,\eta)\nonumber \\
&=&\oint_Cd\lambda n^a(\lambda)
\delta y_{\nu}(\xi(\lambda))\partial_aG_{\nu\mu}(\xi(\lambda),\eta),
\eea
where the contour $C$ has been parametrized with the parameter $\lambda$
and where $n$ is an orthogonal vector to the contour in the tangent
plane to the minimal surface,
\be \lb{3e28}
n^a(\lambda)=g^{ab}|g|^{1/2}\epsilon_{bc}\frac{d\xi^c(\lambda)}
{d\lambda}. 
\ee
\par
The functional derivative of $y_{\mu}(\eta)$ with respect to
$x_{\nu}(\xi(\lambda'))$ then becomes:
\be \lb{3e29}
\frac{\delta y_{\mu}(\eta)}{\delta x_{\nu}(\xi(\lambda'))}=
n^a(\lambda')\partial_aG_{\nu\mu}(\xi,\eta)\Big|_{\xi=\xi(\lambda')}.
\ee
\par
Finally, the derivative of $\frac{\delta y_{\mu}}{\delta x_{\nu}}$
along $n$ is:
\be \lb{3e30}
n^b(\lambda)\partial_{(\eta)b}
\frac{\delta y_{\mu}(\eta)}{\delta x_{\nu}(\xi(\lambda'))}\bigg|_
{\eta=\eta(\lambda)}=n^b(\lambda)n^a(\lambda')\partial_{(\eta)b}
\partial_{(\xi)a}G_{\nu\mu}(\xi,\eta)\Big|_
{\xi=\xi(\lambda'),\eta=\eta(\lambda)}.
\ee
\par
A consequence of Eq. (\rf{3e30}) is the symmetry property
of its right-hand side with respect to the variables $\xi$ and $\eta$
for $\lambda\neq \lambda'$, a feature that is not evident in the
left-hand side. That property is intimately related to the symmetry
property of the second-order variation of the minimal surface with
respect to the order of operations. To exhibit this feature more
explicitly we return to the $(\sigma,\tau)$ parametrization that we were
using above. The derivation $n^a\partial_a$ is equivalent in the
limit $\tau=0$ to the derivation 
$-\sqrt{\frac{x^{\prime 2}}{\dot x^2}}\frac{\partial}{\partial \tau}$. 
Equation (\rf{3e29}) becomes, using the metric (\rf{3e18}):
\be \lb{3e31}
\frac{\delta y^{\lambda}(\sigma,\tau)}{\delta x^{\beta}(\sigma',0))}=
-\sqrt{\frac{x^{\prime 2}}{\dot x^2}}\frac{\partial}{\partial \tau'}
G_{\beta\lambda}(\sigma',\tau';\sigma,\tau)\Big|_{\tau'=0},
\ee
which in turn yields for Eq. (\rf{3e19}) the expression
\bea \lb{3e32}
& &\frac{\delta}{\delta x^{\beta}(\sigma',0)}
\frac{\delta A}{\delta x^{\alpha}(\sigma,0)}=
2\frac{\partial}{\partial \sigma}\delta(\sigma-\sigma')
\int d\sigma^{\alpha\beta}(y)F(y-x(\sigma,0))\nonumber \\
& &\ \ \ \ \ +\frac{x^{\prime 2}}{\dot x^2}
\frac{\partial}{\partial \tau}\frac{\partial}{\partial \tau'}
G_{\beta\lambda}(\sigma',\tau';\sigma,\tau)\Big|_{\tau'=0,\tau=0}
\Big(\delta_{\lambda\alpha}
-\frac{x^{\prime \lambda}(\sigma,0)x^{\prime \alpha}(\sigma,0)}
{x^{\prime 2}}-\frac{\dot x^{\lambda}(\sigma,0)
\dot x^{\alpha}(\sigma,0)}{\dot x^2}\Big).\nonumber \\
& &
\eea
\par
The symmetry property of the second-order derivative of the minimal
area can now be discussed for the case $\sigma\neq\sigma'$. The Green
function can be decomposed along symmetric Lorentz tensors each
multiplied with a scalar function. Because of the boundary conditions 
(\rf{3e25}) the two derivatives $\frac{\partial}{\partial\tau}$ and 
$\frac{\partial}{\partial\tau'}$ must simultaneously act on at least one 
of the scalar functions in order not to yield zero. In that case the product
of the transverse projecter present in the last term of Eq. (\rf{3e32}) 
with the tensor factors of $G_{\beta\lambda}$ gives a symmetric result,
which ensures the commutativity of the two functional derivatives when
acting at two different points of the contour. That symmetry property is
evident when the functional derivatives act on the Wilson loop average;
then the two operators manifestly commute. This study will be completed 
below by also including the case of coinciding points.
\par
Concrete calculations can be done when the Green function is specified
more explicitly. To this aim, one can try to construct the Green function
by means of an iterative procedure with respect to powers and 
derivatives of the curvature of the minimal surface:
\be \lb{3e33a}
G_{\beta\lambda}(\xi,\eta)=\sum_{i=0}^{\infty}G_{\beta\lambda}^{(i)}
(\xi,\eta),
\ee
where the index $i$ is related to the power or the order of the derivative
of the curvature. The lowest-order term of this iteration would
correspond to the case of a plane with zero curvature. The solution to that
problem can be explicitly constructed. Using the metric (\rf{3e18}), one
can use the \textit{ansatz}
\bea \lb{3e33}
G_{\beta\lambda}^{(0)}(\sigma',\tau';\sigma,\tau)&=&
\bigg\{\delta_{\beta\lambda}(z'.y'\dot z.\dot y-z'.\dot y\dot z.y')
\nonumber \\
& &-\dot z.\dot y(z'_{\lambda}y'_{\beta}-z'_{\beta}y'_{\lambda})
+z'.\dot y(\dot z_{\lambda}y'_{\beta}-\dot z_{\beta}y'_{\lambda})\nonumber \\
& &-z'.y'(\dot z_{\lambda}\dot y_{\beta}-
\dot z_{\beta}\dot y_{\lambda})+\dot z.y'(z'_{\lambda}\dot
y_{\beta}-z'_{\beta}\dot y_{\lambda})\bigg\}
\frac{1}{y^{\prime 2}\dot y^2}h^{(0)}(\sigma',\tau';\sigma,\tau),
\nonumber \\
& &
\eea
where we have defined $z\equiv x(\sigma',\tau')$, $y\equiv x(\sigma,\tau)$
and kept in the decomposition of $G_{\beta\lambda}^{(0)}$ tangential
components in order to satisfy at the end the boundary condition
(\rf{3e17}). The function $h^{(0)}$ is assumed to satisfy the 
inhomogeneous Laplace equation
\bea
\lb{3e34}
& &\Delta_{(\xi)}h^{(0)}(\xi;\eta)=\Delta_{(\eta)}h^{(0)}(\xi;\eta)
=\delta^2(\xi-\eta),\ \ \ \ \ 
\xi=(\sigma',\tau'),\ \ \eta=(\sigma,\tau),\\
\lb{3e35}
& &\Delta_{(\eta)}=\frac{1}{\lambda}\frac{\partial^2}{\partial\sigma^2}
+\lambda\frac{\partial^2}{\partial\tau^2},
\eea
and the boundary conditions 
\be \lb{3e36}
h(\xi,\eta)\Big|_{\xi\in C}=h(\xi,\eta)\Big|_{\eta\in C}=0.
\ee
The coordinates $z$ and $y$ represent points on the background minimal
surface and with parametrization (\rf{3e18}) satisfy the Laplace equations
\be \lb{3e37}
\Delta_{(\xi)}z_{\mu}(\xi)=0,\ \ \ \ \ \Delta_{(\eta)}y_{\mu}(\eta)=0.
\ee
Inserting $G_{\beta\lambda}^{(0)}$ in Eqs. (\rf{3e25}) one can
reconstruct $G_{\beta\lambda}^{(1)}$ by appropriately decomposing it along
Lorentz tensors as in Eq. (\rf{3e33}) and determining its Lorentz scalar
parts through inhomogeneous Laplace equations involving as
inhomogeneous terms $h^{(0)}$ and its first-order derivatives.
It should be noticed that the inhomogeneous
part of Eqs. (\rf{3e25}) is already saturated by the function
$G_{\beta\lambda}^{(0)}$ (the factor $|g|^{-1/2}$ being absorbed in the
definition of the Laplace operator (\rf{3e35})) and therefore the
higher-order functions do no longer have contributions to the
delta-functions. The Lorentz tensor parts of the function
$G_{\beta\lambda}^{(1)}$ contain at least one second-order derivative of
$y$ or $z$ and therefore are of first-order in the curvature of
the surface. The procedure can be continued to higher orders; the
function $G_{\beta\lambda}^{(2)}$ will be quadratic in the second-order
derivatives of $y$ and $z$, or will contain third-order derivatives, and
so forth. On technical grounds, it is advantageous to use also a tensor
notation for the parameters of the surface and to simultaneously
decompose the higher-order functions along tensors (defined essentially
by the derivative operators) in parameter space.
We shall not develop here the above procedure, for we shall not need in
the present paper the explicit expressions of the functions
$G_{\beta\lambda}^{(i)}$ ($i\geq 1$). In many instances, the sole
knowledge of the function $G_{\mu\nu}^{(0)}$ is sufficient to obtain
or to check the main properties of the system under study. This is
in particular the case where the background minimal surface from which
deformations are calculated is itself a plane.
\par
Let us now concentrate on the function $G_{\mu\nu}^{(0)}$. We consider
a four-sided contour $C$ as in Fig. \rf{3f1} with the additional
restriction that the four sides correspond to the lines of equations
$\tau=0$, $\sigma=0$, $\tau=1$, $\sigma=1$. The explicit expression
of the function $h^{(0)}$, satisfying the boundary conditions (\rf{3e36})
is then:
\be \lb{3e38}
h^{(0)}(\sigma',\tau';\sigma,\tau)=-\frac{4}{\pi^2}\sum_{n,m=1}^{\infty}
\frac{\sin(n\pi\sigma')\sin(n\pi\sigma)\sin(m\pi\tau')\sin(m\pi\tau)}
{(m^2\lambda+n^2/\lambda)}.
\ee
Calculating, in view of Eq. (\rf{3e31}), the derivative of $h^{(0)}$
with respect to $\tau'$ at $\tau'=0$, one obtains a series in $m$ that 
can be summed into a hyperbolic function:
\be \lb{3e39}
f^{(0)}(\sigma',0;\sigma,\tau)\equiv -\lambda\frac{\partial h^{(0)}}
{\partial \tau'}\bigg|_{\tau'=0}=2\sum_{n=1}^{\infty}
\frac{\sin(n\pi\sigma')\sin(n\pi\sigma)\sinh\Big(n\pi(1-\tau)/\lambda\Big)}
{\sinh(n\pi/\lambda)}.
\ee
One checks that $\frac{\delta y^{\lambda}}{\delta x^{\beta}}$ satisfies,
with $G_{\beta\lambda}^{(0)}$ and through Eq. (\rf{3e31}), the
boundary condition (\rf{3e17}).
\par
We next evaluate the derivative of
$\frac{\delta y^{\lambda}}{\delta x^{\beta}}$ with respect to $\tau$
at $\tau=0$. The derivative acts on $f^{(0)}$ and on the Lorentz tensor
part of $\frac{\delta y^{\lambda}}{\delta x^{\beta}}$. The latter action
yields in Eq. (\rf{3e32}) the finite part of
$x^{\prime \nu}\frac{\partial}{\partial x^{\beta}}
\int d\sigma^{\alpha\nu}(y)F(y-x(\sigma,0))$ multiplied by
$\delta(\sigma-\sigma')$ coming from the restriction of $f^{(0)}$ on the
contour. One finds for the second-order functional derivative of the 
area the result:
\bea \lb{3e40}
& &\frac{\delta}{\delta x^{\beta}(\sigma',0)}
\frac{\delta A}{\delta x^{\alpha}(\sigma,0)}=
2\frac{\partial}{\partial \sigma}\delta(\sigma-\sigma')
\int d\sigma^{\alpha\beta}(u)F(u-y)\nonumber \\
& &\ \ \ \ \ +2\delta(\sigma-\sigma')y^{\prime\nu}
\frac{\partial}{\partial y^{\beta}}
\int d\sigma^{\alpha\nu}(u)F(u-y)\Big|_{\mathrm{fp}}\nonumber \\
& &\ \ \ \ \ -\sqrt{\frac{y^{\prime 2}}{\dot y^2}}
\bigg\{\delta_{\beta\alpha}(z'.y'\dot z.\dot y-z'.\dot y\dot z.y')
\nonumber \\
& &\ \ \ \ \ -(\dot z.\dot yz'_{\alpha}y'_{\beta}
+z'.y'\dot z_{\alpha}\dot y_{\beta}
-z'.\dot y\dot z_{\alpha}y'_{\beta}
-\dot z.y'z'_{\alpha}\dot y_{\beta})\bigg\}
\frac{1}{y^{\prime 2}\dot y^2}\frac{\partial f^{(0)}(\sigma',0;\sigma,\tau)}
{\partial \tau}\bigg|_{\tau=0},\nonumber \\
& &
\eea
where $y\equiv x(\sigma,0)$, $z\equiv x(\sigma',0)$ and the subscript 
``fp'' means finite part. The last term of Eq. (\rf{3e40})
is invariant under the change of the order of derivations, as can
be seen from the tensor terms and with the use of the defining equations 
(\rf{3e38})-(\rf{3e39}) of $f^{(0)}$, reflecting the general property
emphasized after Eq. (\rf{3e32}). The symmetry property of the contact terms
will be established shortly.
\par
The singular part of $\frac{\partial f^{(0)}(\sigma',0;\sigma,\tau)}
{\partial \tau}\Big|_{\tau=0}$ when $(\sigma-\sigma')$ approaches zero
can be calculated from Eq. (\rf{3e39}). It is:
\be \lb{3e41}
\frac{\partial f^{(0)}(\sigma',0;\sigma,\tau)}{\partial\tau}
{\bigg|_{\tau=0}\ \ }_{\stackrel{{\displaystyle =}}{(\sigma-\sigma')
\rightarrow 0}}\ \frac{1}{\pi}\sqrt{\frac{\dot x^2}{x^{\prime 2}}}
(\sigma-\sigma')^{-2}+O((\sigma-\sigma')^0).
\ee
It yields for the contracted second-order derivative
$\frac{\delta^2 A}{\delta x^{\alpha }(\sigma',0)
\delta x^{\alpha}(\sigma,0)}$, in which the contact terms do not 
contribute, the behavior
\be \lb{3e42}
\frac{\delta^2 A}{\delta x^{\alpha}(\sigma',0)
\delta x^{\alpha}(\sigma,0)}\ _{\stackrel{{\displaystyle =}}
{(\sigma-\sigma')\rightarrow 0}}-\frac{2}{\pi}(\sigma-\sigma')^{-2}
+O((\sigma-\sigma')^0).
\ee
It is in agreement with the leading part of the short-distance behavior
obtained in Ref. \cite{lsw}. The contracted second-order derivative
contains also a logarithmic singularity \cite{lsw} which is obtained in
the present approach by considering the contributions of the next two
nonleading terms in the curvature, i.e., the functions
$G_{\mu\nu}^{(i)}$, $i=1,2$ [Eq. (\rf{3e33a})].
\par
We now check the symmetry property between two successive functional 
derivatives. Let us first consider the problem in the Wilson loop case.
The commutator of two functional derivatives on the contour is, using Eqs.
(\rf{2e3}) and (\rf{2e9a}):
\bea \lb{3e43}
& &\left(\frac{\delta}{\delta x^{\beta}(\sigma')}
\frac{\delta}{\delta x^{\alpha}(\sigma)}-
\frac{\delta}{\delta x^{\alpha}(\sigma)}
\frac{\delta}{\delta x^{\beta}(\sigma')}\right)W(C)=\nonumber \\
& &\ \ \ ig\frac{\partial}{\partial\sigma}\delta(\sigma-\sigma')
\langle F_{\beta\alpha}(x(\sigma))\rangle_W
-ig\frac{\partial}{\partial\sigma'}\delta(\sigma-\sigma')
\langle F_{\alpha\beta}(x(\sigma'))\rangle_W\nonumber \\
& &\ \ \ +ig\delta(\sigma-\sigma')x^{\prime \nu}(\sigma)\langle 
\Big(\nabla_{\beta}F_{\nu\alpha}(x(\sigma))-\nabla_{\alpha}
F_{\nu\beta}(x(\sigma))\Big)\rangle_W.
\eea
Expanding in $F_{\alpha\beta}(x(\sigma'))$ $\sigma'$ about $\sigma$
and using in the last expression the Bianchi identity one finds:
\bea \lb{3e44}
& &\left(\frac{\delta}{\delta x^{\beta}(\sigma')}
\frac{\delta}{\delta x^{\alpha}(\sigma)}-
\frac{\delta}{\delta x^{\alpha}(\sigma)}
\frac{\delta}{\delta x^{\beta}(\sigma')}\right)W(C)\nonumber \\
& &\ \ \ =ig\Big[\Big(\frac{\partial}{\partial\sigma'}
\delta(\sigma'-\sigma)\Big)(\sigma'-\sigma)
+\delta(\sigma'-\sigma)\Big]\frac{\partial}
{\partial \sigma}\langle F_{\beta\alpha}(x(\sigma))\rangle_W\nonumber \\
& &\ \ \ =ig\frac{\partial}{\partial\sigma'}\Big(\delta(\sigma'-\sigma)
(\sigma'-\sigma)\Big)\frac{\partial}{\partial \sigma}
\langle F_{\beta\alpha}(x(\sigma))\rangle_W\nonumber \\
& &\ \ \ =0.
\eea
\par
When the Wilson loop average is saturated by the minimal surface
[Eq. (\rf{2e14})] the left-hand side of Eq. (\rf{3e43}) becomes
proportional to the commutator of the functional derivatives applied
on the minimal area. Using Eq. (\rf{3e40}), one obtains:
\bea \lb{3e45}
& &\frac{1}{2}\left(\frac{\delta}{\delta x^{\beta}(\sigma',0)}
\frac{\delta}{\delta x^{\alpha}(\sigma,0)}-
\frac{\delta}{\delta x^{\alpha}(\sigma,0)}
\frac{\delta}{\delta x^{\beta}(\sigma',0)}\right)A(C)=\nonumber \\
& &\ \ \ \frac{\partial}{\partial\sigma}\delta(\sigma'-\sigma)
\int d\sigma^{\alpha\beta}(y)F(y-x(\sigma,0))
-\frac{\partial}{\partial\sigma'}\delta(\sigma'-\sigma)
\int d\sigma^{\beta\alpha}(y)F(y-x(\sigma',0))\nonumber \\
& &\ \ \ +\delta(\sigma-\sigma')x^{\prime \nu}(\sigma,0)\Big[
\frac{\partial}{\partial x^{\beta}(\sigma,0)}\int d\sigma^{\alpha\nu}(y)
F(y-x(\sigma,0))\nonumber \\
& &\ \ \ \ \ \ \ \ -\frac{\partial}{\partial x^{\alpha}(\sigma,0)}
\int d\sigma^{\beta\nu}(y)F(y-x(\sigma,0))\Big]\Big|_{\mathrm{fp}}.
\eea
The right-hand side has the same structure as that of Eq. (\rf{3e43})
and since the minimal surface satisfies the Bianchi identity (\rf{3e9})
the same operations as above lead to the result
\be \lb{3e46}
\left(\frac{\delta}{\delta x^{\beta}(\sigma',0)}
\frac{\delta}{\delta x^{\alpha}(\sigma,0)}-
\frac{\delta}{\delta x^{\alpha}(\sigma,0)}
\frac{\delta}{\delta x^{\beta}(\sigma',0)}\right)A(C)=0.
\ee
[Actually, it is the finite part of the quantity $\frac{\partial}
{\partial x^{\beta}(\sigma,0)}\int d\sigma^{\alpha\nu}(y)F(y-x(\sigma,0))$
that is relevant for the nontrivial content of the Bianchi identity, the
singular part satisfying the Bianchi identity as an identity, 
independent of the type of the surface.] 
\par
The diagonal constant metric (\rf{3e18}) can also be used for the
study of the forms of minimal surfaces in simple cases. In that metric 
the equation of the minimal surface is given by the Laplace equation
(\rf{3e37}). Choosing a plane as a reference surface, the orthogonal 
deformations of the surface with respect to the plane can be represented
with functions of the type (\rf{3e39}), while the parallel deformations
should be constructed according to the orthonormality conditions 
(\rf{3e18}). However, the sole knowledge of the orthogonal deformations
is sufficient to have a rough idea of the form of the surface. We apply
this study to one particular case of interest.
\par
Let us consider the case where the contour $C$ of Fig \rf{3f1} has been
restricted in the following way: the line $x_1x'_1x'_2x_2$ represents
now the three sides of a rectangle (lying in a plane) while the line
$x_2x_1$, denoted $C_{21}$, is arbitrary but lying in an orthogonal plane
to the rectangle and having in it the equation $u(\sigma,0)=a(\sigma)$
(Fig. \rf{3f2}).
\bfg
\vspace*{0.5 cm}
\bc
\input{3f2.pstex_t}
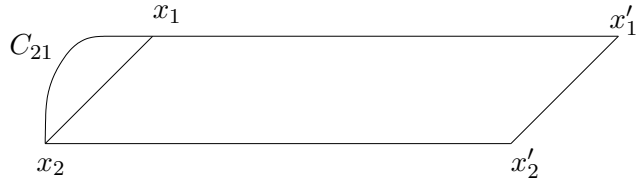
\caption{Contour $C$ made of the three sides $x_1x_1'x_2'x_2$ of a 
rectangle and of the arbitrary line $C_{21}$ lying in an orthogonal 
plane to the rectangle.}
\lb{3f2}
\ec
\efg
\par
Designating by $u(\sigma,\tau)$ the orthogonal deformation of the minimal
surface with respect to the plane, its expression can be obtained from Eq.
(\rf{3e39}) by using the Fourier decomposition of the function $a(\sigma)$:
$a(\sigma)=\sum_{n=1}^{\infty}a_n\sin(n\pi\sigma)$. One obtains:
\be \lb{3e47}
u(\sigma,\tau)=\sum_{n=1}^{\infty}a_n\frac{\sin(n\pi\sigma)
\sinh\Big(n\pi(1-\tau)/\lambda\Big)}{\sinh(n\pi/\lambda)}.
\ee
In general this minimal surface does not have a simple form; $u$ is
different from zero all over the rectangle, but is mainly peaked in
the region $\tau\simeq 0$. Let us next take the limit
$\dot x^2\rightarrow \infty$, or $\lambda\rightarrow 0$. Then $u$ goes
to zero for any $\tau\neq 0$. Therefore, in that limit, the minimal
surface shrinks to the surface made of the union of the rectangle
$x_1x_1'x_2'x_2x_1$ and
of the minimal surface bounded by the straight segment $x_2x_1$ and the
line $C_{21}$. Taking the time interval $T$ proportional to
$\sqrt{\dot x^2}$ in that limit, we see that $T$ is now contained in the
contribution of the rectangle only. Since the energy spectrum is provided
by terms proportional to $T$, we conclude that the energy spectrum
provided by the previous contour is the same as that of the rectangle.
Therefore the shape of the contour $C_{21}$ has no influence on the
energy spectrum and contributes only to the wave functional of the
corresponding bound state, made here of a static quark-antiquark pair.
The meaning of this result is that minimal surfaces, irrespective of 
the shapes of their contours, cannot produce quantum
string-like excitations in the bound state energy spectrum, since 
rectangles do not contain such excitations. This was an expected result,
for the minimal surface is the classical action of the open string
and quantum excitations of it should only be searched for
in fluctuations of surfaces about the classical trajectory \cite{eg,l,lsw};
the present derivation, however, provides another insight into the same 
property.
\par

\section{Normalization of the wave function} \lb{ap2}
\setcounter{equation}{0}

In order to derive the normalization condition satisfied by the wave 
function, we start from Eq. (\rf{6e8}):
\bea \lb{be1}
& &\Big[P_L-(h_{10}+h_{20})\Big]G\nonumber \\
& &\ \ \ \ \ -i\Big[\gamma_{1L}\gamma_1^{\alpha}\int_0^1
d\sigma(1-\sigma)\frac{\delta}{\delta x^{\alpha}(\sigma)}
+\gamma_{2L}\gamma_2^{\beta}\int_0^1d\sigma \sigma
\frac{\delta}{\delta x^{\beta}(\sigma)}\Big]G
\bigg|_{x(\sigma)\in x_1x_2}\nonumber \\
& &\ \ \ \ \ \ \ \ =-i\delta^4(x_1-x_1')\gamma_{1L}\sum_{j=1}^{\infty}G_{0,j}
-i\delta^4(x_2-x_2')\gamma_{2L}\sum_{i=1}^{\infty}G_{i,0}.
\eea
The series $\sum_i G_{i,0}$ and $\sum_j G_{0,j}$ that appear in the 
right-hand side are reminiscent of the series expansions of the 
quark and antiquark propagators in the presence of the gluon field
(Secs. \rf{s4} and \rf{s5}). They have the definitions:
\bea \lb{be2}
\sum_{i=1}^{\infty} G_{i,0}&=&
-\langle\mathrm{tr}_c\,U(x_2,x_1)S_1(x_1,x_1')U(x_1',x_2)\rangle_A,
\nonumber \\
\sum_{j=1}^{\infty} G_{0,j}&=&
\langle\mathrm{tr}_c\,U(x_2,x_1)U(x_1,x_2')S_2(x_2',x_2)\rangle_A.
\eea
They can be interpreted as generalized gauge invariant quark (antiquark)
propagators in the presence of the antiquark (the quark) at position $x_2$
($x_1$). The detailed study of the structure of the series
$\sum_i G_{i,0}$ and $\sum_j G_{0,j}$ by means of the general formula
(\rf{5e4}) shows that they involve surface terms formed by the
propagator lines and the external particle position, which implies that
they contain, in addition to self-energy terms, interaction terms with the
external particles through the functional derivatives of these surfaces.
However, simplifications occur when one considers the equal time limits
$x_{2L}=x_{1L}$ and $x_{2L}'=x_{1L}'$. Because of the presence of the
four-dimensional delta-functions, the corresponding surfaces lie now
in planes orthogonal to the timelike surface lying to infinity and 
containing $x_1x_2$ (or are integrated around such planes). Such types of
surface are generally negligible in front of surfaces that lie to infinity.
In that case the main contributions that arise in the above series are those
coming from the segments representing the boundaries of those surfaces and
producing self-energy corrections. Those are calculated in Sec. \rf{s7}
[Eq. (\rf{7e6})] and are found to be independent of the longitudinal 
momentum. We therefore approximate the right-hand side quantities of Eq.
(\rf{be1}) by the quark propagators with self-energies. We have in momentum
space
\bea \lb{be3}
& &\sum_{i=1}^{\infty} G_{i,0}\simeq -N_cS(p_1,m_1)=-N_c\frac{i}
{\gamma_1.p_1-m_1-\Sigma(p_1^T,m_1)},\nonumber \\
& &\sum_{j=1}^{\infty} G_{0,j}\simeq N_cS(-p_2,m_2)=N_c\frac{i}
{-\gamma_2.p_2-m_2-\Sigma(-p_2^T,m_2)}.
\eea
\par
The equal-time limit amounts to integrating with respect to the realtive
energy variable in momentum space. Defining
\be \lb{be4}
\widetilde G(P,p^T,p^{\prime T})=\int\frac{dp_L}{(2\pi)}\frac{dp_L'}{(2\pi)}
G(P,p,p')
\ee
(after having removed common factors $(2\pi)^4\delta^4(P-P')$), and
integrating Eq. (\rf{be1}) in momentum space with respect to $p_L$ and
$p_L'$ we find:
\be \lb{be5}
\Big[P_L-h_1-h_2-iK\Big]\widetilde G=\frac{i}{2}N_c\left(\frac{h_1}{E_1}
+\frac{h_2}{E_2}\right)\gamma_{1L}\gamma_{2L}
(2\pi)^3\delta^3(p^T-p^{\prime T}).
\ee
In the left-hand side we have designated by $iK$ the kernel that results
from the functional derivatives and which will survive in the bound state
limit; it is essentially represented by the vector potentials calculated
in Sec. \rf{s6}; furthermore, we have extracted from the above
functional derivative contributions, as in Sec. \rf{s7}, the self-energy
parts and included them in the free Dirac hamiltonians $h_1$ and $h_2$
[Eqs. (\rf{7e9})-(\rf{7e11})]. Next, we multiply both sides by
$\widetilde G\gamma_{1L}\gamma_{2L}\frac{1}{2}
(\frac{h_1}{E_1}+\frac{h_2}{E_2})$ and
take the bound state limit, represented in $\widetilde G$ by a pole in
$s\equiv P^2$:
\be \lb{be6}
\widetilde G_{\stackrel{{\displaystyle \simeq}}{s\rightarrow s_0}}
-i\frac{\psi\overline\psi}{s-s_0}.
\ee
Deriving both sides with respect to $s$ and noticing that the left-hand
side operator in Eq. (\rf{be5}) annihilates the bound state wave function
[Eq. (\rf{7e12})], one obtains:
\bea \lb{be7}
& &\psi\Big[\int\frac{d^3p^T}{(2\pi)^3}\mathrm{tr}\psi^{\dagger}(P_L,p^T)
\frac{1}{2}\Big(\frac{h_1}{E_1}+\frac{h_2}{E_2}\Big)\psi(P_L,p^T)\Big]
\overline\psi=\nonumber \\
& &\ \ \ \ \ \ \ \ \ \ 2P_LN_c
\psi\overline\psi\gamma_{1L}\gamma_{2L}\Big[\frac{1}{2}\Big(\frac{h_1}{E_1}
+\frac{h_2}{E_2}\Big)\Big]^2\gamma_{1L}\gamma_{2L}.
\eea
(The trace is over the Dirac spinor indices.) To simplify the
right-hand side, we notice that in the absence of the rotational motion 
of the flux tube, the vector potentials reduce to their longitudinal
components [Eq. (\rf{6e22})]. In that case, the wave function satisfies the 
equation  
\be \lb{be8}
[\epsilon(p_{1L})-\epsilon(p_{2L})]\psi=0,
\ee
similar to the Salpeter equation case \cite{s}, which implies that the
right-hand side operator in Eq. (\rf{be7}),
$\gamma_{1L}\gamma_{2L}\Big[\frac{1}{2}\Big(\frac{h_1}{E_1}
+\frac{h_2}{E_2}\Big)\Big]^2\gamma_{1L}\gamma_{2L}$, can be replaced by 1.
Adopting this approximation, we obtain for the normalization condition:
\be \lb{be9}
\int\frac{d^3p^T}{(2\pi)^3}\mathrm{tr}\psi^{\dagger}\frac{1}{2}
\Big(\frac{h_1}{E_1}+\frac{h_2}{E_2}\Big)\psi=2P_LN_c.
\ee
This formula is the same as that obtained from the Salpeter equation
\cite{s}.
\par

\section{The Breit approximation} \lb{ap3}
\setcounter{equation}{0}

The Breit approximation \cite{b} consists in transforming the wave equation
(\rf{7e12}) into a local equation in $x$-space. This is achieved by first
replacing the energy sign operators appearing in the potentials (\rf{6e19})
by 1; in the nonrelativistic limit this approximation affects only terms of
order $O(1/c^4)$ or higher. Second, the operators $E_a$, $a=1,2$, and
$\mathbf{L}^2/\mathbf{x}^2$ that appear in the potentials are replaced
by their mean values. Third, the self-energy terms are neglected.
The wave equation takes then the form (\rf{6e13}), where 
$h_{a0}$, $a=1,2$, are the free Dirac hamiltonians [Eqs. (\rf{6e7})]
and potentials $A_{a\mu}$, $a=1,2$, have been submitted to the 
approximations described above.
In order not to spoil the property of spontaneous chiral symmetry breakdown
realized by the self-energy parts of the Dirac hamiltonians $h_a$ ($a=1,2$)
[Eqs. (\rf{7e9})] in the massless limit, we replace the total c.m. energy
$P_L$ by $\sqrt{P^2+4E_0^2}$, where $E_0$ is a constant the value of which
is fixed by the requirement that the ground state be massless.
Those approximations do not affect the high 
energy properties of the spectrum, since these are mostly determined
by semi-classical properties (in particular the Regge behavior). 
\par
The resulting equation is now a Breit type
local equation in $\mathbf{x}$ that
can be solved with standard methods. In this form, however, a
difficulty arises coming from the fact that vectorlike local potentials
of the confining type at the classical level do not seem to
confine at the quantum level due to instabilities that appear at large
distances \cite{rnrcfg}. The difficulty, however, is only apparent. The
study of the Pauli--Schr\"odinger equation that results from the Breit
type wave equation shows that the Darwin terms of the equation contain
a repulsive singularity at a finite distance, depending on the energy of
the bound state, which confines the quarks inside a baglike system
with an energy dependent radius and forbids any tunnelling between the
inner domain of the bag and its outer domain. Therefore, confining
normalizable solutions do exist; they are obtained with the boundary
condition that the component $\psi_3$ of the wave function
[Eq. (\rf{8e13})] vanishes at the boundary of the bag. That boundary
condition, breaks, however, implicitly chiral symmetry, since it does not
operate in a symmetric way between chiral partners. This is why, in spite
of the facts that the whole interaction is vectorlike and self-energy
parts have been neglected, the spectrum
displays chiral asymmetry and a massless ground state appears if the
constant $E_0$ is correctly adjusted. 
\par
The existence of that type of confining solutions with vectorlike
potentials was already pointed out by Geffen and Suura \cite{gs}. 
In that work, the masslessness of the ground state is ensured by the 
presence of the short-distance Coulomb-like potential and the adjustment
of the corresponding coupling constant.
\par


\begin{thebibliography}{100}

\bibitem{w}K. G. Wilson, \pr D 10 (1974) 2445.
\bibitem{k}J. B. Kogut, \rmp 55 (1983) 775.
\bibitem{m1}S. Mandelstam, \pr 175 (1968) 1580.
\bibitem{nm}Y. Nambu, \pl 80B (1979) 372.
\bibitem{p}A. M. Polyakov, \np B164 (1979) 171.
\bibitem{mm1}Yu. M. Makeenko and A. A. Migdal, \pl 88B (1979) 135;
97B (1980) 253.
\bibitem{mm2}Yu. M. Makeenko and A. A. Migdal, \np B188 (1981) 269.
\bibitem{mgd}A. A. Migdal, \prp 102 (1983) 199.
\bibitem{mk}Yu. Makeenko, \textit{Large N gauge theories}, hep-th/0001047.
\bibitem{m2}S. Mandelstam, \pr D 19 (1979) 2391.
\bibitem{g}R. Giles, \pr D 24 (1981) 2160.
\bibitem{kkk}V. A. Kazakov and I. K. Kostov, \np B176 (1980) 199;
V. A. Kazakov, \np B179 (1981) 283.
\bibitem{dv}V. S. Dotsenko and S. N. Vergeles, \np B169 (1980) 527.
\bibitem{bnsg}R. A. Brandt, F. Neri and Masa-aki Sato, \pr D 24 (1981) 879;
R. A. Brandt, A. Gocksch, M.-A. Sato and F. Neri, \pr D 26 (1982) 3611.
\bibitem{th1}G. `t Hooft, \np B72 (1974) 461.
\bibitem{ef}E. Eichten and F. Feinberg, \pr D 23 (1981) 2724.
\bibitem{gr}D. Gromes, \zp C 22 (1984) 265; 26 (1984) 401.
\bibitem{bmpbbp}A. Barchielli, E. Montaldi and G. M. Prosperi, \np B296
(1988) 625; B303 (1988) 752 (E);
A. Barchielli, N. Brambilla and G. M. Prosperi, \nc 103 A (1990) 59.
\bibitem{bcpbmp}N. Brambilla, P. Consoli and G. M. Prosperi, \pr D 50
(1994) 5878;
N. Brambilla, E. Montaldi and G. M. Prosperi, \pr D 54 (1996) 3506.
\bibitem{bv1}N. Brambilla and A. Vairo, \textit{Quark confinement and
the hadron spectrum}, hep-ph/9904330.
\bibitem{bp}M. Baldicchi and G. M. Prosperi, \pl B 436 (1998) 145;
\pr D 62 (2000) 114024; D 66 (2002) 074008.
\bibitem{dssdg}H. G. Dosch, \pl B 190 (1987) 177;
Yu. A Simonov, \np B307 (1988) 512;
H. G. Dosch and Yu. A. Simonov, \pl B 205 (1988) 339;
A. Di Giacomo, H. G. Dosch, V. I. Shevchenko and Yu. A. Simonov, \prp
372 (2002) 319.
\bibitem{mwdgopr}J. Maldacena, Adv. Theor. Math. Phys. 2 (1997) 231;
\prl 80 (1998) 4859;
E. Witten, Adv. Theor. Math. Phys. 2 (1998) 505;
N. Drukker, D. G. Gross and H. Ooguri, \pr D 60 (1999) 125006;
A. M. Polyakov and V. S. Rychkov, \np B581 (2000) 116; B594 (2001) 272. 
\bibitem{eg}T. Egushi, \prl 44 (1980) 126.
\bibitem{l}M. L\"uscher, \pl 90B (1980) 277.
\bibitem{lsw}M. L\"uscher, K. Symanzik and P. Weisz, \np B173 (1980) 365.
\bibitem{sb}E. E. Salpeter and H. A. Bethe, \pr 84 (1951) 1232.
\bibitem{gml}M. Gell-Mann and F. Low, \pr 84 (1951) 350.
\bibitem{b}G. Breit, \pr 34 (1929) 553; 36 (1930) 383; 39 (1932) 616.
\bibitem{s}E. E. Salpeter, \pr 87 (1952) 328.
\bibitem{c}R. Courant, \textit{Dirichlet's principle, conformal mapping,
and minimal surfaces}, reprint (Springer-Verlag, New York, 1977).
\bibitem{fschwnwstj}R. P. Feynman, \pr 80 (1950) 440; 84 (1951) 108;
J. Schwinger, \pr 82 (1951) 664;
T. E. Nieuwenhuis, \textit{The Feynman-Schwinger representation of field
theory applied to two-body bound states}, Ph. D. thesis, Utrecht University
(1995), unpublished;
Yu. A. Simonov and J. R. Tjon, \ap 300 (2002) 54.
\bibitem{vwt}G. Veneziano, \np B117 (1976) 519;
E. Witten, \np B160 (1979) 57.
\bibitem{ow}M. G. Olsson and K. Williams, \pr D 48 (1993) 417.
\bibitem{lcooowoo}D. LaCourse and M. G. Olsson, \pr D 39 (1989) 2751;
C. Olson, M. G. Olsson and K. Williams, \pr D 45 (1992) 4307;
C. Olson and M. G. Olsson, \pr D 49 (1994) 4675.
\bibitem{bjl}M. Baker, K. Johnson and B. W. Lee, \pr 133 (1964) B209.
\bibitem{th2}G. `t Hooft, \np B75 (1974) 461.
\bibitem{gsh}I. M. Gel'fand and G. E. Shilov, \textit{Generalized functions},
Vol. I (Academic Press, New York, 1964).
\bibitem{njl}Y. Nambu and G. Jona-Lasinio, \pr 122 (1961) 345; 124 (1961)
246.
\bibitem{fgmw}J. Finger, J. E. Mandula and J. Weyers, \pl 96B (1980) 367;
J. Finger and J. E. Mandula, \np B199 (1982) 168;
J. Govaerts, J. E. Mandula and J. Weyers, \np B237 (1984) 59.
\bibitem{alyopro}A. Amer, A. Le Yaouanc, L. Oliver, O. P\`ene and
J.-C. Reynal, \prl 50 (1983) 87;
A. Le Yaouanc, L. Oliver, O. P\`ene and J.-C. Reynal, \pr D 29 (1984)
1233;
A. Le Yaouanc, L. Oliver, S. Ono, O. P\`ene and J.-C. Reynal, \pr D 31
(31) 87.
\bibitem{ad}S. L. Adler and A. C. Davis, \np B244 (1984) 469.
\bibitem{aa}R. Alkofer and P. A. Amundsen, \np B306 (1988) 305. 
\bibitem{lg}J.-F. Laga\"e, \pr D 45 (1992) 305; 317.
\bibitem{pl}H. D. Politzer, \np B117 (1976) 397.
\bibitem{gmor}M. Gell-Mann, R. J. Oakes and B. Renner, \pr 175 (1968)
2195.
\bibitem{n}S. Narison, \textit{QCD spectral sum rules} (World Scientific,
Singapore, 1989), and references therein.
\bibitem{bm}W. Buchm\"uller, \pl 112B (1982) 479.
\bibitem{hm}A. Huntley and C. Michael, \np B286 (1987) 211.
\bibitem{bsw}G. S. Bali, K Schilling and A. Wachter, \pr D 56 (1997) 2566.
\bibitem{lsg}W. Lucha, F. F. Sch\"oberl and D. Gromes, \prp 200 (1991) 127.
\bibitem{bl}G. S. Bali, \prp 343 (2001) 1.
\bibitem{id}M. Ida, \ptp 59 (1978) 1661.
\bibitem{rnrcfg}M. E. Rose and R. R. Newton, \pr 82 (1951) 470;
B. Ram, Am. J. Phys. 50 (1982) 549;
A. Z. Capri and R. Ferrari, Can. J. Phys. 63 (1985) 1029;
H. Gali\'c, Am. J. Phys. 56 (1988) 312.
\bibitem{gs}D. A. Geffen and H. Suura, \pr D 16 (1977) 3305;
H. Suura, \pr D 17 (1978) 469.

\end{thebibliography}
\end{document}